\pdfoutput=1
\documentclass[prd,showpacs,letterpaper,10pt,aps,notitlepage,nofootinbib,superscriptaddress]{revtex4-1}

\usepackage[utf8]{inputenc}
\usepackage{amsfonts}
\usepackage{amsmath}
\usepackage{graphicx}
\usepackage{hyperref}
\usepackage{placeins}
\usepackage{mathrsfs}
\usepackage{bm}
\usepackage[utf8]{inputenc}
\usepackage{multirow}
\usepackage{slashed}
\usepackage[export]{adjustbox}
\hypersetup{colorlinks, linkcolor = [rgb]{0,0.0,0.75}, citecolor = [rgb]{0,0.0,0.75}, urlcolor = [rgb]{0,0.0,0.75}}

\renewcommand{\tilde}{\widetilde}

\newcommand{\nb}{\phantom{0}}
\newcommand{\wm}{\phantom{-}}
\newcommand{\bs}[1]{\ensuremath{{\boldsymbol{#1}}}}

\begin{document}

\title{\texorpdfstring{$\bm{\Lambda_b \to \Lambda_c^*(2595,2625)\ell^-\bar{\nu}}$}{Lambdab to Lambdac*} form factors from lattice QCD}

\author{Stefan Meinel}
\affiliation{Department of Physics, University of Arizona, Tucson, AZ 85721, USA}

\author{Gumaro Rendon}
\affiliation{Physics Department, Brookhaven National Laboratory, Upton, NY 11973, USA}

\date{April 15, 2021}

\begin{abstract}
We present the first lattice-QCD determination of the form factors describing the semileptonic decays $\Lambda_b \to \Lambda_c^*(2595)\ell^-\bar{\nu}$ and $\Lambda_b \to \Lambda_c^*(2625)\ell^-\bar{\nu}$, where the $\Lambda_c^*(2595)$ and $\Lambda_c^*(2625)$ are the lightest charm baryons with $J^P=\frac12^-$ and $J^P=\frac32^-$, respectively. These decay modes provide new opportunities to test lepton flavor universality and also play an important role in global analyses of the strong interactions in $b\to c$ semileptonic decays. We determine the full set of vector, axial vector, and tensor form factors for both decays, but only in a small kinematic region near the zero-recoil point. The lattice calculation uses three different ensembles of gauge-field configurations with $2+1$ flavors of domain-wall fermions, and we perform extrapolations of the form factors to the continuum limit and physical pion mass. We present Standard-Model predictions for the differential decay rates and angular observables. In the kinematic region considered, the differential decay rate for the $\frac12^-$ final state is found to be approximately 2.5 times larger than the rate for the $\frac32^-$ final state. We also test the compatibility of our form-factor results with zero-recoil sum rules.
\end{abstract}

\maketitle

\FloatBarrier
\section{Introduction}
\FloatBarrier

Semileptonic $b\to c \ell^- \bar{\nu}$ decays are used to determine the CKM matrix element $V_{cb}$ and to search for deviations from lepton flavor universality \cite{Gambino:2020jvv,Bifani:2018zmi,Bernlochner:2021vlv}. They also provide an important testing ground for heavy-quark effective theory \cite{Neubert:1993mb}. In recent years, the operation of the
Large Hadron Collider has provided new opportunities for measurements involving $b$ baryons. The simplest baryonic $b\to c \ell^- \bar{\nu}$ process is $\Lambda_b \to \Lambda_c \ell^- \bar{\nu}$, in which both the initial and final hadrons are the ground states with $J^P=\frac12^+$. This mode has been used in combination with $\Lambda_b \to p \ell^- \bar{\nu}$ to determine $|V_{ub}/V_{cb}|$ \cite{Detmold:2015aaa,Aaij:2015bfa} and offers the prospect of measuring the $\tau$-versus-$\mu$ ratio $R(\Lambda_c)$ and related observables \cite{Bifani:2018zmi,Bernlochner:2021vlv}. The baryonic decays can provide complementary information on physics beyond the Standard Model when compared with mesonic decays \cite{Dutta:2015ueb,Li:2016pdv,Albrecht:2017odf,Datta:2017aue,Alioli:2017ces,Ray:2018hrx,Boer:2019zmp,Penalva:2019rgt,Ferrillo:2019owd,Mu:2019bin,Hu:2020axt}. The $\Lambda_b \to \Lambda_c$ form factors have been computed using lattice QCD \cite{Bowler:1997ej,Gottlieb:2003yb,Detmold:2015aaa,Datta:2017aue}, and the lattice results predict a shape for the $\Lambda_b \to \Lambda_c \mu^- \bar{\nu}$ differential decay rate in the Standard Model that is consistent with the LHCb measurement \cite{Aaij:2017svr}. Heavy-quark symmetry provides strong constraints on $\Lambda_b \to \Lambda_c \mu^- \bar{\nu}$, in which the light hadronic degrees of freedom have total angular momentum zero. In the heavy-quark-effective-theory (HQET) description of this decay, no subleading order-$\Lambda_{\rm QCD}/m_{c,b}$ Isgur-Wise functions occur and only two sub-subleading Isgur-Wise functions enter at order $\Lambda_{\rm QCD}^2/m_c^2$; the available lattice and LHCb results are well described by a fit of this order \cite{Bernlochner:2018kxh,Bernlochner:2018bfn}.

In addition to $\Lambda_b \to \Lambda_c \mu^- \bar{\nu}$, the LHCb detector has also collected (and will continue to collect) large numbers of $\Lambda_b \to \Lambda_c^*(2595) \mu^- \bar{\nu}$ and $\Lambda_b \to \Lambda_c^*(2625) \mu^- \bar{\nu}$ samples \cite{Aaij:2017svr}; the relative branching fractions of these modes have been measured by the CDF Collaboration to be $\mathcal{B}(\Lambda_b \to \Lambda_c^*(2595) \mu^- \bar{\nu})/\mathcal{B}(\Lambda_b \to \Lambda_c \mu^- \bar{\nu})=0.126\pm0.033^{+0.047}_{-0.038}$ and $\mathcal{B}(\Lambda_b \to \Lambda_c^*(2625) \mu^- \bar{\nu})/\mathcal{B}(\Lambda_b \to \Lambda_c \mu^- \bar{\nu})=0.210\pm0.042^{+0.071}_{-0.050}$ \cite{Aaltonen:2008eu}. The $\Lambda_c^*(2595)$ and $\Lambda_c^*(2625)$ are the lightest charm baryons with $J^P=\frac 12^-$ and $J^P=\frac 32^-$, respectively, and are very narrow resonances decaying to $\Lambda_c\pi\pi$ \cite{Zyla:2020zbs}. It has been projected that $R(\Lambda_c^*)=\mathcal{B}(\Lambda_b \to \Lambda_c^* \tau^- \bar{\nu})/\mathcal{B}(\Lambda_b \to \Lambda_c^* \mu^- \bar{\nu})$ can be measured using LHCb data with approximately 17 percent uncertainty at the end of LHC Run 3, and as low as 5 percent uncertainty at the end of Run 6 \cite{Bernlochner:2021vlv}. To predict $R(\Lambda_c^*)$ in the Standard Model and beyond, the $\Lambda_b \to \Lambda_c^*$ form factors are needed. A calculation of these form factors may also improve the control of the backgrounds in a measurement of $R(\Lambda_c)$. Another potential impact will be on zero-recoil sum rules \cite{Mannel:2015osa,Boer:2018vpx} and on global analyses of $b\to c \ell^- \bar{\nu}$ form factors using dispersion relations \cite{Cohen:2019zev}. The authors of Ref.~\cite{Cohen:2019zev} wrote ``Given the large fractional saturation of the unitarity bounds by $\Lambda_b \to \Lambda_c$, the inclusion of $\Lambda_b \to \Lambda_c^*$ could be particularly fruitful once such data is available.'' Finally, we note that there is significant interest in the structure and strong decays of the $\Lambda_c^*(2595)$ and $\Lambda_c^*(2625)$, in part due to the closeness of the $\Sigma_c^{(*)} \pi$ thresholds \cite{Blechman:2003mq,Guo:2016wpy,Arifi:2018yhr,Nieves:2019kdh,Nieves:2019nol}.

In the limit of heavy charm quarks, the light degrees of freedom in $\Lambda_c^*(2595)$ and $\Lambda_c^*(2625)$ have total angular momentum 1 and these two baryons become degenerate. Note that there is no heavy-quark spin-symmetry relation between the $\Lambda_c^*$ and the $\Lambda_c$ due to the different quantum numbers of the light degrees of freedom. This difference also means that the normalization of the leading Isgur-Wise function for $\Lambda_b \to \Lambda_c^*$ remains unconstrained in the heavy-quark limit, and the matrix elements vanish at zero recoil \cite{Roberts:1992xm,Leibovich:1997az}. The HQET relations for the
$\Lambda_b \to \Lambda_c^*(2595)$ and $\Lambda_b \to \Lambda_c^*(2625)$ vector and axial vector form factors including the subleading order-$\Lambda_{\rm QCD}/m_{c,b}$ contributions were derived in Refs.~\cite{Roberts:1992xm}, \cite{Leibovich:1997az}, and \cite{Boer:2018vpx}; the authors of the latter reference specifically studied the possibility of using HQET fits to LHCb data for the muonic decay $\Lambda_b \to \Lambda_c^* \mu^- \bar{\nu}$ to make Standard-Model predictions for $R(\Lambda_c^*)$. It is still an open question how well HQET at this order can describe these transitions.

Quark-model studies of the $\Lambda_b \to \Lambda_c^*(2595)$ and $\Lambda_b \to \Lambda_c^*(2625)$ form factors can be found in Refs.~\cite{Pervin:2005ve,Gutsche:2017wag,Gutsche:2018nks,Becirevic:2020nmb}. In the following, we present the first lattice-QCD determination of these form factors. Our calculation follows
that of the $\Lambda_b \to \Lambda^*(1520)$ form factors in Ref.~\cite{Meinel:2020owd} and uses the same ensembles of gauge-field configurations. We observe that the $\Lambda_c^*(2595)$ and $\Lambda_c^*(2625)$ energy levels for our simulation parameters are below all potential strong-decay thresholds, although they come quite close at the lowest pion mass. As in Ref.~\cite{Meinel:2020owd}, we work in the rest frame of the final-state baryon to avoid mixing between $J=\frac32$ and $J=\frac12$ and between negative and positive parity. This again limits the kinematic coverage to the region near $q^2_{\rm max}$.

Our definitions of the form factors are given in Sec.~\ref{sec:FFdefs}. Following a brief summary of the lattice parameters in Sec.~\ref{sec:latticeparams}, we discuss the baryon interpolating fields, two-point functions, and the results for the masses in Sec.~\ref{sec:twopoint}. The extraction of the form factors from three-point functions is described in Sec.~\ref{sec:threept}, and their extrapolation to the physical pion mass and continuum limit is discussed in Sec.~\ref{sec:FFextrap}. We test the compatibility with zero-recoil sum rules in Sec.~\ref{sec:ZRSR} and present the Standard-Model predictions for $\Lambda_b \to \Lambda_c^*(2595)\ell^-\bar{\nu}$ and $\Lambda_b \to \Lambda_c^*(2625)\ell^-\bar{\nu}$ in Sec.~\ref{sec:observables}. Our conclusions are given in Sec.~\ref{sec:conclusions}, and Appendix \ref{sec:FFrelations} contains relations to other form factor definitions used in the literature.

\FloatBarrier
\section{Definitions of the form factors}
\label{sec:FFdefs}
\FloatBarrier

In the following, we denote the $\Lambda_c^*(2595)$ and $\Lambda_c^*(2625)$ as $\Lambda_{c,1/2}^*$ and $\Lambda_{c,3/2}^*$, respectively. The masses and total decay widths determined by experiments are $m_{\Lambda_{c,1/2}^*}=2592.25(28) \:{\rm MeV}$, $m_{\Lambda_{c,3/2}^*}=2628.11(19) \:{\rm MeV}$, $\Gamma_{\Lambda_{c,1/2}^*}=2.6(0.6) \:{\rm MeV}$, $\Gamma_{\Lambda_{c,3/2}^*}<0.97 \:{\rm MeV}$ (${\rm CL}=90\%$) \cite{Zyla:2020zbs}. We neglect the decay widths throughout this work. In our lattice calculations at heavier-than-physical pion masses, the strong decays are in fact kinematically forbidden, except perhaps at the lightest pion mass; the hadron masses we find on the lattice are given in Sec.~\ref{sec:twopoint}.

We normalize the baryon states as
\begin{eqnarray}
 \langle \Lambda_b(\mathbf{k},r) |  \Lambda_b(\mathbf{p},s)\rangle &=&  \delta_{rs} 2E_{\Lambda_b} (2\pi)^3 \delta^3(\mathbf{k}-\mathbf{p}), \\
 \langle \Lambda_{c,1/2}^*(\mathbf{k}^\prime,r^\prime) |  \Lambda_{c,1/2}^*(\mathbf{p}^\prime,s^\prime)\rangle &=&  \delta_{r^\prime s^\prime} 2E_{\Lambda_{c,1/2}^*} (2\pi)^3 \delta^3(\mathbf{k}^\prime-\mathbf{p}^\prime), \\
 \langle \Lambda_{c,3/2}^*(\mathbf{k}^\prime,r^\prime) |  \Lambda_{c,3/2}^*(\mathbf{p}^\prime,s^\prime)\rangle &=&  \delta_{r^\prime s^\prime} 2E_{\Lambda_{c,3/2}^*} (2\pi)^3 \delta^3(\mathbf{k}^\prime-\mathbf{p}^\prime),
\end{eqnarray}
and work with Dirac and Rarita-Schwinger spinors satisfying
\begin{eqnarray}
 \sum_s u(m_{\Lambda_b},\mathbf{p}, s) \bar{u}(m_{\Lambda_b},\mathbf{p}, s) &=& m_{\Lambda_b} + \slashed{p}, \\
  \sum_{s^\prime} u(m_{\Lambda_{c,1/2}^*},\mathbf{p}^\prime, s^\prime) \bar{u}(m_{\Lambda_{c,1/2}^*},\mathbf{p}^\prime, s^\prime) &=& m_{\Lambda_{c,1/2}^*} + \slashed{p}^\prime, \\
 \nonumber \sum_{s^\prime} u_\mu(m_{\Lambda_{c,3/2}^*},\mathbf{p}^\prime, s^\prime) \bar{u}_\nu(m_{\Lambda_{c,3/2}^*},\mathbf{p}^\prime, s^\prime) &=& -(m_{\Lambda_{c,3/2}^*}+\slashed{p}^\prime)\left(g_{\mu\nu}-\frac13\gamma_\mu\gamma_\nu - \frac{2}{3m_{\Lambda_{c,3/2}^*}^2}p^\prime_\mu p^\prime_\nu - \frac{1}{3m_{\Lambda_{c,3/2}^*}}(\gamma_\mu p^\prime_\nu - \gamma_\nu p^\prime_\mu)\right). \\
\end{eqnarray}
In the equations throughout this paper, Minkowski-space gamma matrices and the metric $(g_{\mu\nu})={\rm diag}(1,-1,-1,-1)$ are used, except where indicated otherwise. We introduce the notation
\begin{eqnarray}
 \langle \Lambda_{c,1/2}^*(\mathbf{p^\prime},s^\prime) |\, \bar{c}\Gamma b \, |  \Lambda_b(\mathbf{p},s) \rangle &=& \bar{u}(m_{\Lambda_{c,1/2}^*},\mathbf{p^\prime}, s^\prime) \:\gamma_5\:\mathscr{G}^{(\frac12^-)}[\Gamma]\:  u(m_{\Lambda_b},\mathbf{p}, s), \label{eq:Gscr12}\\
 \langle \Lambda_{c,3/2}^*(\mathbf{p^\prime},s^\prime) |\, \bar{c}\Gamma b \, |  \Lambda_b(\mathbf{p},s) \rangle &=& \bar{u}_\lambda(m_{\Lambda_{c,3/2}^*},\mathbf{p^\prime}, s^\prime)  \:\mathscr{G}^{\lambda(\frac32^-)}[\Gamma]\:  u(m_{\Lambda_b},\mathbf{p}, s)
\end{eqnarray}
and 
\begin{equation}
 s_\pm = (m_{\Lambda_b}\pm m_{\Lambda_c^*})^2 - q^2,
\end{equation}
where $q=p-p^\prime$. We use a helicity basis for all form factors. For the $J^P=\frac12^-$ final state, our definition follows the one introduced previously for $J^P=\frac12^+$ final states \cite{Feldmann:2011xf} except for the changes resulting from the opposite parity [note the $\gamma_5$ in Eq.~(\ref{eq:Gscr12})]:
\begin{eqnarray}
\nonumber \mathscr{G}^{(\frac12^-)}[\gamma^\mu] &=& f_0^{(\frac{1}{2}^-)}\: (m_{\Lambda_b}+m_{\Lambda_{c,1/2}^*})\frac{q^\mu}{q^2} \\
 \nonumber && + f_+^{(\frac{1}{2}^-)}\frac{m_{\Lambda_b}-m_{\Lambda_{c,1/2}^*}}{s_-}\left( p^\mu + p^{\prime \mu} - (m_{\Lambda_b}^2-m_{\Lambda_{c,1/2}^*}^2)\frac{q^\mu}{q^2}  \right) \\
 && + f_\perp^{(\frac{1}{2}^-)} \left(\gamma^\mu + \frac{2m_{\Lambda_{c,1/2}^*}}{s_-} p^\mu - \frac{2\,m_{\Lambda_b}}{s_-} p^{\prime \mu} \right),
\end{eqnarray}
\begin{eqnarray}
\nonumber \mathscr{G}^{(\frac12^-)}[\gamma^\mu\gamma_5] &=& -g_0^{(\frac{1}{2}^-)}\gamma_5\: (m_{\Lambda_b}-m_{\Lambda_{c,1/2}^*})\frac{q^\mu}{q^2} \\
 \nonumber && - g_+^{(\frac{1}{2}^-)}\gamma_5 \frac{m_{\Lambda_b}+m_{\Lambda_{c,1/2}^*}}{s_+}\left( p^\mu + p^{\prime \mu} - (m_{\Lambda_b}^2-m_{\Lambda_{c,1/2}^*}^2)\frac{q^\mu}{q^2}  \right) \\
 && - g_\perp^{(\frac{1}{2}^-)}\gamma_5 \left(\gamma^\mu - \frac{2m_{\Lambda_{c,1/2}^*}}{s_+} p^\mu - \frac{2\,m_{\Lambda_b}}{s_+} p^{\prime \mu} \right),
\end{eqnarray}
\begin{eqnarray}
\nonumber \mathscr{G}^{(\frac12^-)}[i\sigma^{\mu\nu} q_\nu] &=&  -h_+^{(\frac{1}{2}^-)} \, \frac{q^2}{s_-} \left( p^\mu + p^{\prime \mu} -  (m_{\Lambda_b}^2-m_{\Lambda_{c,1/2}^*}^2) \frac{q^\mu}{q^2} \right) \\
 &&   - h_\perp^{(\frac{1}{2}^-)}\,  (m_{\Lambda_b}-m_{\Lambda_{c,1/2}^*}) \left( \gamma^\mu +  \frac{2\,m_{\Lambda_{c,1/2}^*}}{s_-} \, p^\mu - \frac{2\,m_{\Lambda_b}}{s_-} \, p^{\prime \mu}  \right),
\end{eqnarray}
\begin{eqnarray}
\nonumber \mathscr{G}^{(\frac12^-)}[i\sigma^{\mu\nu}\gamma_5 q_\nu] &=& -\widetilde{h}_+^{(\frac{1}{2}^-)}\gamma_5 \frac{q^2}{s_+} \left( p^\mu + p^{\prime \mu} - (m_{\Lambda_b}^2-m_{\Lambda_{c,1/2}^*}^2)\frac{q^\mu}{q^2} \right) \\
 &&  - \widetilde{h}_\perp^{(\frac{1}{2}^-)}\gamma_5\, (m_{\Lambda_b}+m_{\Lambda_{c,1/2}^*}) \left( \gamma^\mu -  \frac{2  m_{\Lambda_{c,1/2}^*}}{s_+} \, p^\mu - \frac{2m_{\Lambda_b}}{s_+} \, p^{\prime \mu}   \right).
\end{eqnarray}
For the $J^P=\frac32^-$ final state, we use the definition introduced by us in Ref.~\cite{Meinel:2020owd}, which reads
\begin{eqnarray}
\nonumber \mathscr{G}^{\lambda(\frac32^-)}[\gamma^\mu] &=&
 f_0^{(\frac{3}{2}^-)} \frac{ m_{\Lambda_{c,3/2}^*}}{s_+}\,\frac{(m_{\Lambda_b}-m_{\Lambda_{c,3/2}^*})\,p^\lambda q^\mu}{q^2}   \\
\nonumber &&  + f_+^{(\frac{3}{2}^-)} \frac{m_{\Lambda_{c,3/2}^*}}{s_-} \,\frac{(m_{\Lambda_b}+m_{\Lambda_{c,3/2}^*})\, p^\lambda ( q^2(p^\mu+p^{\prime \mu}) - (m_{\Lambda_b}^2-m_{\Lambda_{c,3/2}^*}^2) q^\mu   )}{q^2\, s_+}   \\
 \nonumber && + f_\perp^{(\frac{3}{2}^-)} \frac{m_{\Lambda_{c,3/2}^*}}{s_-} \left(p^\lambda \gamma^\mu - \frac{2\, p^\lambda(m_{\Lambda_b}p^{\prime \mu} + m_{\Lambda_{c,3/2}^*} p^\mu)}{s_+}    \right)   \\
 && + f_{\perp^\prime}^{(\frac{3}{2}^-)} \frac{m_{\Lambda_{c,3/2}^*}}{s_-} \left( p^\lambda \gamma^\mu - \frac{2\, p^\lambda p^{\prime \mu}}{m_{\Lambda_{c,3/2}^*}}
 + \frac{2\, p^\lambda(m_{\Lambda_b}p^{\prime \mu} + m_{\Lambda_{c,3/2}^*} p^\mu  )}{s_+} + \frac{s_-\,  g^{\lambda\mu}}{m_{\Lambda_{c,3/2}^*}}\right), \label{eq:Ggmu}
\end{eqnarray}
\begin{eqnarray}
\nonumber \mathscr{G}^{\lambda(\frac32^-)}[\gamma^\mu\gamma_5] &=&
  - g_0^{(\frac{3}{2}^-)}\gamma_5\,\frac{m_{\Lambda_{c,3/2}^*}}{s_-}\frac{(m_{\Lambda_b}+m_{\Lambda_{c,3/2}^*})\,p^\lambda q^\mu}{q^2}   \\
\nonumber &&  - g_+^{(\frac{3}{2}^-)}\gamma_5\,\frac{m_{\Lambda_{c,3/2}^*}}{s_+}\frac{(m_{\Lambda_b}-m_{\Lambda_{c,3/2}^*})\, p^\lambda ( q^2(p^\mu+p^{\prime \mu}) - (m_{\Lambda_b}^2-m_{\Lambda_{c,3/2}^*}^2) q^\mu   )}{q^2\, s_-}   \\
 \nonumber && - g_\perp^{(\frac{3}{2}^-)}\gamma_5 \frac{m_{\Lambda_{c,3/2}^*}}{s_+}\left(p^\lambda \gamma^\mu - \frac{2\, p^\lambda(m_{\Lambda_b}p^{\prime \mu} - m_{\Lambda_{c,3/2}^*} p^\mu  )}{s_-}    \right)   \\
 && - g_{\perp^\prime}^{(\frac{3}{2}^-)}\gamma_5 \frac{m_{\Lambda_{c,3/2}^*}}{s_+}\left(p^\lambda \gamma^\mu + \frac{2\, p^\lambda p^{\prime \mu}}{m_{\Lambda_{c,3/2}^*}}
 + \frac{2\, p^\lambda(m_{\Lambda_b}p^{\prime \mu}  - m_{\Lambda_{c,3/2}^*} p^\mu )}{s_-} - \frac{s_+\,  g^{\lambda\mu}}{m_{\Lambda_{c,3/2}^*}}\right),
\end{eqnarray}
\begin{eqnarray}
\nonumber \mathscr{G}^{\lambda(\frac32^-)}[i\sigma^{\mu\nu}q_\nu] &=&
 - h_+^{(\frac{3}{2}^-)}\frac{m_{\Lambda_{c,3/2}^*}}{s_-}\,\frac{ p^\lambda ( q^2(p^\mu+p^{\prime \mu}) - (m_{\Lambda_b}^2-m_{\Lambda_{c,3/2}^*}^2) q^\mu   )}{s_+}   \\
 \nonumber && - h_\perp^{(\frac{3}{2}^-)}\frac{m_{\Lambda_{c,3/2}^*}}{s_-} (m_{\Lambda_b}+m_{\Lambda_{c,3/2}^*}) \left(p^\lambda \gamma^\mu - \frac{2\, p^\lambda(m_{\Lambda_b}p^{\prime \mu} + m_{\Lambda_{c,3/2}^*} p^\mu)}{s_+}    \right)   \\
 \nonumber && - h_{\perp^\prime}^{(\frac{3}{2}^-)}\frac{m_{\Lambda_{c,3/2}^*}}{s_-} (m_{\Lambda_b}+m_{\Lambda_{c,3/2}^*}) \left( p^\lambda \gamma^\mu - \frac{2\, p^\lambda p^{\prime \mu}}{m_{\Lambda_{c,3/2}^*}}
 + \frac{2\, p^\lambda(m_{\Lambda_b}p^{\prime \mu} + m_{\Lambda_{c,3/2}^*} p^\mu  )}{s_+} + \frac{s_-\,  g^{\lambda\mu}}{m_{\Lambda_{c,3/2}^*}}\right), \\
\end{eqnarray}
\begin{eqnarray}
\nonumber \mathscr{G}^{\lambda(\frac32^-)}[i\sigma^{\mu\nu}q_\nu\gamma_5] &=&
 - \tilde{h}_+^{(\frac{3}{2}^-)}\gamma_5\frac{m_{\Lambda_{c,3/2}^*}}{s_+}\,\frac{ p^\lambda ( q^2(p^\mu+p^{\prime \mu}) - (m_{\Lambda_b}^2-m_{\Lambda_{c,3/2}^*}^2) q^\mu   )}{s_-}   \\
 \nonumber && - \tilde{h}_\perp^{(\frac{3}{2}^-)}\gamma_5\frac{m_{\Lambda_{c,3/2}^*}}{s_+} (m_{\Lambda_b}-m_{\Lambda_{c,3/2}^*}) \left(p^\lambda \gamma^\mu - \frac{2\, p^\lambda(m_{\Lambda_b}p^{\prime \mu} - m_{\Lambda_{c,3/2}^*} p^\mu)}{s_-}    \right)   \\
  \nonumber && - \tilde{h}_{\perp^\prime}^{(\frac{3}{2}^-)}\gamma_5\frac{m_{\Lambda_{c,3/2}^*}}{s_+} (m_{\Lambda_b}-m_{\Lambda_{c,3/2}^*}) \left( p^\lambda \gamma^\mu + \frac{2\, p^\lambda p^{\prime \mu}}{m_{\Lambda_{c,3/2}^*}}
 + \frac{2\, p^\lambda(m_{\Lambda_b}p^{\prime \mu} - m_{\Lambda_{c,3/2}^*} p^\mu  )}{s_-} - \frac{s_+\,  g^{\lambda\mu}}{m_{\Lambda_{c,3/2}^*}}\right). \\ \label{eq:Gsmunug5}
\end{eqnarray}
Only the vector and axial-vector form factors are needed to describe $\Lambda_b \to \Lambda_c^*\ell^-\bar{\nu}$ decays in the Standard Model, but we also compute the tensor form factors. Above, $\sigma^{\mu\nu}=\frac{i}{2}(\gamma^\mu\gamma^\nu-\gamma^\nu\gamma^\mu)$. Note that the overall sign of the form factors for each decay mode depends on the phase conventions of the states. This means that also the relative overall sign between the two different final states is left undetermined. Relations between our form-factor definitions and alternative definitions used in the literature are given in Appendix \ref{sec:FFrelations}.

\FloatBarrier
\section{Lattice actions and parameters}
\label{sec:latticeparams}
\FloatBarrier

The lattice actions and parameters used in this work are the same as in our calculation of $\Lambda_b \to \Lambda^*(1520)$ form factors \cite{Meinel:2020owd}, except that here the valence strange quark is replaced by a valence charm quark. For the latter, we employ the same form of action and analogous tuning conditions as for the bottom quark \cite{Aoki:2012xaa}, i.e., an anisotropic clover action with bare parameters $a m_Q^{(c)}$, $\nu^{(c)}$, $c_{E,B}^{(c)}$ tuned to obtain the correct $D_s$ meson kinetic mass, rest mass, and hyperfine splitting (our notation for the bare parameters follows Ref.~\cite{Brown:2014ena}, while Ref.~\cite{Aoki:2012xaa} uses $m_0=m_Q$, $\zeta=\nu$, $c_P=c_E=c_B$). The values of these parameters are given in Table \ref{tab:latticeparams}. The gauge-field ensembles with $2+1$ flavors of domain-wall fermions were generated by the RBC and UKQCD Collaborations \cite{Aoki:2010dy, Blum:2014tka}. For the up and down valence quarks, we reuse the domain-wall propagators computed for Ref.~\cite{Meinel:2020owd}. Our computation utilizes all-mode averaging \cite{Blum:2012uh,Shintani:2014vja}, in which unbiased estimates with small statistical uncertainties are obtained at reduced cost by combining ``exact'' and ``sloppy'' samples.

\begin{table}[h]
 \begin{tabular}{lccccccccccccccc}
\hline\hline
Label & $N_s^3\times N_t$ & $\beta$  & $a$ [fm] & $am_{u,d}$ & $am_s$ 
& $a m_Q^{(b)}$ & $\nu^{(b)}$ & $c_{E,B}^{(b)}$  & $\wm a m_Q^{(c)}$ & $\nu^{(c)}$ & $c_{E,B}^{(c)}$    & $N_{\rm ex}$ & $N_{\rm sl}$ \\
\hline
C01  & $24^3\times64$ & $2.13$   & $0.1106(3)$  & $0.01\nb$  & $0.04$ & $7.3258$ & $3.1918$ & $4.9625$ & $\wm0.1541$ & $1.2004$ & $1.8407$  & 283 & 9056  \\
C005 & $24^3\times64$ & $2.13$   & $0.1106(3)$  & $0.005$    & $0.04$ & $7.3258$ & $3.1918$ & $4.9625$ & $\wm0.1541$ & $1.2004$ & $1.8407$  & 311 & 9952  \\
F004 & $32^3\times64$ & $2.25$   & $0.0828(3)$  & $0.004$    & $0.03$ & $3.2823$ & $2.0600$ & $2.7960$ & $-0.0517$   & $1.1021$ & $1.4483$  & 251 & 8032  \\
\hline\hline
\end{tabular}
\caption{\label{tab:latticeparams} Parameters of the lattice actions, lattice spacings, and numbers of exact (ex) and sloppy (sl) samples computed for the correlation functions. The light-quark and gluon actions and the determination of the lattice spacings are described in Refs.~\cite{Aoki:2010dy, Blum:2014tka}. The form of the heavy-quark action is given in Ref.~\cite{Aoki:2012xaa}, where $m_0=m_Q$, $\zeta=\nu$, $c_P=c_E=c_B$.}
\end{table}

\FloatBarrier
\section{Two-point functions and hadron masses}
\label{sec:twopoint}
\FloatBarrier

\begin{table}
	\begin{tabular}{lcccccccccccccc} \hline \hline 
    & \multicolumn{4}{c}{Up and down quarks} & \hspace{2ex} & \multicolumn{4}{c}{Bottom quarks} & \hspace{2ex} & \multicolumn{4}{c}{Charm quarks} \\
               & $N_\textrm{Gauss}$ & $\sigma_\textrm{Gauss}/a$ & $N_\textrm{APE}$ & $\alpha_\textrm{APE}$ && $N_\textrm{Gauss}$ & $\sigma_\textrm{Gauss}/a$ & $N_\textrm{Stout}$ & $\rho_\textrm{Stout}$  && $N_\textrm{Gauss}$ & $\sigma_\textrm{Gauss}/a$ & $N_\textrm{Stout}$ & $\rho_\textrm{Stout}$  \\ \hline
    Coarse     & $\phantom{0}30$ & $4.350$ & $25$ & $2.5$           && $10$ & $2.000$ & $10$ & $0.08$    && $20$ & $3.000$ & $10$ & $0.08$ \\
    Fine       & $\phantom{0}60$ & $5.728$ & $25$ & $2.5$           && $10$ & $2.000$ & $10$ & $0.08$    && $20$ & $3.000$ & $10$ & $0.08$ \\ \hline \hline
	\end{tabular}
	\caption{\label{tab:smearingparams}Paramters of the quark-field smearing used in the baryon interpolating fields. See Ref.~\cite{Meinel:2020owd} for explanations.}
\end{table}

We now move to the discussion of the baryon interpolating fields, two-point functions, and results for the masses. For the $\Lambda_b$, everything is identical to Ref.~\cite{Meinel:2020owd}. The $\Lambda_c^*(2625)$ has the same isospin and spin-parity quantum numbers as the $\Lambda^*(1520)$ ($I=0$, $J^P=\frac32^-$), but with a charm quark instead of a strange quark. We therefore use the interpolating field
\begin{equation}
 (O_{\Lambda_c^*})_{j\gamma} = \epsilon^{abc}\:(C\gamma_5)_{\alpha\beta}\Big(\frac{1+\gamma_0}{2}\Big)_{\gamma\delta}\left[ \tilde{c}^a_\alpha\:\tilde{d}^b_\beta\: (\tilde{\nabla}_j \tilde{u})^c_\delta  -  \tilde{c}^a_\alpha\:\tilde{u}^b_\beta\: (\tilde{\nabla}_j \tilde{d})^c_\delta  + \tilde{u}^a_\alpha\:(\tilde{\nabla}_j \tilde{d})^b_\beta\: \tilde{c}^c_\delta - \tilde{d}^a_\alpha\:(\tilde{\nabla}_j \tilde{u})^b_\beta\: \tilde{c}^c_\delta  \right], \label{eq:Lambda2}
\end{equation}
which differs from Eq.~(18) of Ref.~\cite{Meinel:2020owd} only by the replacement $s\to c$. As before, this form will work only at zero momentum. The tilde indicates gauge-covariant Gaussian smearing of the quark fields with the parameters given in Table \ref{tab:smearingparams}. The field (\ref{eq:Lambda2}) actually has nonzero overlap with both the $\Lambda_c^*(2595)$ and the $\Lambda_c^*(2625)$,
\begin{eqnarray}
 \langle 0 | (O_{\Lambda_c^*})_j  | \Lambda_{c,1/2}^*(\mathbf{0},s^\prime) \rangle &=& Z_{\Lambda_{c,1/2}^*}\:\frac{1+\gamma_0}{2} \,\gamma_j \gamma_5\,  u(m_{\Lambda_{c,1/2}^*}, \mathbf{0}, s^\prime), \\
 \langle 0 | (O_{\Lambda_c^*})_j  | \Lambda_{c,3/2}^*(\mathbf{0},s^\prime) \rangle &=& Z_{\Lambda_{c,3/2}^*}\:\frac{1+\gamma_0}{2}\, u_j(m_{\Lambda_{c,3/2}^*}, \mathbf{0}, s^\prime),
\end{eqnarray}
and we can isolate the $J=\frac12$ and $J=\frac32$ components\footnote{At zero momentum, the continuum $J^P=\frac12^-$ and $J^P=\frac32^-$ irreducible representations subduce identically to the $G_{1}^{u}$ and $H^{u}$ irreducible representations of the double-cover of the cubic group \cite{Johnson:1982yq}; the next-higher values of $J^P$ that subduce to the same cubic irreps are $\frac72^-$ and $J^P=\frac52^-$, respectively, and such states will have higher energies. It is therefore safe to refer to only the continuum quantum numbers in this case.} using the projectors
\begin{eqnarray}
 P^{kj}_{(1/2)} &=& \frac13\gamma^k\gamma^j, \\
 P^{kj}_{(3/2)} &=& g^{kj}-\frac13\gamma^k\gamma^j.
\end{eqnarray}
The zero-momentum $\Lambda_c^*$ two-point functions are defined like those for the $\Lambda^*$ in Ref.~\cite{Meinel:2020owd}, and after applying the above projectors their spectral decomposition reads
\begin{eqnarray}
\nonumber P^{jl}_{(1/2)} C^{(2,\Lambda_c^*)}_{lk}(t)  &=&  -\frac12  Z_{\Lambda_{c,1/2}^*}^2 (1+\gamma_0)\, \gamma^j\gamma_k \, e^{-m_{\Lambda_{c,1/2}^*} t} \\
&& + \:\:(\text{excited-state contributions}), \\
\nonumber P^{jl}_{(3/2)} C^{(2,\Lambda_c^*)}_{lk}(t)  &=&  -\frac12 Z_{\Lambda_{c,3/2}^*}^2 (1+\gamma_0) \left(g^j_{\:\:k}-\frac13\gamma^j\gamma_k \right) \, e^{-m_{\Lambda_{c,3/2}^*} t} \\
&& + \:\:(\text{excited-state contributions}).
\end{eqnarray}
At this point the reader may wonder why we did not analyze the $\Lambda^*(1405)$ with $J^P=\frac12^-$ in Ref.~\cite{Meinel:2020owd}, despite being able to project to $J^P=\frac12^-$ with the available data. The reason is that we do not trust the single-hadron/narrow-width approximation for the $\Lambda^*(1405)$, which has a larger decay width than the $\Lambda^*(1520)$ and likely a two-pole structure \cite{Oller:2000fj}.

The masses extracted from single-exponential fits to our results for $P^{jl}_{(1/2)} C^{(2,\Lambda_c^*)}$ and $P^{jl}_{(3/2)} C^{(2,\Lambda_c^*)}$ in the plateau regions are given in Table \ref{tab:hadronmasses}, along with the masses of potential decay products. The latter are not used in our determination of the form factors but are included to assess whether the $\Lambda_c^*$ baryons are stable under the strong interactions for our quark masses. We find that both $m_{\Lambda_{c,1/2}^*}$ and $m_{\Lambda_{c,3/2}^*}$ are lower than all of the following: $m_{\Lambda_c}+m_\pi+m_\pi$, $m_{\Sigma_c}+m_\pi$, $m_D+m_N$, although the difference $m_{\Lambda_{c,3/2}^*}-m_{\Sigma_c}-m_\pi$ becomes consistent with zero for the F004 ensemble within the statistical uncertainties. The results are of course affected by the finite volume to some degree, but it appears likely that both the $\Lambda_{c,1/2}^*$ and the $\Lambda_{c,3/2}^*$ are stable hadrons at least on the C01 and C005 ensembles, where the energies are well below all thresholds.

\begin{figure}
 \includegraphics[width=0.47\linewidth]{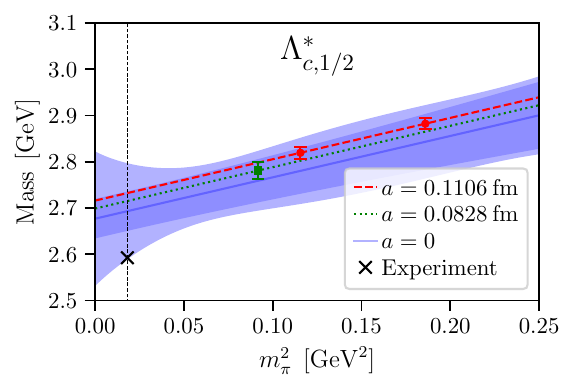} \hfill \includegraphics[width=0.47\linewidth]{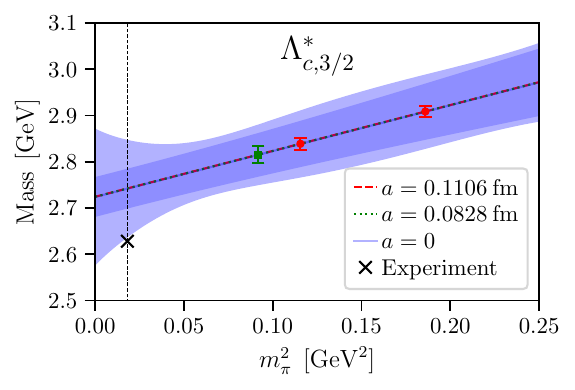}
 \caption{\label{fig:massesextrap}Chiral and continuum extrapolations of our results for the $\Lambda_{c,1/2}^*$ and $\Lambda_{c,3/2}^*$ masses. The inner error bands are statistical only and the outer bands include estimates of the systematic uncertainties associated with these extrapolations. The experimental values from Ref.~\cite{Zyla:2020zbs} are also shown.}
\end{figure}

We also performed simple chiral-continuum extrapolations of $m_{\Lambda_{c,1/2}^*}$ and $m_{\Lambda_{c,3/2}^*}$ of the form
\begin{equation}
 m_{\Lambda_{c,J}^*} = m_{\Lambda_{c,J}^*}^{(\rm phys)} \left[1+c_J\:\frac{m_{\pi}^2-m_{\pi,\rm phys}^2}{(4\pi f_{\pi})^2}+d_J\: a^2 \Lambda^2 \right]
\end{equation}
with fit parameters $m_{\Lambda_{c,J}^*}^{(\rm phys)}$, $c_J$, $d_J$, and constants $f_{\pi}=132\,\text{MeV}$, $\Lambda=300\,\text{MeV}$. These fits yield $ m_{\Lambda_{c,1/2}^*}^{(\rm phys)}=2693(43)$ MeV, $ m_{\Lambda_{c,3/2}^*}^{(\rm phys)}=2742(43)$ MeV. To estimate systematic uncertainties associated with the choice of fit model, we additionally performed higher-order fits of the form 
\begin{equation}
 m_{\Lambda_{c,J}^*} = m_{\Lambda_{c,J}^*,{\rm HO}}^{(\rm phys)} \left[1+c_{J,{\rm HO}}\:\frac{m_{\pi}^2-m_{\pi,\rm phys}^2}{(4\pi f_{\pi})^2}+h_{J,{\rm HO}}\:\frac{m_{\pi}^3-m_{\pi,\rm phys}^3}{(4\pi f_{\pi})^3}+d_{J,{\rm HO}}\: a^2 \Lambda^2  +g_{J,{\rm HO}}\: a^3 \Lambda^3 \right],
\end{equation}
with Gaussian priors $h_{J,{\rm HO}}=0\pm 10$ and $g_{J,{\rm HO}}=0\pm 10$, and computed the systematic uncertainties using
\begin{equation}
 \sigma_{m,{\rm syst}} = {\rm max}\left( |m_{\rm HO}-m|,\: \sqrt{|\sigma_{m,{\rm HO}}^2-\sigma_m^2|}  \right), \label{eq:sigmasystm}
\end{equation}
where $m$, $\sigma_m$ denote the central value and uncertainty obtained using the parameter values and covariance matrix of the nominal fit and $m_{\rm HO}$, $\sigma_{m,{\rm HO}}^2$ denote the central value and uncertainty obtained using the parameter values and covariance matrix of the higher-order fit. In this way we finally obtain
\begin{eqnarray}
 m_{\Lambda_{c,1/2}^*}^{(\rm phys)}&=& (2693 \pm 43_{\,\rm stat} \pm 95_{\,\rm syst})\:{\rm MeV}, \\
 m_{\Lambda_{c,3/2}^*}^{(\rm phys)}&=& (2742 \pm 43_{\,\rm stat} \pm 96_{\,\rm syst})\:{\rm MeV},
\end{eqnarray}
which are consistent with the experimental values of $m_{\Lambda_{c,1/2}^*}=2592.25(28) \:{\rm MeV}$, $m_{\Lambda_{c,3/2}^*}=2628.11(19) \:{\rm MeV}$ \cite{Zyla:2020zbs}. Plots of the extrapolations are shown in Fig.~\ref{fig:massesextrap}. Note that we do not use the chiral-continuum extrapolations of the baryon masses in our determination of the form factors; we use the lattice baryon masses when computing the form factors on each ensemble, and then extrapolate only the form factors themselves. The mass extrapolations merely provide a test of our methodology. Finally, in Table \ref{tab:hadronmasses} we also list the hyperfine splittings $ m_{\Lambda_{c,3/2}^*}-m_{\Lambda_{c,1/2}^*}$ computed on each ensemble. Their relative uncertainties are too large to obtain a useful chiral-continuum extrapolation, but the results are consistent within  $<2\sigma$ with the experimental value of $35.86(34)$ MeV on each ensemble.

\begin{table}
 \begin{tabular}{lccccccccc}
\hline\hline
Label & $m_\pi$  & $m_D$   & $m_N$   & $m_{\Lambda_c}$ & $m_{\Sigma_c}^{(\rm est)}$   & $m_{\Lambda_{c,1/2}^*}$ & $m_{\Lambda_{c,3/2}^*}$ & $m_{\Lambda_{c,3/2}^*}-m_{\Lambda_{c,1/2}^*}$ & $m_{\Lambda_b}$ \\
\hline
C01  & $0.4312(13)$    & $1.9119(54)$   & $1.2647(51)$   &  $2.4652(82)$         & $2.617(10)$      &  $2.882(12)$                  &  $2.909(12)$      & $0.0265(85)$           & $5.793(17)$ \\
C005 & $0.3400(11)$    & $1.8942(54)$   & $1.1649(58)$   &  $2.4038(75)$         & $2.565(12)$      &  $2.819(13)$                  &  $2.839(13)$      & $0.0185(97)$           & $5.726(17)$ \\
F004 & $0.3030(12)$    & $1.8880(70)$   & $1.1197(59)$   &  $2.367(12)\nb$       & $2.550(19)$      &  $2.781(18)$                  &  $2.815(18)$      & $0.033(17)\nb$            & $5.722(23)$ \\
\hline\hline
\end{tabular}
\caption{\label{tab:hadronmasses} Hadron masses in GeV. We did not compute $\Sigma_c$ two-point functions in this work and the $\Sigma_c$ masses were estimated by adding the $\Sigma_c-\Lambda_c$ mass differences computed in Ref.~\cite{Brown:2014ena} on the same ensembles with a slightly different tuning of the charm-quark action to the $\Lambda_c$ masses computed here.}
\end{table}

\FloatBarrier
\section{Three-point functions and form factors}
\label{sec:threept}
\FloatBarrier

As in Ref.~\cite{Meinel:2020owd}, we compute forward and backward three-point functions
\begin{eqnarray}
C^{(3,{\rm fw})}_{j\,\gamma\,\delta}(\mathbf{p},\Gamma, t, t^\prime) &=& \sum_{\mathbf{y},\mathbf{z}} e^{-i\mathbf{p}\cdot(\mathbf{y}-\mathbf{z})} \left\langle (O_{\Lambda_c^*})_{j\gamma}(x_0,\mathbf{x})\:\: J_\Gamma(x_0-t+t^\prime,\mathbf{y})\:\: (\overline{O_{\Lambda_b}})_{\delta} (x_0-t,\mathbf{z}) \right\rangle, \label{eq:threeptfw} \\
C^{(3,\mathrm{bw})}_{j\,\delta\,\gamma}(\mathbf{p},\Gamma, t, t-t^\prime) &=& \sum_{\mathbf{y},\mathbf{z}}
e^{-i\mathbf{p}\cdot(\mathbf{z}-\mathbf{y})} \Big\langle (O_{\Lambda_b})_{\delta}(x_0+t,\mathbf{z})\:\: J_\Gamma^\dag(x_0+t^\prime,\mathbf{y})
\:\: (\overline{O_{\Lambda_c^*}})_{j\gamma} (x_0,\mathbf{x}) \Big\rangle, \label{eq:threeptbw}
\end{eqnarray}
where $\mathbf{p}$ is the $\Lambda_b$ momentum, $\Gamma$ is the Dirac matrix in the $b\to c$ weak current, $t$ is the source-sink separation, and $t^\prime$ is the current-insertion time. With both the $b$ and $c$ quarks implemented using anisotropic clover actions, the current now includes $\mathcal{O}(a)$-improvement terms for both quarks:
\begin{equation}
 J_\Gamma=\rho_\Gamma\sqrt{Z_V^{(cc)} Z_V^{(bb)}} \left[ \bar{c}\: \Gamma\: b + a\, d_1^{(b)}\,\bar{c}\: \Gamma\: \bs{\gamma}_{\rm E}\cdot\overrightarrow{\bs{\nabla}}  b - a\, d_1^{(c)}\,\bar{c}\,\overleftarrow{\bs{\nabla}}\cdot\bs{\gamma}_{\rm E}\: \Gamma\:   b  \right]. \label{eq:JGamma}
\end{equation}
Here, $\bs{\gamma}_{\rm E}=(\gamma_{\rm E}^j)=(-i\gamma^j)$ are the Euclidean spatial gamma matrices, and $\overrightarrow{\bs{\nabla}}$ are the gauge-covariant symmetric lattice derivatives. The overall matching factors in the current are written as $\rho_\Gamma\sqrt{Z_V^{(cc)} Z_V^{(bb)}}$ \cite{Hashimoto:1999yp, ElKhadra:2001rv}, where $Z_V^{(QQ)}$ are the matching factors for the flavor-conserving temporal vector currents $\bar{Q}\gamma^0 Q$. We determined the values of $Z_V^{(QQ)}$ nonperturbatively using the charge-conservation condition for three-point functions with $D_s$ and $B_s$ meson interpolating fields; the results are given in Table \ref{tab:matching}. With this choice, the residual matching factors $\rho_\Gamma$ are equal to 1 at tree level and can be computed in perturbation theory without introducing large uncertainties. For the vector and axial-vector currents, we use the one-loop results given in Table III of Ref.~\cite{Detmold:2015aaa}. Here we use more accurately tuned parameters in the $b$- and $c$-quark actions, but we expect the resulting change in the matching factors to be negligible. For the tensor currents, one-loop results are not presently available so we set $\rho_{\sigma_{\mu\nu}}=1$ and estimate the resulting systematic uncertainty at $\mu=m_b$ to be 4.04\% as in Ref.~\cite{Datta:2017aue}. The values of the $\mathcal{O}(a)$-improvement coefficients for all currents are also computed at tree level and are given in Table \ref{tab:matching}.

\begin{table}
 \begin{tabular}{llccccc}
\hline\hline
       & & $Z_V^{(bb)}$  & $Z_V^{(cc)}$      & $d_1^{(b)}$ & $d_1^{(c)}$  \\
\hline
Coarse lattice (C01, C005) & & $9.0631(84)$  & $1.35761(16)\nb$  & $0.0728$   & $0.0412$     \\
Fine lattice (F004)   & & $4.7449(21)$  & $1.160978(74)$    & $0.0696$    & $0.0301$        \\
\hline\hline
\end{tabular}
\caption{\label{tab:matching} The values of the nonperturbative matching factors $Z_V^{(bb)}$ and $Z_V^{(cc)}$, determined using charge-conservation from ratios of zero-momentum $B_s$ and $D_s$ two-point and three-point functions, as well as the values of the $\mathcal{O}(a)$-improvement coefficients, computed at tree level in mean-field-improved perturbation theory.}
\end{table}

We generated data for the same two choices of $\Lambda_b$ momenta as in Ref.~\cite{Meinel:2020owd}, $\mathbf{p}=(0,0,2)\frac{2\pi}{L}$ and $\mathbf{p}=(0,0,3)\frac{2\pi}{L}$, and for slightly larger source-sink separations: $t/a=6...14$ at the coarse lattice spacing and $t/a=8...16$ at the fine lattice spacing. Here we project the $\Lambda_c^*$ field in the three-point functions to both $J=\frac12$ and $J=\frac32$, and the spectral decompositions read

\begin{eqnarray}
\nonumber P_{(1/2)}^{jl} \:  C^{(3,{\rm fw})}_l(\mathbf{p},\Gamma, t, t^\prime) &=&  \frac{1}{v^0} Z_{\Lambda_{c,1/2}^*} \frac{1+\gamma_0}{2}  \gamma^j\:\mathscr{G}^{(\frac12^-)}[\Gamma]\:  \frac{1+\slashed{v}}{2} (Z_{\Lambda_b}^{(1)}+Z_{\Lambda_b}^{(2)}\gamma^0)  \:e^{-m_{\Lambda_{c,1/2}^*} (t-t^\prime)} e^{-E_{\Lambda_b}t^\prime} \\
&& + \:\:(\text{excited-state contributions}), \\
\nonumber P_{(3/2)}^{jl} \:  C^{(3,{\rm fw})}_l(\mathbf{p},\Gamma, t, t^\prime) &=&  -\frac{1}{v^0}  Z_{\Lambda_{c,3/2}^*} \frac{1+\gamma_0}{2} \left(g^j_{\:\:\lambda}-\frac13\gamma^j\gamma_\lambda - \frac{1}{3}\gamma^j g_{0\lambda} \right) \:\mathscr{G}^{\lambda(\frac32^-)}[\Gamma]\: \frac{1+\slashed{v}}{2} (Z_{\Lambda_b}^{(1)}+Z_{\Lambda_b}^{(2)}\gamma^0) \\
\nonumber && \times\:e^{-m_{\Lambda_{c,3/2}^*} (t-t^\prime)} e^{-E_{\Lambda_b}t^\prime} \\
&& + \:\:(\text{excited-state contributions}),
\end{eqnarray}
where $v^\mu=p^\mu/m_{\Lambda_b}$, and $\mathscr{G}^{(\frac12^-)}[\Gamma]$, $\mathscr{G}^{\lambda(\frac32^-)}[\Gamma]$ contain the form factors as explained in Sec.~\ref{sec:FFdefs}.

In the following, we introduce a label $X\in\{V, A, TV, TA\}$ denoting the type of weak current, such that the matrix $\Gamma$ in Eq.~(\ref{eq:JGamma}) is equal to
\begin{equation}
 \Gamma_X^\mu = \left\{\begin{array}{ll}  \gamma^\mu & {\rm for}\:\:X=V, \\
                                          \gamma^\mu\gamma_5 & {\rm for}\:\:X=A, \\
                                          i\sigma^{\mu\nu}q_\nu  & {\rm for}\:\:X=TV, \\
                                          i\sigma^{\mu\nu}q_\nu\gamma_5  & {\rm for}\:\:X=TA.
                       \end{array}\right.
\end{equation}
We also introduce a label $\lambda \in \{0, +, \perp, \perp^\prime\}$ for the different helicities. As in Ref.~\cite{Meinel:2020owd}, we compute the quantities
\begin{align}
F^{(J^P)X}_\lambda(\mathbf{p},t)=\frac{S^{(J^P)X}_\lambda(\mathbf{p},t,t/2)}{S^{(J^P)X_{\rm ref}}_{\lambda_{\rm ref}}(\mathbf{p},t,t/2)}\sqrt{R_{\lambda_{\rm ref}}^{(J^P)X_{\rm ref}}(\mathbf{p})}, \label{eq:newratiomethod}
\end{align}
where $J^P\in\{\frac12^-,\frac32^-\}$ are the quantum numbers of the $\Lambda_c^*$. Here, $R_{\lambda_{\rm ref}}^{(J^P)X_{\rm ref}}(\mathbf{p})$ denotes a constant fit at large $t$ to a ratio $R_{\lambda_{\rm ref}}^{(J^P)X_{\rm ref}}(\mathbf{p}, t)$ of three-point and two-point functions that is constructed such that at large $t$ it becomes equal to the square of the form factor associated with current $X_{\rm ref}$ and helicity $\lambda_{\rm ref}$. The quantities $S^{(J^P)X}_\lambda(\mathbf{p},t,t/2)$ are linear projections of the three-point functions proportional to the form factor with current $X$ and helicity $\lambda$. In this way, the relative signs of the form factors are preserved, and $F^{(J^P)X}_\lambda(\mathbf{p},t)$ becomes equal to the form factor of interest at large $t$, which is then extracted from a constant fit. The choice of reference form factor $(X_{\rm ref},\lambda_{\rm ref})$ is arbitrary in principle, and we select it based on the signal-to-noise ratio and quality of the ground-state plateau.

The equations for $J^P=\frac32^-$ were given in Ref.~\cite{Meinel:2020owd} and we do not repeat them here. For $J^P=\frac12^-$, the construction of $R_{\lambda}^{(\frac12^-)X}(\mathbf{p}, t)$ is similar to that used previously for $J^P=\frac12^+$ in Refs.~\cite{Detmold:2015aaa,Detmold:2016pkz}. We define
\begin{eqnarray}
\mathscr{R}_{0}^{(\frac12^-)X}(\mathbf{p},t,t^\prime) &=& \frac{ q_\mu \: q_\nu \:
\mathrm{Tr}\Big[   \gamma_l\,P_{(1/2)}^{li} \:  C^{(3,{\rm fw})}_i(\mathbf{p},\:\Gamma_X^\mu, t, t^\prime) \:(1+\slashed{v})\:  C^{(3,{\rm bw})}_n (\mathbf{p},\:\Gamma_X^\nu, t, t-t^\prime)P_{(1/2)}^{nm}\,\gamma_m  \Big] }
{\mathrm{Tr}\Big[ P^{jk}_{(1/2)} C^{(2,\Lambda_c^*)}_{kj}(t) \Big]\mathrm{Tr}\Big[(1+\slashed{v})\: C^{(2,\:\Lambda_b)}(\mathbf{p},t)  \Big]}, \\
\nonumber \mathscr{R}_{+}^{(\frac12^-)X}(\mathbf{p},t,t^\prime) &=& r_\mu[(1,\mathbf{0})] \: r_\nu[(1,\mathbf{0})] \\
&& \times \frac{ \mathrm{Tr}\Big[   \gamma_l\,P_{(1/2)}^{li} \:  C^{(3,{\rm fw})}_i(\mathbf{p},\:\Gamma_X^\mu, t, t^\prime) \:(1+\slashed{v})\:    C^{(3,{\rm bw})}_n (\mathbf{p},\:\Gamma_X^\nu, t, t-t^\prime)P_{(1/2)}^{nm}\,\gamma_m  \Big] }
{\mathrm{Tr}\Big[ P^{jk}_{(1/2)} C^{(2,\Lambda_c^*)}_{kj}(t) \Big]\mathrm{Tr}\Big[(1+\slashed{v})\: C^{(2,\:\Lambda_b)}(\mathbf{p},t)  \Big]},  \\
\nonumber \mathscr{R}_{\perp}^{(\frac12^-)X}(\mathbf{p},t,t^\prime) &=& r_\mu[(0,\mathbf{e}_j\times \mathbf{p})] \:   r_\nu[(0,\mathbf{e}_k\times \mathbf{p})] \\
&& \times \frac{\mathrm{Tr}\Big[  \gamma_l\,P_{(1/2)}^{li} \:  C^{(3,{\rm fw})}_i(\mathbf{p},\:\Gamma_X^\mu, t, t^\prime) \gamma_5 \gamma^j \:(1+\slashed{v})\:    C^{(3,{\rm bw})}_n (\mathbf{p},\:\Gamma_X^\nu, t, t-t^\prime)P_{(1/2)}^{nm}\,\gamma_m \gamma_5 \gamma^k  \Big] }
{\mathrm{Tr}\Big[ P^{jk}_{(1/2)} C^{(2,\Lambda_c^*)}_{kj}(t) \Big]\mathrm{Tr}\Big[(1+\slashed{v})\: C^{(2,\:\Lambda_b)}(\mathbf{p},t)  \Big]}, 
\end{eqnarray}
where
\begin{equation}
r[n]=n-\frac{(q\cdot n)}{q^2}q
\end{equation}
for any four-vector $n$, and $\mathbf{e}_j$ denotes the three-dimensional unit vector in direction $j$. Above, repeated Greek indices are summed over from 0 to 3, while Latin indices are summed only over the spatial directions. The ratios $\mathscr{R}_{\lambda}^{(\frac12^-)X}(\mathbf{p},t,t^\prime)$ are equal to  kinematic factors depending on the baryon energies times the squares of individual helicity form factors, up to excited-state contamination that decays exponentially for $t$ and $t-t^\prime$ both large. We then set $t^\prime=t/2$ [or average over $(t+a)/2$ and $(t-a)/2$ in the case of odd $t/a$] and divide out the kinematic factors to obtain
\begin{eqnarray}
 {R_0^{(\frac12^-)V}(\mathbf{p}, t)} &=& \frac{4\,E_{\Lambda_b}}{3 (m_{\Lambda_b} + m_{\Lambda_{c,1/2}^*})^2 (E_{\Lambda_b}-m_{\Lambda_b}) }  \:\mathscr{R}_{0}^{(\frac12^-)V}(\mathbf{p},t,t/2)  \:\: \nonumber\\
 &=&\:\: [f_0^{(\frac12^-)}]^2 \:+\: (\text{excited-state contributions}), \hspace{2ex} \label{eq:R0V} \\\nonumber\\
 {R_+^{(\frac12^-)V}(\mathbf{p}, t)} &=&  \frac{4\, E_{\Lambda_b}\, q^4}{3 (E_{\Lambda_b}+m_{\Lambda_b})^2 (E_{\Lambda_b}-m_{\Lambda_b})(m_{\Lambda_b} - m_{\Lambda_{c,1/2}^*})^2 }   \:\mathscr{R}_{+}^{(\frac12^-)V}(\mathbf{p},t,t/2) \:\:\nonumber\\
 &=&\:\: [f_+^{(\frac12^-)}]^2 \:+\: (\text{excited-state contributions}), \hspace{2ex} \label{eq:RplusV} \\\nonumber\\
 {R_\perp^{(\frac12^-)V}(\mathbf{p}, t)} &=&  \frac{E_{\Lambda_b}}{3 (E_{\Lambda_b}+m_{\Lambda_b})^2 (E_{\Lambda_b}-m_{\Lambda_b}) }     \:\mathscr{R}_{\perp}^{(\frac12^-)V}(\mathbf{p},t,t/2)  \:\:\nonumber\\
 &=&\:\: [f_\perp^{(\frac12^-)}]^2 \:+\: (\text{excited-state contributions}),
 \end{eqnarray}
 \begin{eqnarray}
 {R_0^{(\frac12^-)A}(\mathbf{p}, t)} &=&    \frac{4\,E_{\Lambda_b}}{3 (m_{\Lambda_b} - m_{\Lambda_{c,1/2}^*})^2 (E_{\Lambda_b}+m_{\Lambda_b}) }     \:\mathscr{R}_{0}^{(\frac12^-)A}(\mathbf{p},t,t/2) \:\: \nonumber\\
 &=&\:\: [g_0^{(\frac12^-)}]^2 \:+\: (\text{excited-state contributions}), \hspace{2ex} \label{eq:R0A} \\\nonumber\\
 {R_+^{(\frac12^-)A}(\mathbf{p}, t)} &=&  \frac{4\, E_{\Lambda_b}\, q^4}{3 (E_{\Lambda_b}-m_{\Lambda_b})^2 (E_{\Lambda_b}+m_{\Lambda_b})(m_{\Lambda_b} + m_{\Lambda_{c,1/2}^*})^2 }     \:\mathscr{R}_{+}^{(\frac12^-)A}(\mathbf{p},t,t/2) \:\:\nonumber\\
 &=&\:\: [g_+^{(\frac12^-)}]^2 \:+\: (\text{excited-state contributions}), \hspace{2ex} \\\nonumber\\
 {R_\perp^{(\frac12^-)A}(\mathbf{p}, t)} &=&   -\frac{E_{\Lambda_b}}{3 (E_{\Lambda_b}-m_{\Lambda_b})^2 (E_{\Lambda_b}+m_{\Lambda_b}) }    \:\mathscr{R}_{\perp}^{(\frac12^-)A}(\mathbf{p},t,t/2)  \:\:\nonumber\\
 &=&\:\: [g_\perp^{(\frac12^-)}]^2 \:+\: (\text{excited-state contributions}),
 \end{eqnarray}
 \begin{eqnarray}
  {R_+^{(\frac12^-)TV}(\mathbf{p}, t)} &=&   \frac{4\, E_{\Lambda_b}}{3 (E_{\Lambda_b}+m_{\Lambda_b})^2 (E_{\Lambda_b}-m_{\Lambda_b}) }   \:\mathscr{R}_{+}^{(\frac12^-)TV}(\mathbf{p},t,t/2) \:\:\nonumber\\
 &=&\:\: [h_+^{(\frac12^-)}]^2 \:+\: (\text{excited-state contributions}), \hspace{2ex} \\\nonumber\\
 {R_\perp^{(\frac12^-)TV}(\mathbf{p}, t)} &=&    \frac{E_{\Lambda_b}}{3 (E_{\Lambda_b}+m_{\Lambda_b})^2 (E_{\Lambda_b}-m_{\Lambda_b})(m_{\Lambda_b} - m_{\Lambda_{c,1/2}^*})^2 }  \:\mathscr{R}_{\perp}^{(\frac12^-)TV}(\mathbf{p},t,t/2)  \:\:\nonumber\\
 &=&\:\: [h_\perp^{(\frac12^-)}]^2 \:+\: (\text{excited-state contributions}),
 \end{eqnarray}
 \begin{eqnarray}
 {R_+^{(\frac12^-)TA}(\mathbf{p}, t)} &=&   \frac{4\, E_{\Lambda_b}}{3 (E_{\Lambda_b}-m_{\Lambda_b})^2 (E_{\Lambda_b}+m_{\Lambda_b}) }  \:\mathscr{R}_{+}^{(\frac12^-)TA}(\mathbf{p},t,t/2) \:\:\nonumber\\
 &=&\:\: [\tilde{h}_+^{(\frac12^-)}]^2 \:+\: (\text{excited-state contributions}), \hspace{2ex} \\\nonumber \\
 {R_\perp^{(\frac12^-)TA}(\mathbf{p}, t)} &=& -\frac{E_{\Lambda_b}}{3 (E_{\Lambda_b}-m_{\Lambda_b})^2 (E_{\Lambda_b}+m_{\Lambda_b})(m_{\Lambda_b} + m_{\Lambda_{c,1/2}^*})^2 }   \:\mathscr{R}_{\perp}^{(\frac12^-)TA}(\mathbf{p},t,t/2)  \:\:\nonumber\\
 &=&\:\: [\tilde{h}_\perp^{(\frac12^-)}]^2 \:+\: (\text{excited-state contributions}). \label{eq:RperprimeTA}
\end{eqnarray}
The linear projections of the three-point functions are constructed using
\begin{align}
 {\mathscr{S}_\lambda^{(\frac12^-)V,TV}(\mathbf{p},t,t^\prime)} &= \mathrm{Tr}\Big[ M^{(\lambda)}_{\mu j} P_{(1/2)}^{jl} \:  C^{(3,{\rm fw})}_l(\mathbf{p},\Gamma_{V,TV}^\mu, t, t^\prime) \:\: \frac{(1+\slashed{v})}{2} \Big] , \\
 {\mathscr{S}_\lambda^{(\frac12^-)A,TA}(\mathbf{p},t,t^\prime)} &= \mathrm{Tr}\Big[\gamma_5 M^{(\lambda)}_{\mu j} P_{(1/2)}^{jl} \:  C^{(3,{\rm fw})}_l(\mathbf{p},\Gamma_{A,TA}^\mu, t, t^\prime) \:\: \frac{(1+\slashed{v})}{2} \Big] ,
\end{align}
where
\begin{align}
M^{(0)}_{\mu j}&=\epsilon_{\mu}^{(0)} \epsilon_{j}^{(0)}, \\
M^{(+)}_{\mu j}&=\epsilon_{\mu}^{(+)} \epsilon_{j}^{(0)}, \\
M^{(\perp)}_{\mu j}&=\sum^3_{i=1}\epsilon_{\mu}^{(\perp,i)} \epsilon_{j}^{(\perp,i)},
\end{align}
with the polarization vectors
\begin{equation}
 \epsilon^{(0)} = (\,q^0,\: \mathbf{q}\,), \hspace{3ex}
 \epsilon^{(+)} = (\,|\mathbf{q}|,\: (q^0/|\mathbf{q}|)\mathbf{q}\,), \hspace{3ex}
 \epsilon^{(\perp,\,j)} = (\,0,\: \mathbf{e}_j \times \mathbf{q}\,).
\end{equation}
To improve the signals, we use the average of the forward three-point function and the Dirac adjoint of the backward three-point function instead of just $C^{(3,{\rm fw})}$. We then divide out appropriate kinematic factors to obtain
\begin{align}
  {S_0^{(\frac12^-)V}(\mathbf{p},t,t^\prime)}&=-\frac{ E_{\Lambda_b} m_{\Lambda_b}}{ (E_{\Lambda_b}-m_{\Lambda_b}) (E_{\Lambda_b}+m_{\Lambda_b}) (m_{\Lambda_b}+m_{\Lambda_{c,1/2}^*})}\:{\mathscr{S}_0^{(\frac12^-)V}(\mathbf{p},t,t^\prime)}\cr
 &=f_0^{(\frac{1}{2}^-)}  \: Z_{\Lambda_{c,1/2}^*} ( Z^{(1)}_{\Lambda_b} m_{\Lambda_b} + Z^{(2)}_{\Lambda_b} E_{\Lambda_b} ) e^{-m_{\Lambda_{c,1/2}^*}(t-t')} e^{-E_{\Lambda_b} t'} \cr
 &\hspace{3ex}+ \text{(excited-state contributions)}, \\ \nonumber \\
 {S_+^{(\frac12^-)V}(\mathbf{p},t,t^\prime)}&=-\frac{E_{\Lambda_b} m_{\Lambda_b}}{ (E_{\Lambda_b}-m_{\Lambda_b})^{1/2} (E_{\Lambda_b}+m_{\Lambda_b})^{3/2} (m_{\Lambda_b}-m_{\Lambda_{c,1/2}^*})}\:{\mathscr{S}_+^{(\frac12^-)V}(\mathbf{p},t,t^\prime)}\cr
 &=f_+^{(\frac{1}{2}^-)}  \: Z_{\Lambda_{c,1/2}^*} ( Z^{(1)}_{\Lambda_b} m_{\Lambda_b} + Z^{(2)}_{\Lambda_b} E_{\Lambda_b} ) e^{-m_{\Lambda_{c,1/2}^*}(t-t')} e^{-E_{\Lambda_b} t'} \cr
 &\hspace{3ex}+ \text{(excited-state contributions)},  \\ \nonumber \\
  {S_\perp^{(\frac12^-)V}(\mathbf{p},t,t^\prime)}&=-\frac{E_{\Lambda_b} m_{\Lambda_b}}{ 2(E_{\Lambda_b}-m_{\Lambda_b}) (E_{\Lambda_b}+m_{\Lambda_b})^2}\:{\mathscr{S}_\perp^{(\frac12^-)V}(\mathbf{p},t,t^\prime)}\cr
 &=f_{\perp}^{(\frac{1}{2}^-)}  \: Z_{\Lambda_{c,1/2}^*} ( Z^{(1)}_{\Lambda_b} m_{\Lambda_b} + Z^{(2)}_{\Lambda_b} E_{\Lambda_b} ) e^{-m_{\Lambda_{c,1/2}^*}(t-t')} e^{-E_{\Lambda_b} t'} \cr
 &\hspace{3ex}+ \text{(excited-state contributions)}, 
\end{align}
\begin{align}
  {S_0^{(\frac12^-)A}(\mathbf{p},t,t^\prime)}&=-\frac{ E_{\Lambda_b} m_{\Lambda_b}}{ (E_{\Lambda_b}-m_{\Lambda_b}) (E_{\Lambda_b}+m_{\Lambda_b}) (m_{\Lambda_b}-m_{\Lambda_{c,1/2}^*})}\:{\mathscr{S}_0^{(\frac12^-)A}(\mathbf{p},t,t^\prime)}\cr
 &=g_0^{(\frac{1}{2}^-)}  \: Z_{\Lambda_{c,1/2}^*} ( Z^{(1)}_{\Lambda_b} m_{\Lambda_b} + Z^{(2)}_{\Lambda_b} E_{\Lambda_b} ) e^{-m_{\Lambda_{c,1/2}^*}(t-t')} e^{-E_{\Lambda_b} t'} \cr
 &\hspace{3ex}+ \text{(excited-state contributions)},  \\ \nonumber \\
 {S_+^{(\frac12^-)A}(\mathbf{p},t,t^\prime)}&=-\frac{E_{\Lambda_b} m_{\Lambda_b}}{ (E_{\Lambda_b}-m_{\Lambda_b})^{3/2} (E_{\Lambda_b}+m_{\Lambda_b})^{1/2} (m_{\Lambda_b}+m_{\Lambda_{c,1/2}^*})}\:{\mathscr{S}_+^{(\frac12^-)A}(\mathbf{p},t,t^\prime)}\cr
 &=g_+^{(\frac{1}{2}^-)}  \: Z_{\Lambda_{c,1/2}^*} ( Z^{(1)}_{\Lambda_b} m_{\Lambda_b} + Z^{(2)}_{\Lambda_b} E_{\Lambda_b} ) e^{-m_{\Lambda_{c,1/2}^*}(t-t')} e^{-E_{\Lambda_b} t'} \cr
 &\hspace{3ex}+ \text{(excited-state contributions)},  \\ \nonumber \\
  {S_\perp^{(\frac12^-)A}(\mathbf{p},t,t^\prime)}&=\frac{E_{\Lambda_b} m_{\Lambda_b}}{ 2(E_{\Lambda_b}-m_{\Lambda_b})^2 (E_{\Lambda_b}+m_{\Lambda_b})}\:{\mathscr{S}_\perp^{(\frac12^-)A}(\mathbf{p},t,t^\prime)}\cr
 &=g_{\perp}^{(\frac{1}{2}^-)}  \: Z_{\Lambda_{c,1/2}^*} ( Z^{(1)}_{\Lambda_b} m_{\Lambda_b} + Z^{(2)}_{\Lambda_b} E_{\Lambda_b} ) e^{-m_{\Lambda_{c,1/2}^*}(t-t')} e^{-E_{\Lambda_b} t'} \cr
 &\hspace{3ex}+ \text{(excited-state contributions)},  
\end{align}
\begin{align}
 {S^{(\frac12^-)TV}_+(\mathbf{p},t,t^\prime)}&=\frac{E_{\Lambda_b} m_{\Lambda_b}}{ (E_{\Lambda_b}-m_{\Lambda_b})^{1/2} (E_{\Lambda_b}+m_{\Lambda_b})^{3/2} q^2}\:{\mathscr{S}_+^{(\frac12^-)TV}(\mathbf{p},t,t^\prime)}\cr
 &=h_+^{(\frac{1}{2}^-)}  \: Z_{\Lambda_{c,1/2}^*} ( Z^{(1)}_{\Lambda_b} m_{\Lambda_b} + Z^{(2)}_{\Lambda_b} E_{\Lambda_b} ) e^{-m_{\Lambda_{c,1/2}^*}(t-t')} e^{-E_{\Lambda_b} t'} \cr
 &\hspace{3ex}+ \text{(excited-state contributions)},  \\ \nonumber \\
  {S^{(\frac12^-)TV}_\perp(\mathbf{p},t,t^\prime)}&=\frac{E_{\Lambda_b} m_{\Lambda_b}}{ 2(E_{\Lambda_b}-m_{\Lambda_b}) (E_{\Lambda_b}+m_{\Lambda_b})^2 (m_{\Lambda_b}-m_{\Lambda_{c,1/2}^*}) }\:{\mathscr{S}^{(\frac12^-)TV}_\perp(\mathbf{p},t,t^\prime)}\cr
 &=h_{\perp}^{(\frac{1}{2}^-)}  \: Z_{\Lambda_{c,1/2}^*} ( Z^{(1)}_{\Lambda_b} m_{\Lambda_b} + Z^{(2)}_{\Lambda_b} E_{\Lambda_b} ) e^{-m_{\Lambda_{c,1/2}^*}(t-t')} e^{-E_{\Lambda_b} t'} \cr
 &\hspace{3ex}+ \text{(excited-state contributions)}, 
\end{align}
\begin{align}
 {S^{(\frac12^-)TA}_+(\mathbf{p},t,t^\prime)}&=-\frac{E_{\Lambda_b} m_{\Lambda_b}}{ (E_{\Lambda_b}+m_{\Lambda_b})^{1/2} (E_{\Lambda_b}-m_{\Lambda_b})^{3/2} q^2}\:{\mathscr{S}_+^{(\frac12^-)TA}(\mathbf{p},t,t^\prime)}\cr
 &=\widetilde{h}_+^{(\frac{1}{2}^-)}  \: Z_{\Lambda_{c,1/2}^*} ( Z^{(1)}_{\Lambda_b} m_{\Lambda_b} + Z^{(2)}_{\Lambda_b} E_{\Lambda_b} ) e^{-m_{\Lambda_{c,1/2}^*}(t-t')} e^{-E_{\Lambda_b} t'} \cr
 &\hspace{3ex}+ \text{(excited-state contributions)},  \\ \nonumber \\
  {S^{(\frac12^-)TA}_\perp(\mathbf{p},t,t^\prime)}&=\frac{E_{\Lambda_b} m_{\Lambda_b}}{ 2(E_{\Lambda_b}+m_{\Lambda_b}) (E_{\Lambda_b}-m_{\Lambda_b})^2 (m_{\Lambda_b}+m_{\Lambda_{c,1/2}^*}) }\:{\mathscr{S}^{(\frac12^-)TA}_\perp(\mathbf{p},t,t^\prime)}\cr
 &=\widetilde{h}_{\perp}^{(\frac{1}{2}^-)}  \: Z_{\Lambda_{c,1/2}^*} ( Z^{(1)}_{\Lambda_b} m_{\Lambda_b} + Z^{(2)}_{\Lambda_b} E_{\Lambda_b} ) e^{-m_{\Lambda_{c,1/2}^*}(t-t')} e^{-E_{\Lambda_b} t'} \cr
 &\hspace{3ex}+ \text{(excited-state contributions)},
\end{align}
such that the unwanted factors of $Z_{\Lambda_{c,1/2}^*} ( Z^{(1)}_{\Lambda_b} m_{\Lambda_b} + Z^{(2)}_{\Lambda_b} E_{\Lambda_b} ) e^{-m_{\Lambda_{c,1/2}^*}(t-t')} e^{-E_{\Lambda_b} t'}$ cancel in Eq.~(\ref{eq:newratiomethod}) at large $t$.

\begin{figure}
\flushleft

\includegraphics[width=0.245\linewidth,valign=t]{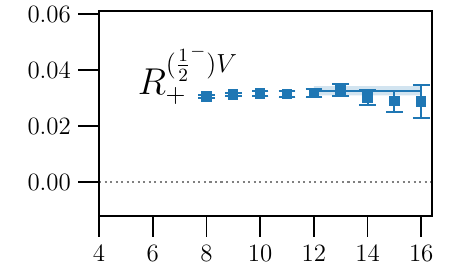} \includegraphics[width=0.245\linewidth,valign=t]{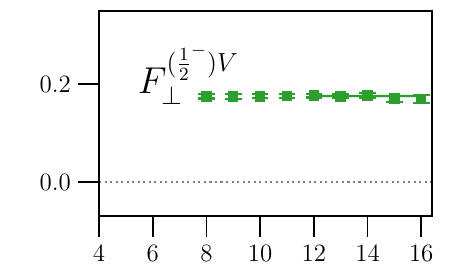}  \includegraphics[width=0.245\linewidth,valign=t]{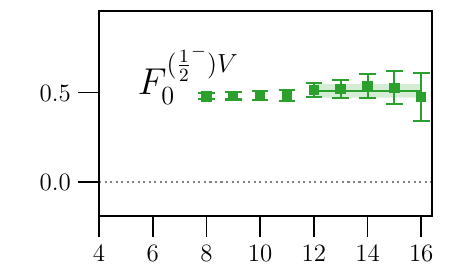}

\includegraphics[width=0.245\linewidth,valign=t]{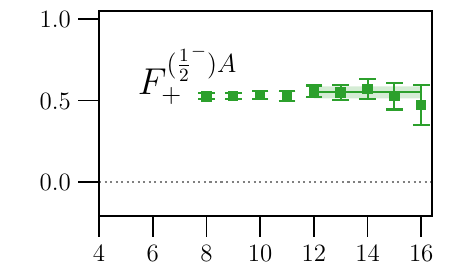} \includegraphics[width=0.245\linewidth,valign=t]{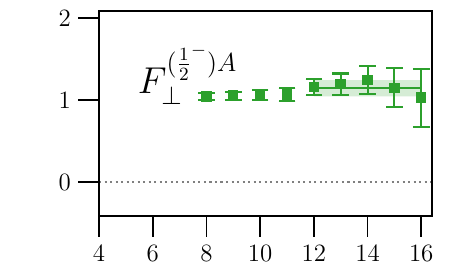}  \includegraphics[width=0.245\linewidth,valign=t]{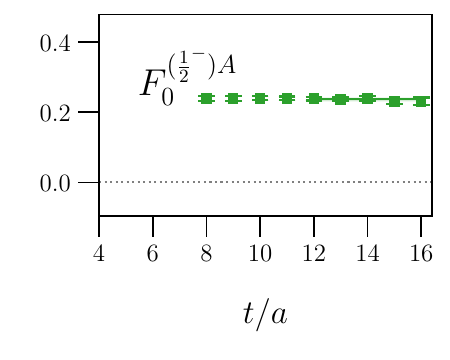}

\vspace{-6.0ex}

\includegraphics[width=0.245\linewidth,valign=t]{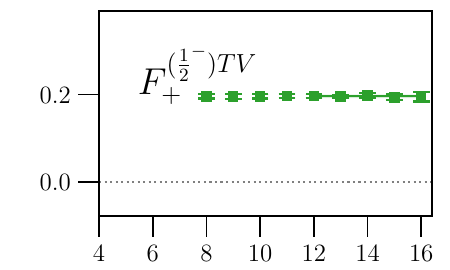} \includegraphics[width=0.245\linewidth,valign=t]{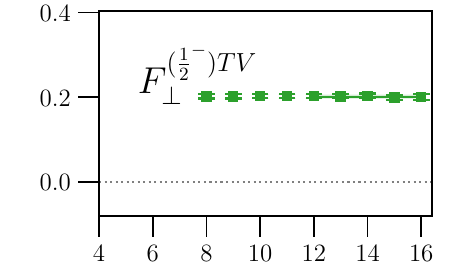}

\includegraphics[width=0.245\linewidth,valign=t]{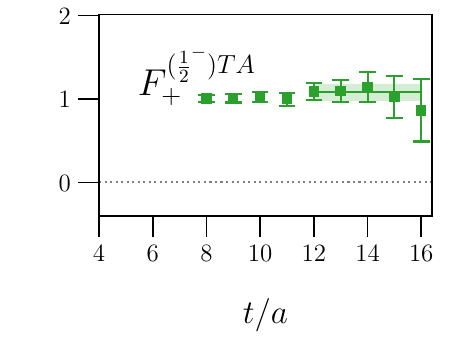} \includegraphics[width=0.245\linewidth,valign=t]{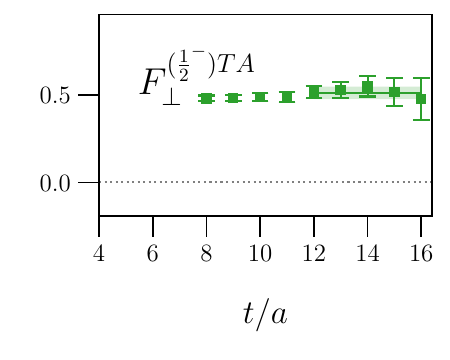} \

\vspace{-1ex}
 \caption{\label{fig:ratios12}Numerical results for the quantities $F^{(\frac12^-)X}_\lambda(\mathbf{p},t)$, defined in Eq.~(\protect\ref{eq:newratiomethod}), as a function of the source-sink separation, for $\mathbf{p}=(0,0,2)\frac{2\pi}{L}$ and for the F004 ensemble. Also shown is $R_+^{(\frac12^-)V}(\mathbf{p}, t)$, which is used to extract the square of the reference form factor $f_+^{(\frac12^-)}$. The horizontal lines indicate the ranges and extracted values of constant fits.}
\end{figure}

\begin{figure}
\flushleft

\includegraphics[width=0.245\linewidth,valign=t]{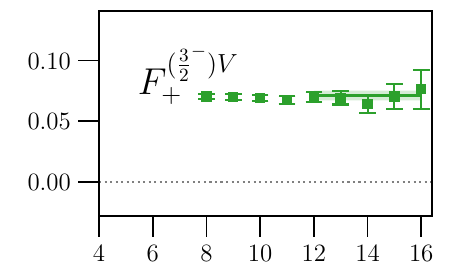} \includegraphics[width=0.245\linewidth,valign=t]{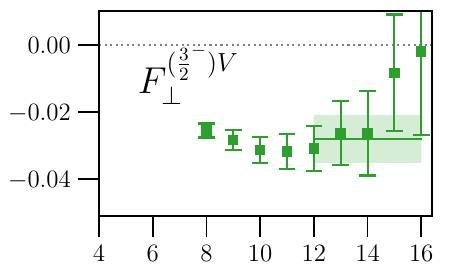} \includegraphics[width=0.245\linewidth,valign=t]{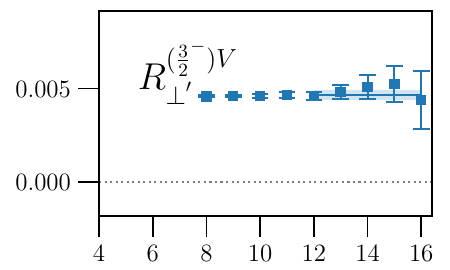} \includegraphics[width=0.245\linewidth,valign=t]{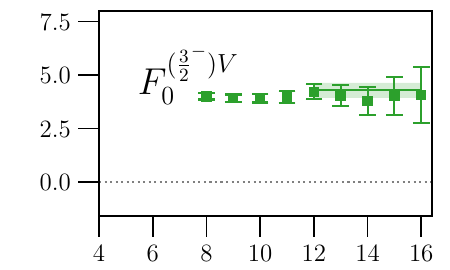}

\includegraphics[width=0.245\linewidth,valign=t]{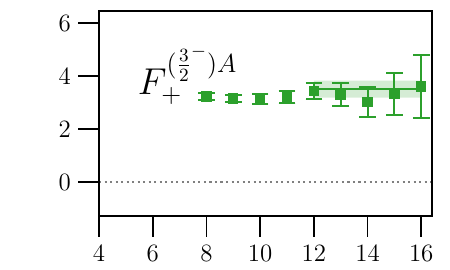} \includegraphics[width=0.245\linewidth,valign=t]{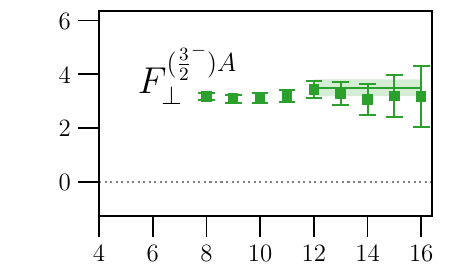} \includegraphics[width=0.245\linewidth,valign=t]{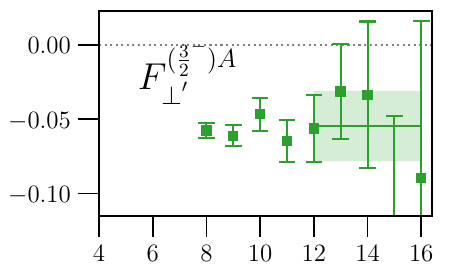} \includegraphics[width=0.245\linewidth,valign=t]{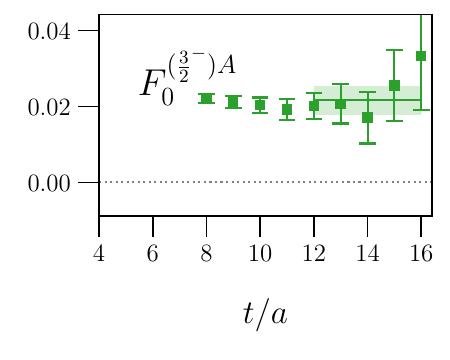}

\vspace{-6.0ex}

\includegraphics[width=0.245\linewidth,valign=t]{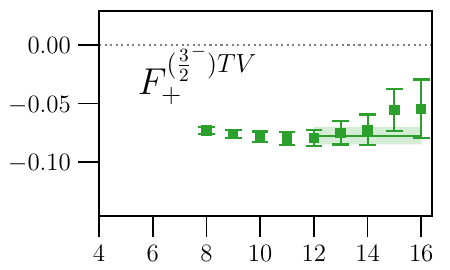} \includegraphics[width=0.245\linewidth,valign=t]{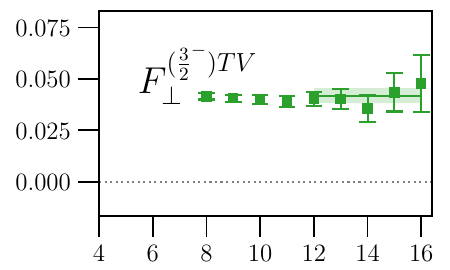} \includegraphics[width=0.245\linewidth,valign=t]{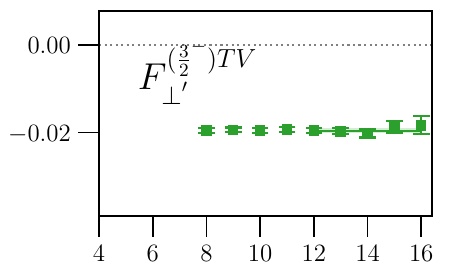}

\includegraphics[width=0.245\linewidth,valign=t]{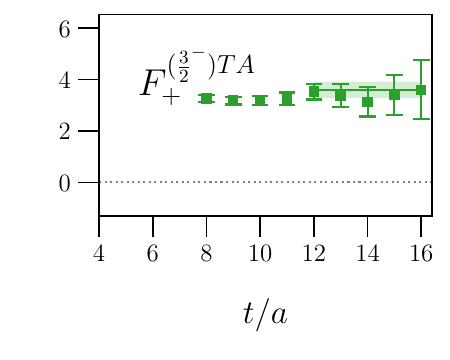} \includegraphics[width=0.245\linewidth,valign=t]{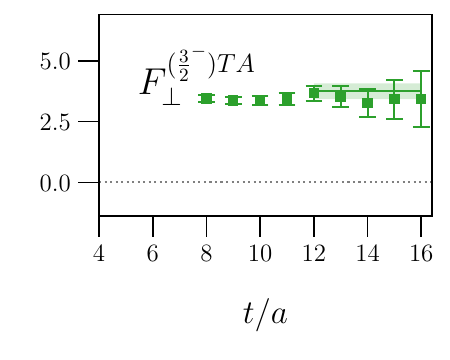} \includegraphics[width=0.245\linewidth,valign=t]{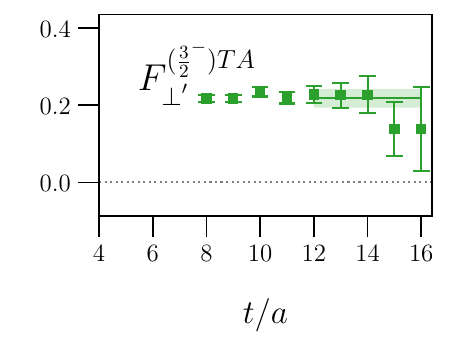}

\vspace{-1ex}
 \caption{\label{fig:ratios32}Numerical results for the quantities $F^{(\frac32^-)X}_\lambda(\mathbf{p},t)$, defined in Eq.~(\protect\ref{eq:newratiomethod}), as a function of the source-sink separation, for $\mathbf{p}=(0,0,2)\frac{2\pi}{L}$ and for the F004 ensemble. Also shown is $R_{\perp^\prime}^{(\frac32^-)V}(\mathbf{p}, t)$, which is used to extract the square of the reference form factor $f^{(\frac32^-)}_{\perp^\prime}$. The horizontal lines indicate the ranges and extracted values of constant fits.}
\end{figure}

For $J^P=\frac12^-$, we use $X_{\rm ref}=V$, $\lambda_{\rm ref}=+$. Sample results for $F^{(\frac12^-)X}_\lambda(\mathbf{p},t)$ and our constant fits thereof are shown in Fig.~\ref{fig:ratios12}. For $J^P=\frac32^-$, we use $X_{\rm ref}=V$, $\lambda_{\rm ref}=\perp^\prime$ as in Ref.~\cite{Meinel:2020owd}. Sample results for $F^{(\frac32^-)X}_\lambda(\mathbf{p},t)$ and our constant fits thereof are shown in Fig.~\ref{fig:ratios32}. The values of the form factors obtained from the constant fits are listed in Tables \ref{tab:FFvalues12} and \ref{tab:FFvalues32}. The fits were done individually for each form factor and take into account the correlations between the data at different $t$. The values of $\chi^2/{\rm d.o.f.}$ range between approximately 0.5 and 1.0, where typically ${\rm d.o.f.}=4$. The correlations between the results for different form factors and different momenta on a given ensemble were evaluated using bootstrap resampling.

\begin{table}
 \begin{tabular}{cccllllll}
  \hline\hline
  Form factor                & & $|\mathbf{p}|/(2\pi/L)$ & & \hspace{2ex}C01 & & \hspace{2ex}C005 & & \hspace{2ex}F004 \\
  \hline
  $f_0^{(\frac12^-)}$   &&   2   &&   $\wm   0.592(43)$   &&   $\wm   0.550(54)$   &&   $\wm   0.510(38)$   \\
                        &&   3   &&   $\wm   0.536(31)$   &&   $\wm   0.496(38)$   &&   $\wm   0.483(29)$   \\
  $f_+^{(\frac12^-)}$   &&   2   &&   $\wm  0.1843(51)$   &&   $\wm  0.1743(59)$   &&   $\wm  0.1804(47)$   \\
                        &&   3   &&   $\wm  0.2005(68)$   &&   $\wm  0.1887(80)$   &&   $\wm  0.1990(65)$   \\
  $f_{\perp}^{(\frac12^-)}$              &&   2   &&   $\wm  0.1728(39)$   &&   $\wm  0.1692(47)$   &&   $\wm  0.1748(37)$   \\
                        &&   3   &&   $\wm  0.1781(49)$   &&   $\wm  0.1735(58)$   &&   $\wm  0.1837(48)$   \\
  $g_0^{(\frac12^-)}$   &&   2   &&   $\wm  0.2414(55)$   &&   $\wm  0.2324(67)$   &&   $\wm  0.2366(53)$   \\
                        &&   3   &&   $\wm  0.2521(73)$   &&   $\wm  0.2433(88)$   &&   $\wm  0.2511(71)$   \\
  $g_+^{(\frac12^-)}$   &&   2   &&   $\wm   0.624(38)$   &&   $\wm   0.601(49)$   &&   $\wm   0.549(36)$   \\
                        &&   3   &&   $\wm   0.571(29)$   &&   $\wm   0.542(35)$   &&   $\wm   0.522(28)$   \\
  $g_{\perp}^{(\frac12^-)}$              &&   2   &&   $\wm    1.35(11)$   &&   $\wm    1.27(14)$   &&   $\wm    1.14(10)$   \\
                        &&   3   &&   $\wm   1.205(80)$   &&   $\wm    1.12(10)$   &&   $\wm   1.063(72)$   \\
  $h_+^{(\frac12^-)}$   &&   2   &&   $\wm  0.1935(42)$   &&   $\wm  0.1896(52)$   &&   $\wm  0.1956(40)$   \\
                        &&   3   &&   $\wm  0.1957(52)$   &&   $\wm  0.1908(63)$   &&   $\wm  0.2028(51)$   \\
  $h_{\perp}^{(\frac12^-)}$              &&   2   &&   $\wm  0.2065(50)$   &&   $\wm  0.1955(59)$   &&   $\wm  0.2013(47)$   \\
                        &&   3   &&   $\wm  0.2203(67)$   &&   $\wm  0.2081(79)$   &&   $\wm  0.2172(64)$   \\
  $\widetilde{h}_+^{(\frac12^-)}$        &&   2   &&   $\wm    1.32(11)$   &&   $\wm    1.24(14)$   &&   $\wm    1.08(10)$   \\
                        &&   3   &&   $\wm   1.182(82)$   &&   $\wm    1.09(10)$   &&   $\wm   1.011(74)$   \\
  $\widetilde{h}_{\perp}^{(\frac12^-)}$  &&   2   &&   $\wm   0.576(39)$   &&   $\wm   0.555(49)$   &&   $\wm   0.513(36)$   \\
                        &&   3   &&   $\wm   0.528(29)$   &&   $\wm   0.503(35)$   &&   $\wm   0.486(27)$   \\
  \hline\hline
 \end{tabular}
 \caption{\label{tab:FFvalues12}Values of the $\Lambda_b\to \Lambda_{c,1/2}^*$ form factors for each ensemble and for the two different $\Lambda_b$ momenta.}
\end{table}

\begin{table}
 \begin{tabular}{cccllllll}
  \hline\hline
  Form factor                & & $|\mathbf{p}|/(2\pi/L)$ & & \hspace{2ex}C01 & & \hspace{2ex}C005 & & \hspace{2ex}F004 \\
  \hline
  $f_0^{(\frac32^-)}$            &&   2   &&   $\wm    5.24(40)$   &&   $\wm    4.68(47)$   &&   $\wm    4.28(35)$   \\
                                 &&   3   &&   $\wm    4.70(34)$   &&   $\wm    4.05(35)$   &&   $\wm    3.91(28)$   \\
  $f_+^{(\frac32^-)}$            &&   2   &&   $\wm  0.0784(45)$   &&   $\wm  0.0670(50)$   &&   $\wm  0.0711(40)$   \\
                                 &&   3   &&   $\wm  0.1074(72)$   &&   $\wm  0.0904(76)$   &&   $\wm  0.0949(60)$   \\
  $f_{\perp}^{(\frac32^-)}$      &&   2   &&   $   - 0.0127(79)$   &&   $   - 0.0295(90)$   &&   $   - 0.0280(72)$   \\
                                 &&   3   &&   $\wm   0.046(10)$   &&   $\wm   0.020(11)$   &&   $\wm  0.0205(88)$   \\
  $f_{\perp^{\prime}}^{(\frac32^-)}$              &&   2   &&   $\wm  0.0708(24)$   &&   $\wm  0.0693(28)$   &&   $\wm  0.0682(20)$   \\
                                 &&   3   &&   $\wm  0.0658(32)$   &&   $\wm  0.0634(37)$   &&   $\wm  0.0639(27)$   \\
  $g_0^{(\frac32^-)}$            &&   2   &&   $\wm  0.0305(41)$   &&   $\wm  0.0194(48)$   &&   $\wm  0.0216(38)$   \\
                                 &&   3   &&   $\wm  0.0605(60)$   &&   $\wm  0.0451(65)$   &&   $\wm  0.0454(52)$   \\
  $g_+^{(\frac32^-)}$            &&   2   &&   $\wm    4.41(36)$   &&   $\wm    3.86(42)$   &&   $\wm    3.50(32)$   \\
                                 &&   3   &&   $\wm    3.94(30)$   &&   $\wm    3.33(32)$   &&   $\wm    3.16(25)$   \\
  $g_{\perp}^{(\frac32^-)}$      &&   2   &&   $\wm    4.34(36)$   &&   $\wm    3.86(42)$   &&   $\wm    3.50(31)$   \\
                                 &&   3   &&   $\wm    3.90(29)$   &&   $\wm    3.36(31)$   &&   $\wm    3.19(24)$   \\
  $g_{\perp^{\prime}}^{(\frac32^-)}$              &&   2   &&   $   -  0.037(29)$   &&   $   -  0.048(31)$   &&   $   -  0.055(24)$   \\
                                 &&   3   &&   $   -  0.029(21)$   &&   $   -  0.044(23)$   &&   $   -  0.041(17)$   \\
  $h_+^{(\frac32^-)}$            &&   2   &&   $   - 0.0609(81)$   &&   $   - 0.0733(93)$   &&   $   - 0.0776(74)$   \\
                                 &&   3   &&   $   -  0.004(10)$   &&   $   -  0.024(11)$   &&   $   - 0.0296(87)$   \\
  $h_{\perp}^{(\frac32^-)}$      &&   2   &&   $\wm  0.0490(40)$   &&   $\wm  0.0379(46)$   &&   $\wm  0.0419(36)$   \\
                                 &&   3   &&   $\wm  0.0784(62)$   &&   $\wm  0.0621(66)$   &&   $\wm  0.0652(52)$   \\
  $h_{\perp^{\prime}}^{(\frac32^-)}$              &&   2   &&   $   -0.01943(68)$   &&   $   -0.01839(75)$   &&   $   -0.01954(59)$   \\
                                 &&   3   &&   $   - 0.0188(10)$   &&   $   - 0.0172(10)$   &&   $   -0.01925(87)$   \\
  $\widetilde{h}_+^{(\frac32^-)}$&&   2   &&   $\wm    4.43(36)$   &&   $\wm    3.97(42)$   &&   $\wm    3.60(32)$   \\
                                 &&   3   &&   $\wm    3.98(30)$   &&   $\wm    3.45(31)$   &&   $\wm    3.27(25)$   \\
  $\widetilde{h}_{\perp}^{(\frac32^-)}$           &&   2   &&   $\wm    4.64(37)$   &&   $\wm    4.06(43)$   &&   $\wm    3.75(32)$   \\
                                 &&   3   &&   $\wm    4.16(31)$   &&   $\wm    3.52(32)$   &&   $\wm    3.40(25)$   \\
  $\widetilde{h}_{\perp^{\prime}}^{(\frac32^-)}$  &&   2   &&   $\wm   0.249(30)$   &&   $\wm   0.223(31)$   &&   $\wm   0.219(24)$   \\
                                 &&   3   &&   $\wm   0.237(25)$   &&   $\wm   0.198(24)$   &&   $\wm   0.218(20)$   \\
  \hline\hline
 \end{tabular}
 \caption{\label{tab:FFvalues32}Values of the $\Lambda_b\to \Lambda_{c,3/2}^*$ form factors for each ensemble and for the two different $\Lambda_b$ momenta.}
\end{table}

\FloatBarrier
\section{Chiral and continuum extrapolations of the form factors}
\label{sec:FFextrap}
\FloatBarrier

\begin{figure}
 \centerline{\includegraphics[width=0.6\linewidth]{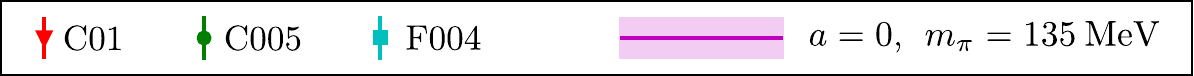}}
 
 \vspace{1ex}

 \includegraphics[width=0.47\linewidth]{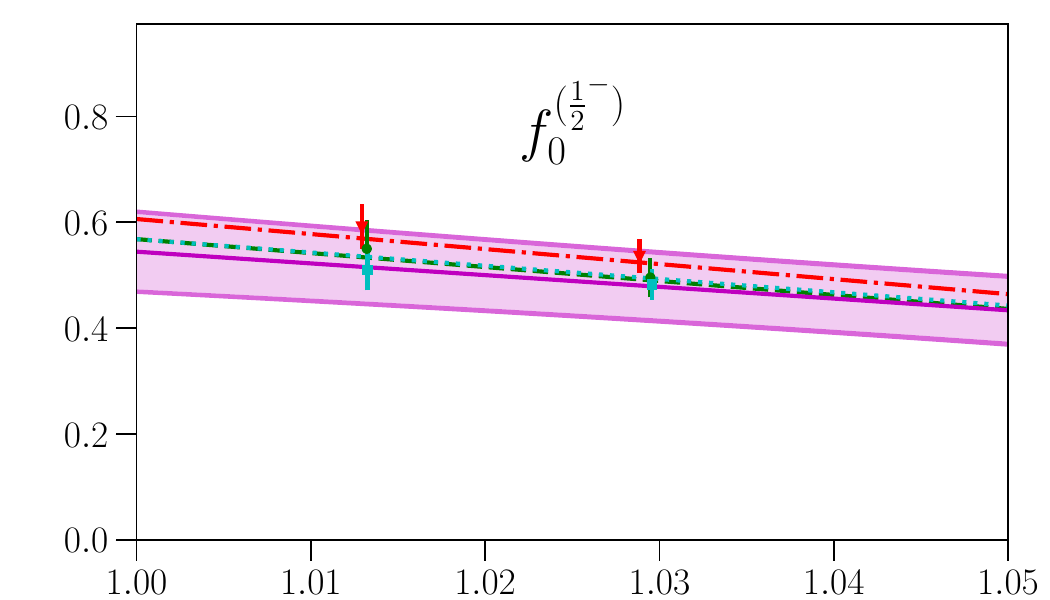} \hfill \includegraphics[width=0.47\linewidth]{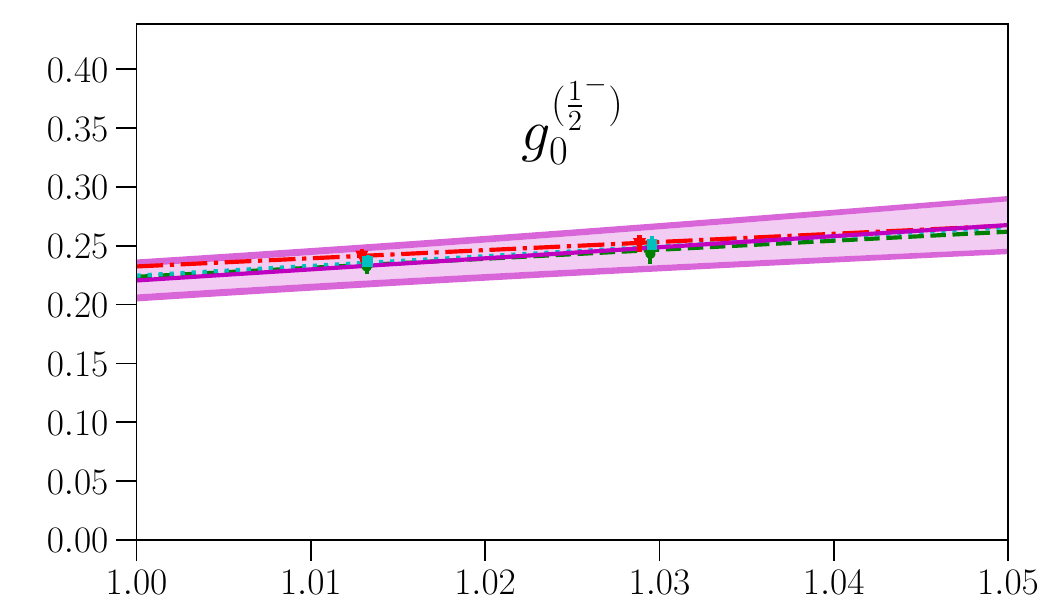} \\
 \includegraphics[width=0.47\linewidth]{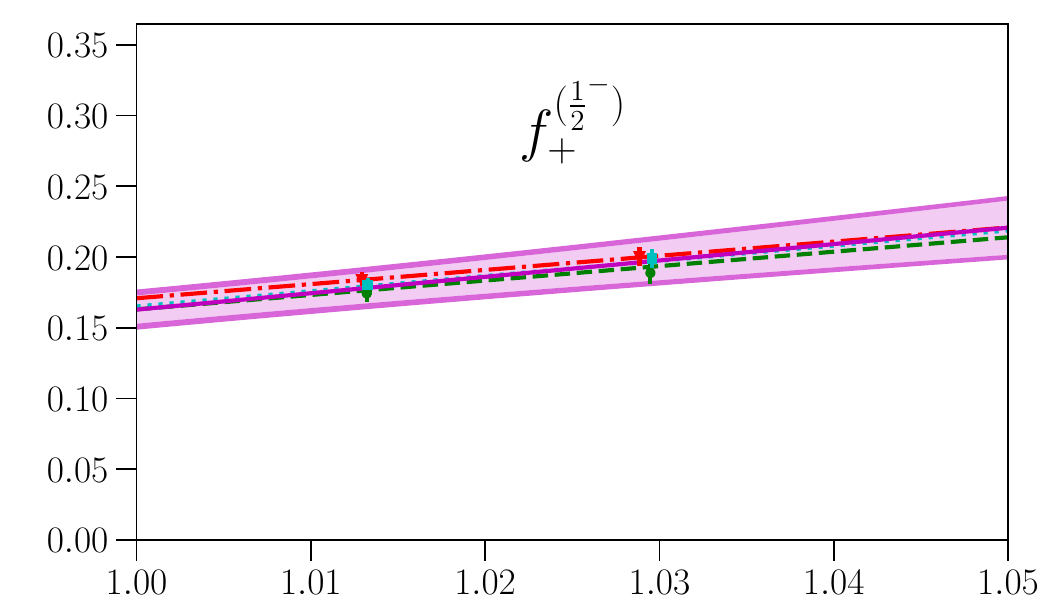} \hfill \includegraphics[width=0.47\linewidth]{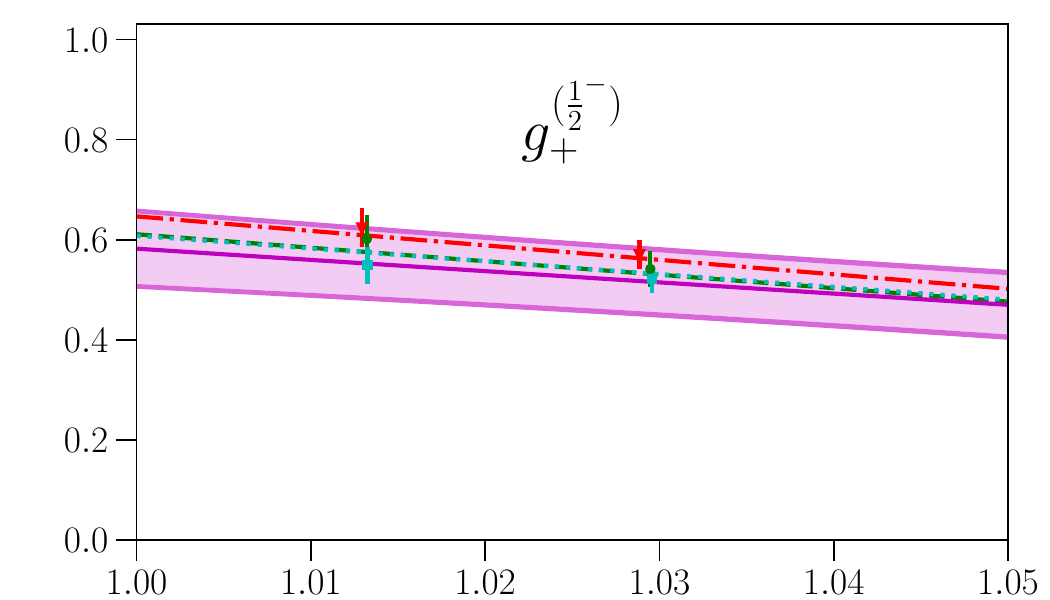} \\
 \includegraphics[width=0.47\linewidth]{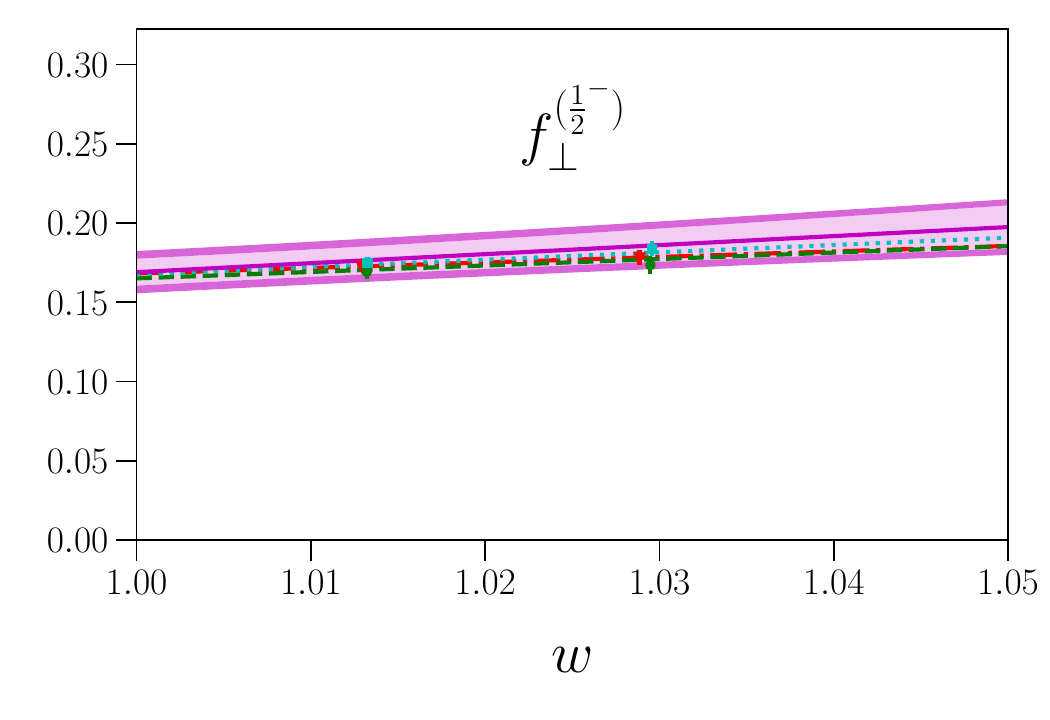} \hfill \includegraphics[width=0.47\linewidth]{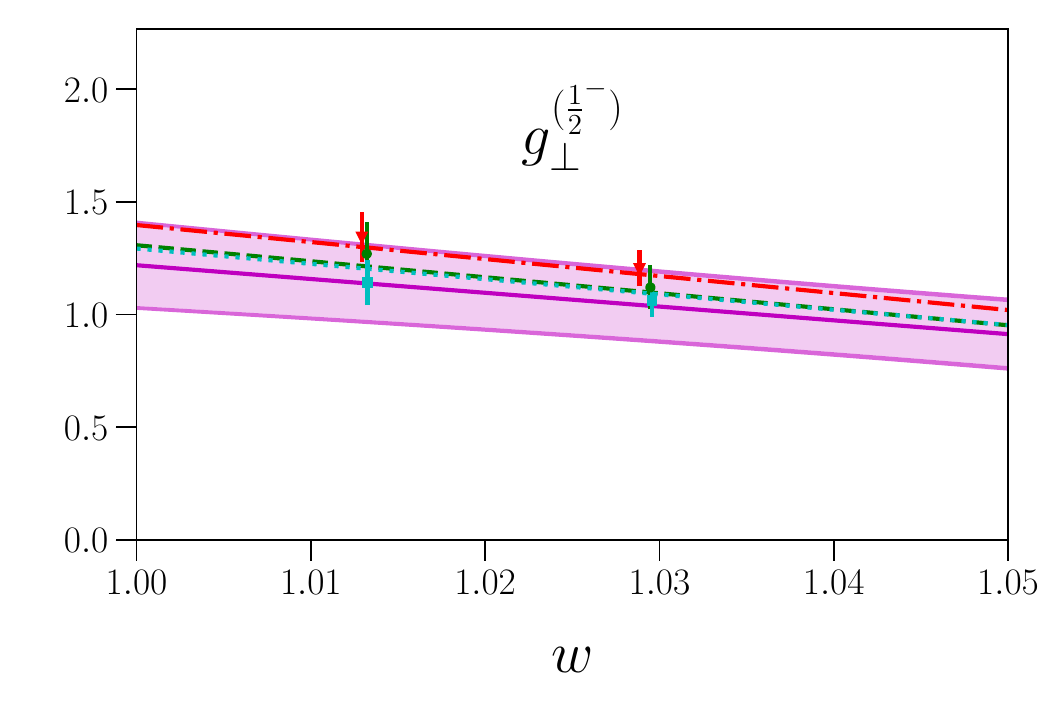} \\
 
 \caption{\label{fig:FFextrapJ12VA}Chiral and continuum extrapolations of the $\Lambda_b \to \Lambda_c^*(2595)$ vector and axial vector form factors. The solid magenta curves show the form factors in the physical limit $a=0$, $m_\pi=135\:{\rm MeV}$, with inner light magenta bands indicating the statistical uncertainties and outer dark magenta bands indicating the total uncertainties. The dashed-dotted, dashed, and dotted curves show the fit functions evaluated at the lattice spacings and pion masses of the individual data sets C01, C005, and F004, respectively, with uncertainty bands omitted for clarity.}
\end{figure}

\begin{figure}
 \centerline{\includegraphics[width=0.6\linewidth]{figures/legend_datasets.pdf}}
 
 \vspace{1ex}

 \includegraphics[width=0.47\linewidth]{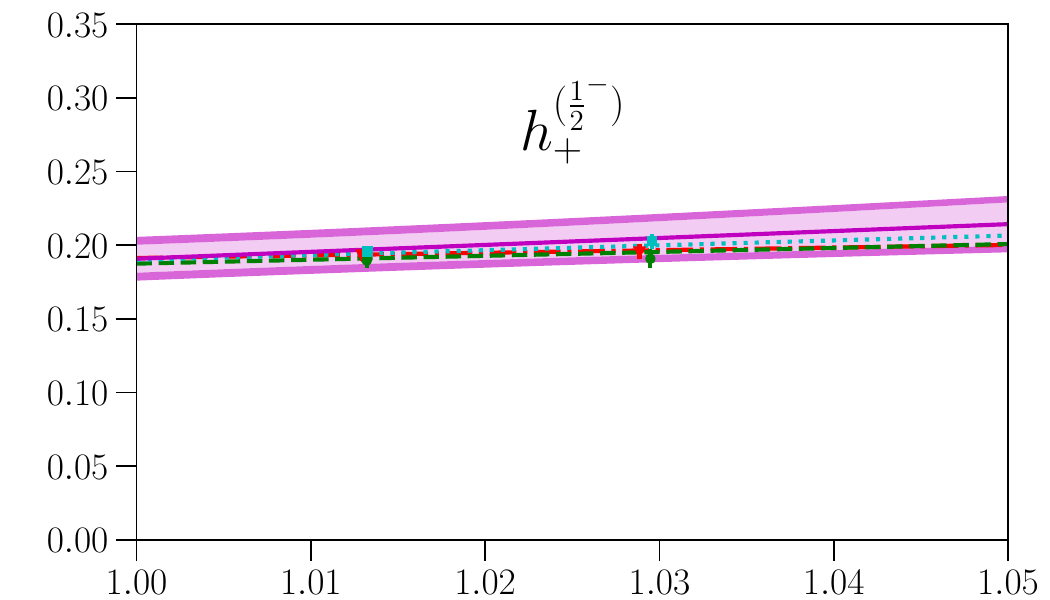} \hfill \includegraphics[width=0.47\linewidth]{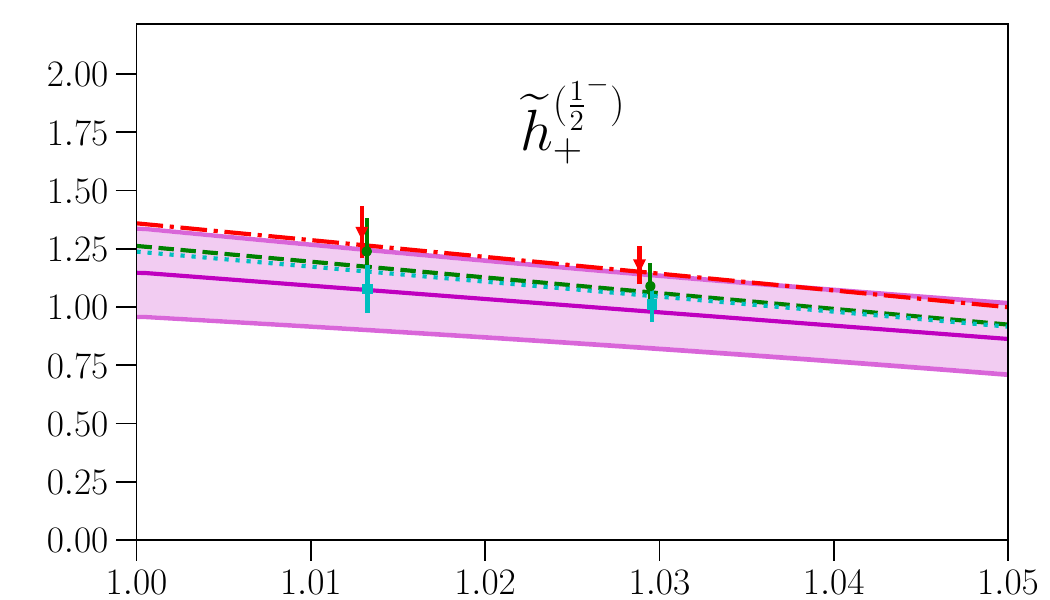} \\
 \includegraphics[width=0.47\linewidth]{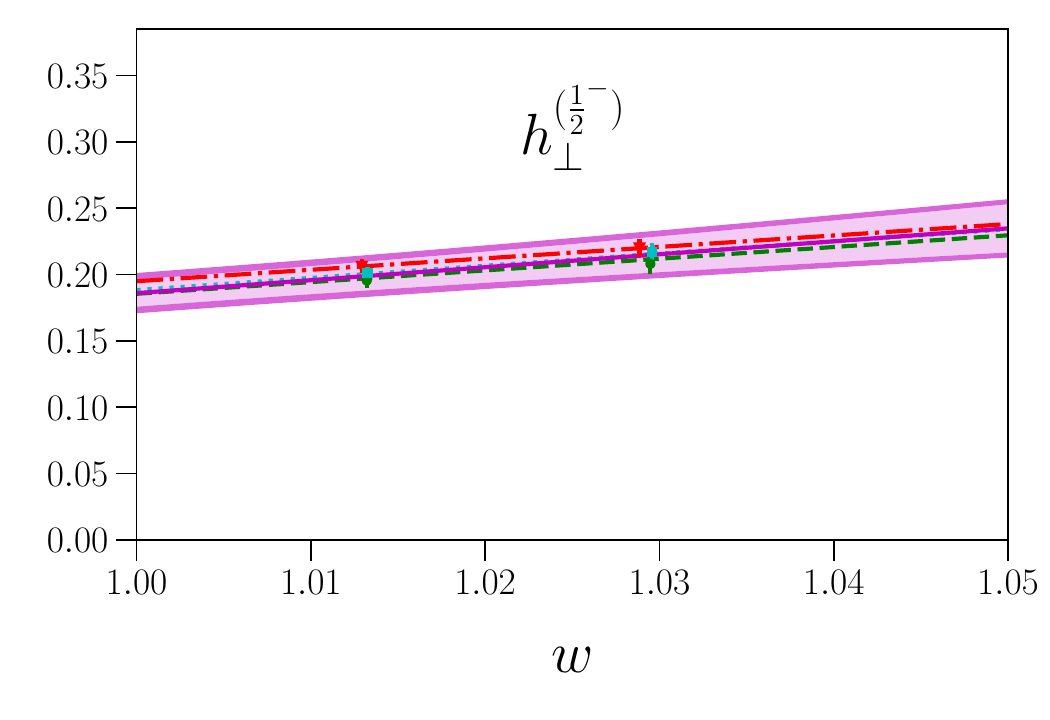} \hfill \includegraphics[width=0.47\linewidth]{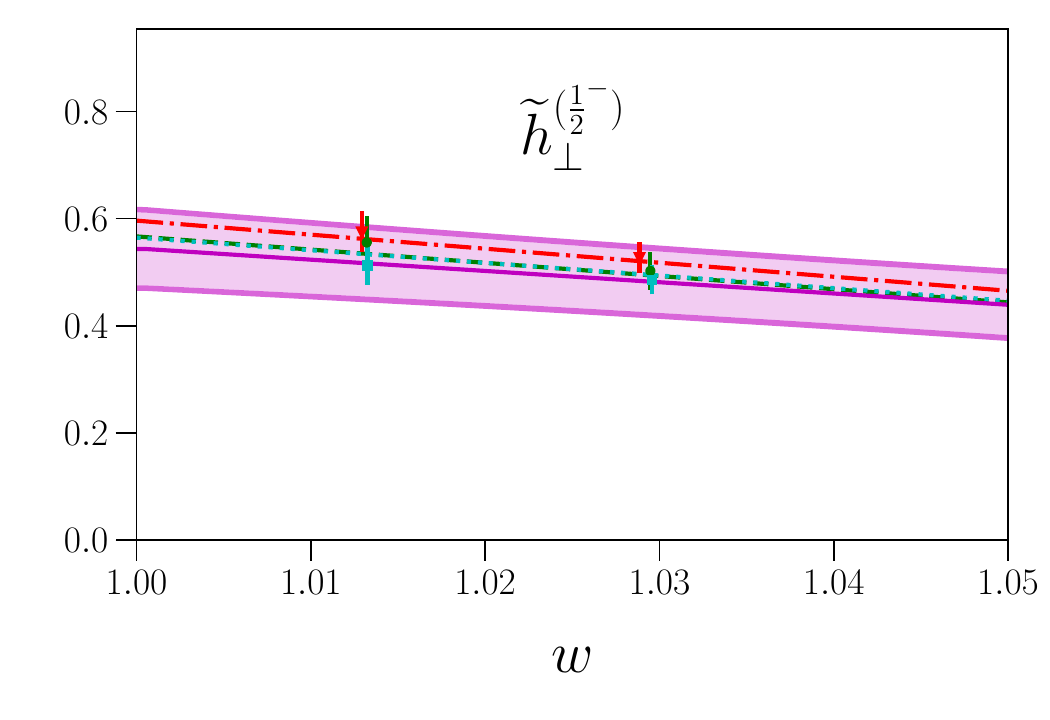} \\
 
 \caption{\label{fig:FFextrapJ12T}Like Fig.~\protect\ref{fig:FFextrapJ12VA}, but for the $\Lambda_b \to \Lambda_c^*(2595)$ tensor form factors. }
\end{figure}

\begin{figure}
 \centerline{\includegraphics[width=0.6\linewidth]{figures/legend_datasets.pdf}}
 
 \vspace{1ex}

 \includegraphics[width=0.47\linewidth]{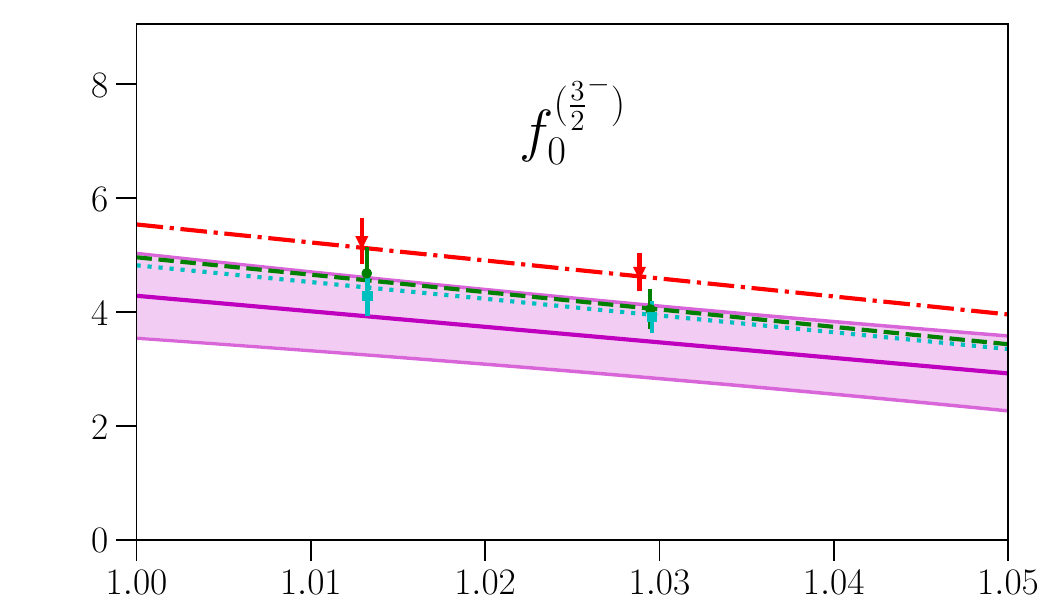} \hfill \includegraphics[width=0.47\linewidth]{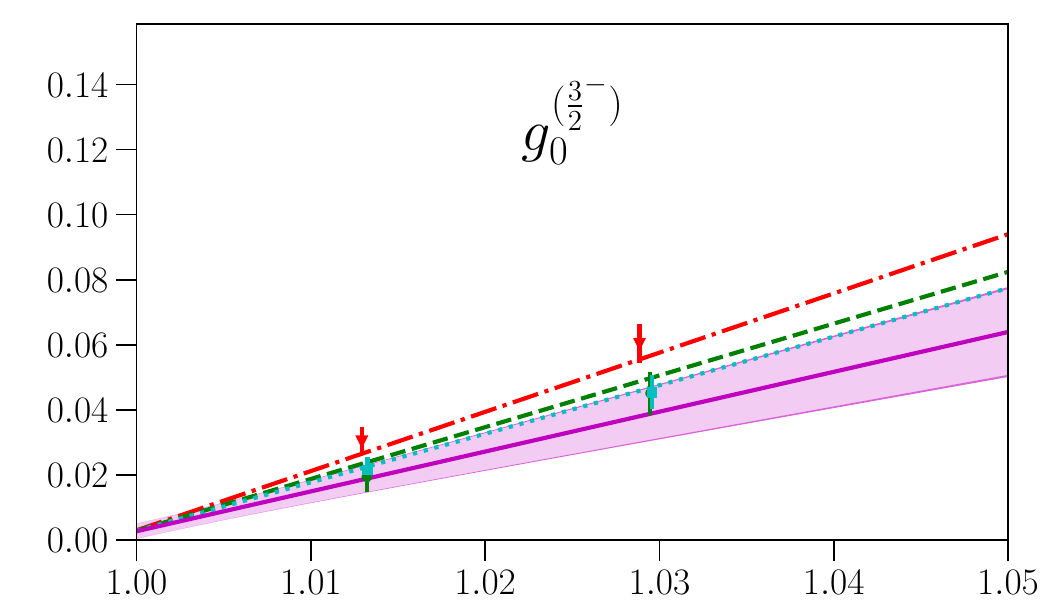} \\
 \includegraphics[width=0.47\linewidth]{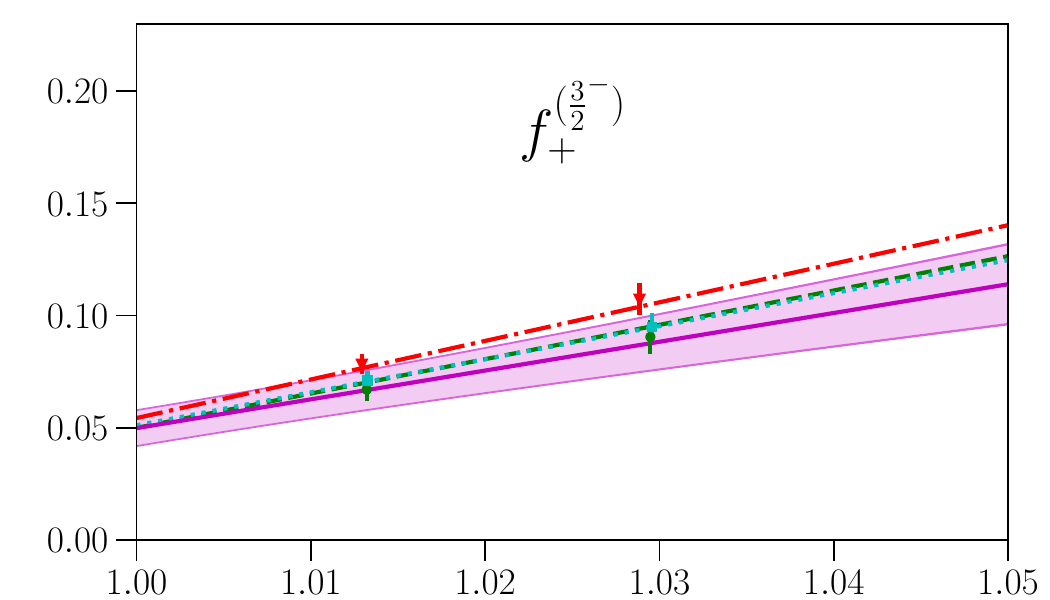} \hfill \includegraphics[width=0.47\linewidth]{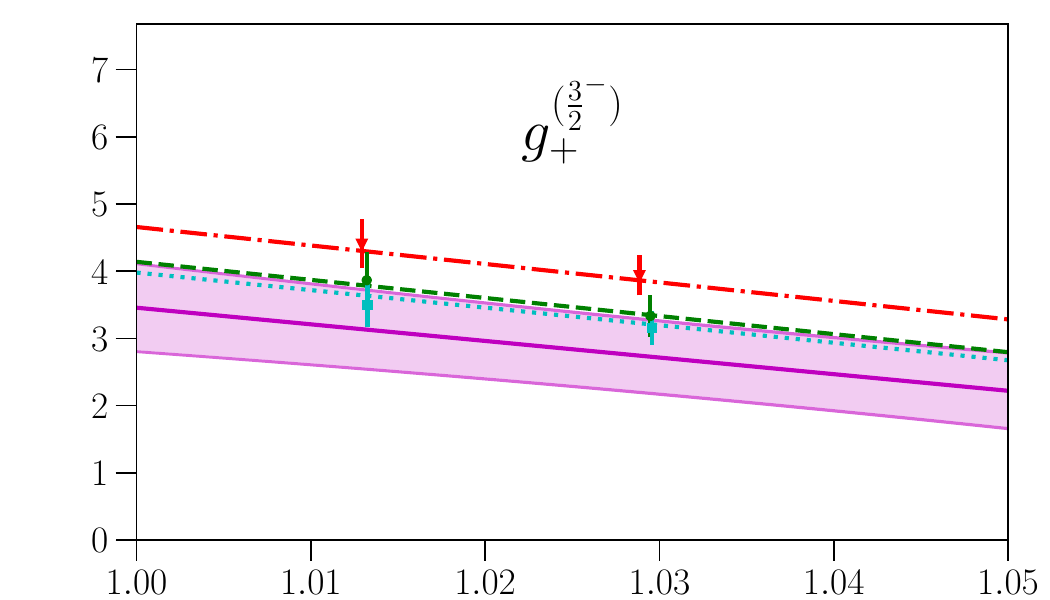} \\
 \includegraphics[width=0.47\linewidth]{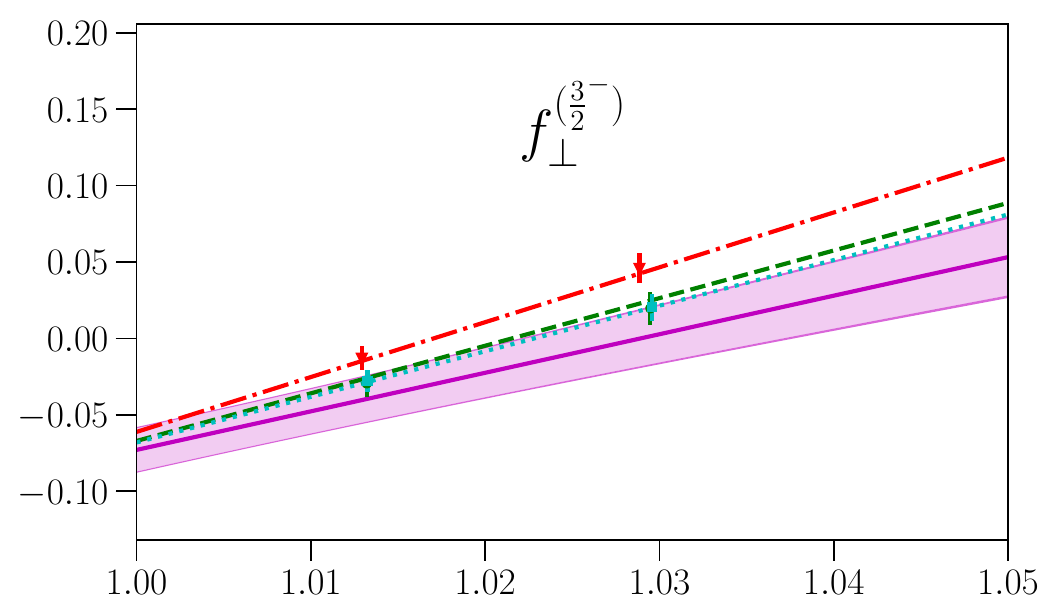} \hfill \includegraphics[width=0.47\linewidth]{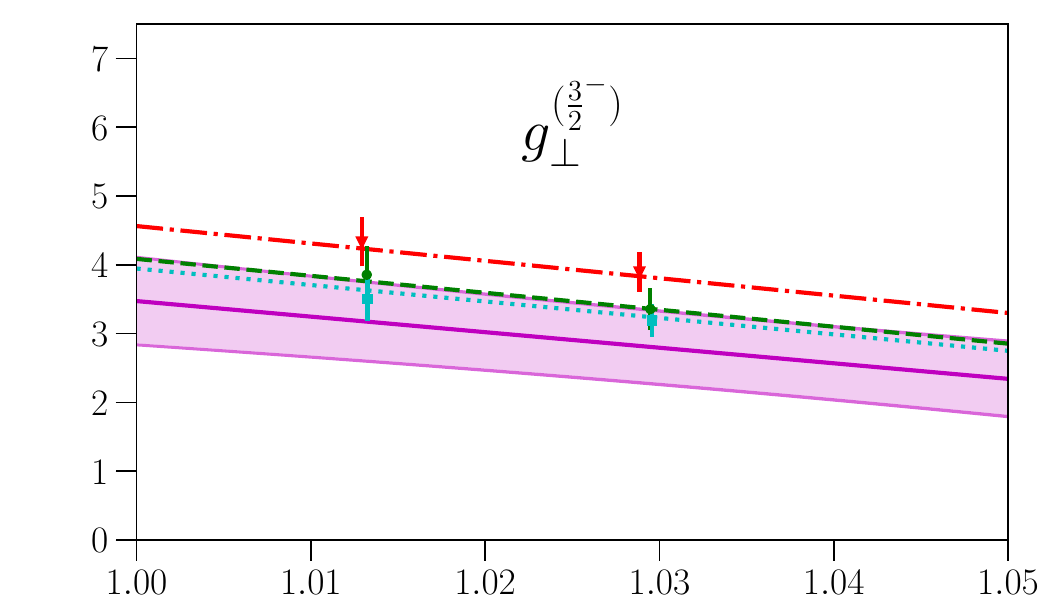} \\
 \includegraphics[width=0.47\linewidth]{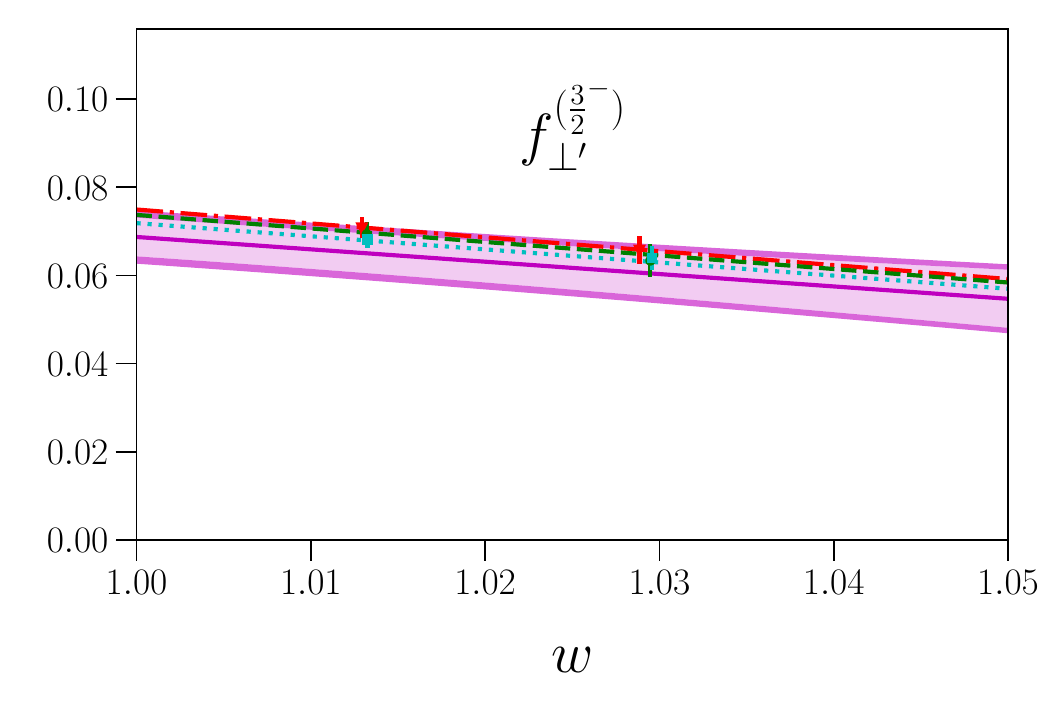} \hfill \includegraphics[width=0.47\linewidth]{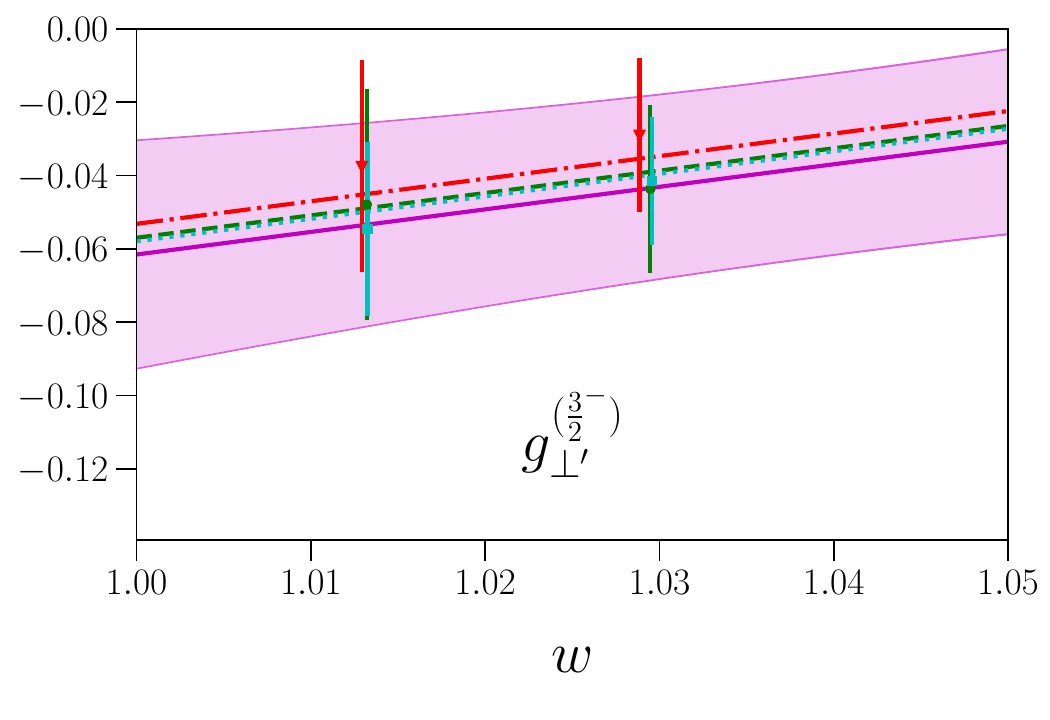} \\
 
 \caption{\label{fig:FFextrapJ32VA}Like Fig.~\protect\ref{fig:FFextrapJ12VA}, but for the $\Lambda_b \to \Lambda_c^*(2625)$ vector and axial vector form factors. }
\end{figure}

\begin{figure}
 \centerline{\includegraphics[width=0.6\linewidth]{figures/legend_datasets.pdf}}
 
 \vspace{1ex}

 \includegraphics[width=0.47\linewidth]{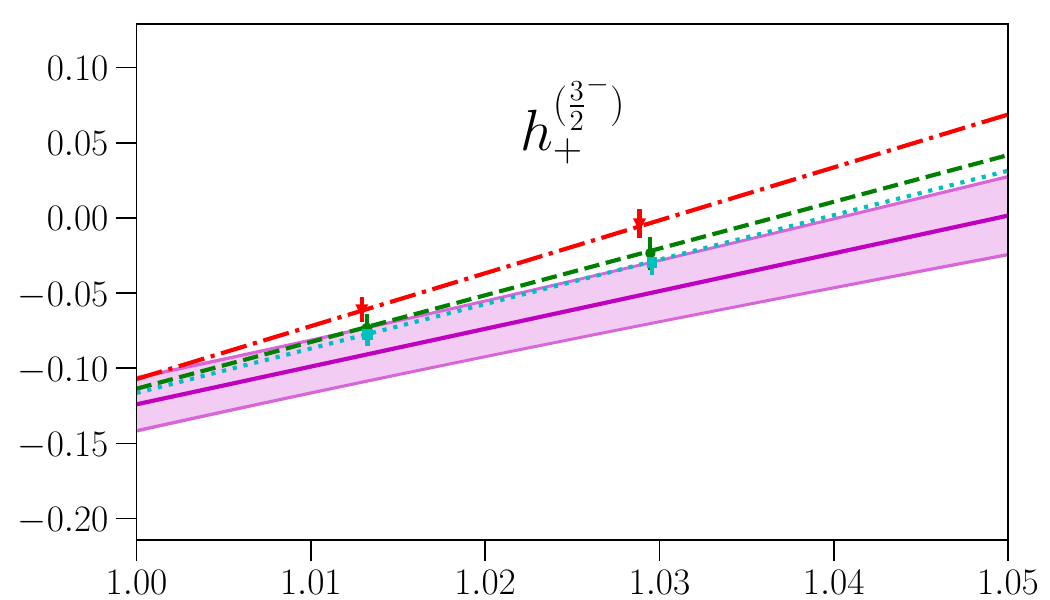} \hfill \includegraphics[width=0.47\linewidth]{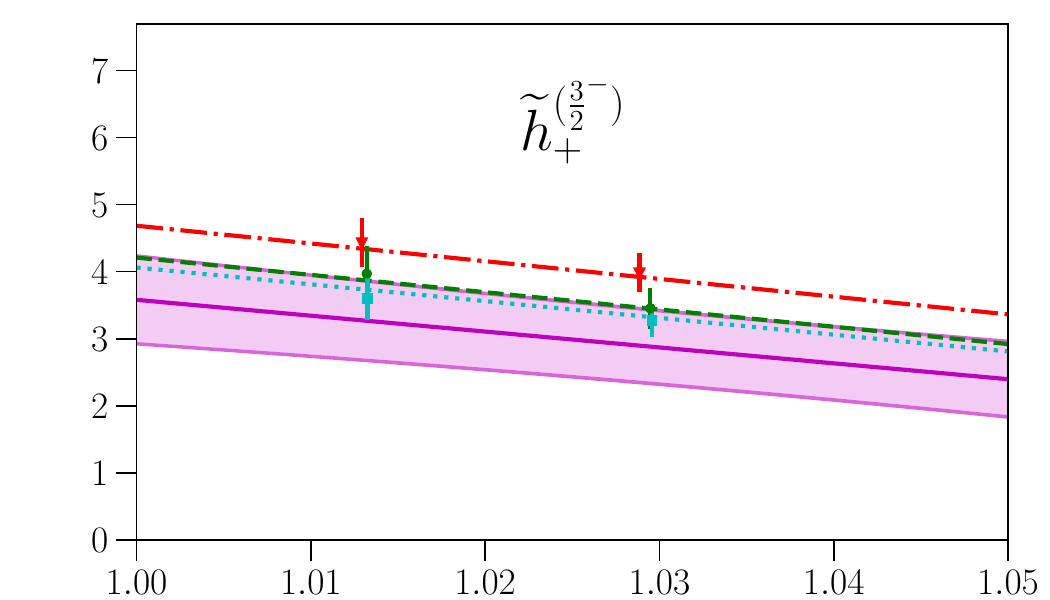} \\
 \includegraphics[width=0.47\linewidth]{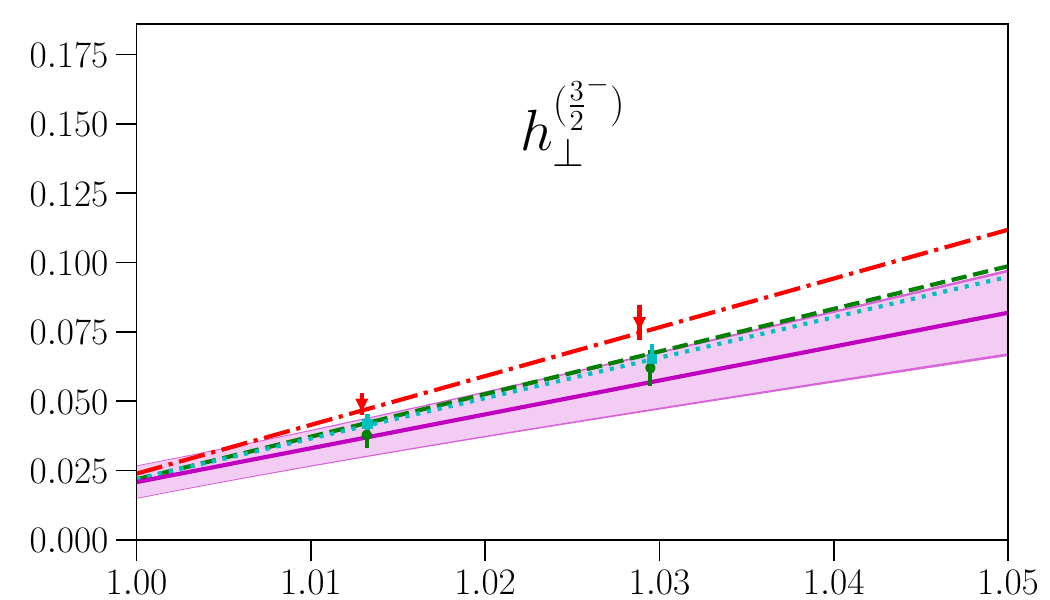} \hfill \includegraphics[width=0.47\linewidth]{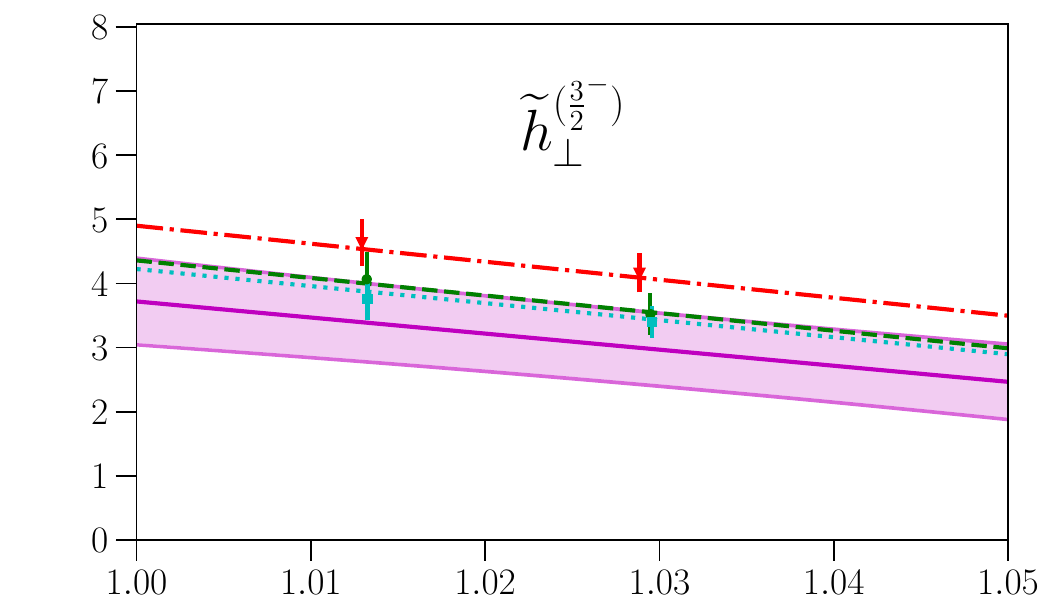} \\
 \includegraphics[width=0.47\linewidth]{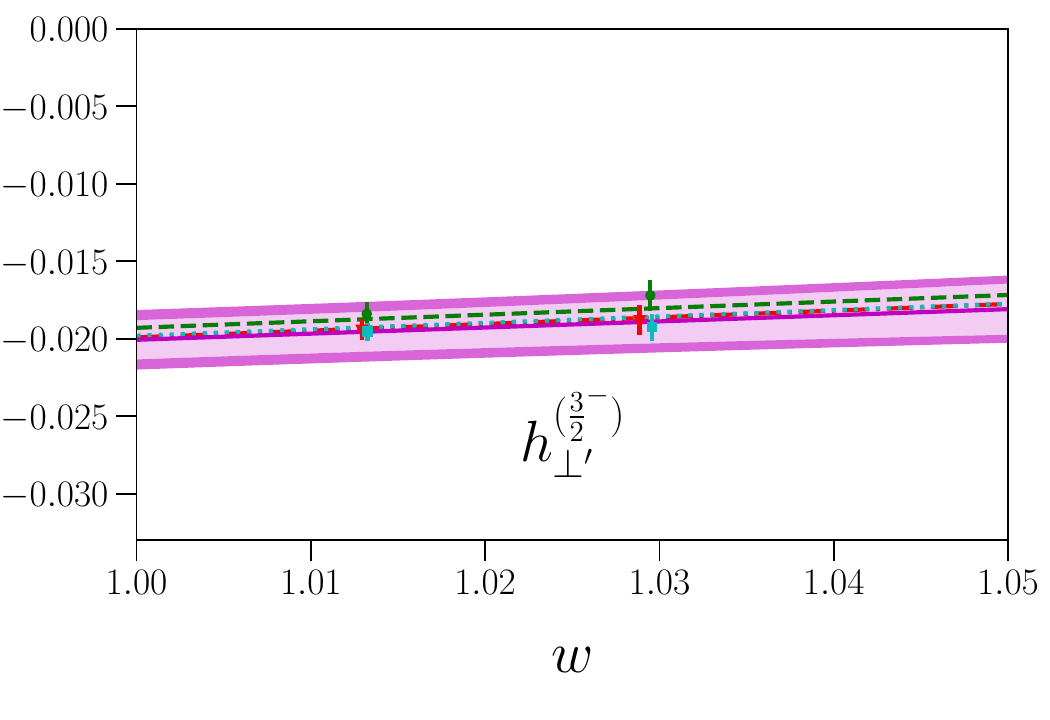} \hfill \includegraphics[width=0.47\linewidth]{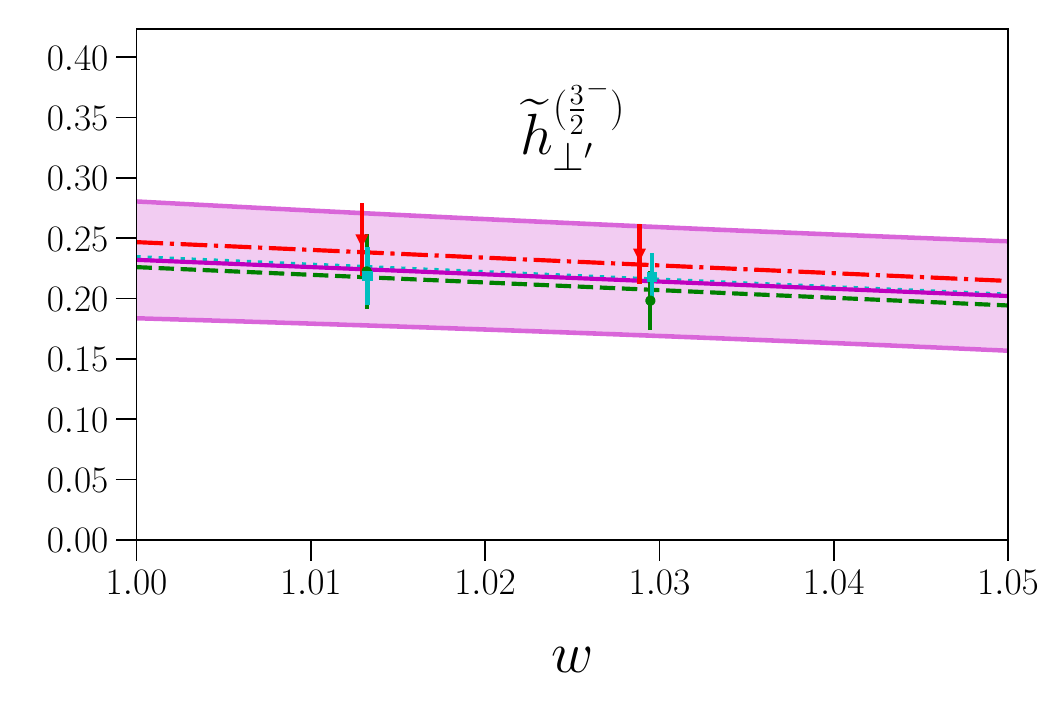} \\
 
 \caption{\label{fig:FFextrapJ32T}Like Fig.~\protect\ref{fig:FFextrapJ12VA}, but for the $\Lambda_b \to \Lambda_c^*(2625)$ tensor form factors. }
\end{figure}

As in Ref.~\cite{Meinel:2020owd}, we extrapolate the lattice results for the form factors to the continuum limit and the physical pion mass using the model
\begin{equation}\label{eq:extrapolation}
f(q^2)=F^{f}\left[1+C^{f}\frac{m_{\pi}^2-m_{\pi,\rm phys}^2}{(4\pi f_{\pi})^2}+D^{f}a^2\Lambda^2\right]+A^{f}\left[1+\tilde{C}^{f}\frac{m_{\pi}^2-m_{\pi,\rm phys}^2}{(4\pi f_{\pi})^2}+\tilde{D}^{f}a^2\Lambda^2\right](w-1)
\end{equation}
with fit parameters $F^f$, $A^f$, $C^f$, $D^f$, $\tilde{C}^f$, $\tilde{D}^f$ for each form factor $f$, and using the kinematic variable

\begin{equation}
w(q^2)=v\cdot v^\prime=\frac{m_{\Lambda_b}^2+m_{\Lambda_c^*}^2-q^2}{2m_{\Lambda_b}m_{\Lambda_c^*}},
\end{equation}
where $m_{\Lambda_c^*}=m_{\Lambda_{c,1/2}^*}$ or $m_{\Lambda_c^*}=m_{\Lambda_{c,3/2}^*}$ depending on the final state considered. In the physical limit $m_\pi=m_{\pi,\rm phys}$, $a=0$, the functions reduce to
\begin{equation}\label{eq:physicalFF}
f(q^2)=F^{f}+A^{f}(w-1).
\end{equation}
This parametrization corresponds to a Taylor expansion of the shape of the form factors around the endpoint $w=1$, i.e. an expansion in powers of $(w-1)$; because we have lattice results for only two different kinematic points near $w=1.01$ and $w=1.03$, we work only to first order, and we expect the parametrization to become unreliable for large $(w-1)$. Our results for $F^{f}$ and $A^{f}$ from fits using Eq.~(\ref{eq:extrapolation}) are given in the first two columns of Table \ref{tab:FFparams}, and the values and full covariance matrices (evaluated using bootstrap) are also provided as supplemental files. As can be seen in Figs.~\ref{fig:FFextrapJ12VA}, \ref{fig:FFextrapJ12T}, \ref{fig:FFextrapJ32VA}, and \ref{fig:FFextrapJ32T}, the lattice data are well described by the model. The fits of the individual form factors have $\chi^2/{\rm d.o.f.}$ in the range from approximately 0.5 to 1.5, where we count $F^f$, $A^f$, $C^f$, and $D^f$ as parameters that are primarily constrained by the data, such that ${\rm d.o.f.}=6-4=2$.

Again following Ref.~\cite{Meinel:2020owd}, to estimate systematic uncertainties associated with the chiral and continuum extrapolations, we also performed ``higher-order'' fits including additional terms describing the dependence on the lattice spacing and pion mass,
\begin{eqnarray}
\nonumber f_{\rm HO}(q^2)&=&F_{\rm HO}^{f}\left[1+C_{\rm HO}^{f}\frac{m_{\pi}^2-m_{\pi,\rm phys}^2}{(4\pi f_{\pi})^2}+H_{\rm HO}^{f}\frac{m_{\pi}^3-m_{\pi,\rm phys}^3}{(4\pi f_{\pi})^3}+D_{\rm HO}^{f}a^2\Lambda^2+E_{\rm HO}^{f}a \Lambda +G_{\rm HO}^{f}a^3\Lambda^3 \right] \\
 &&+A_{\rm HO}^{f}\left[1+\tilde{C}_{\rm HO}^{f}\frac{m_{\pi}^2-m_{\pi,\rm phys}^2}{(4\pi f_{\pi})^2}+\tilde{H}_{\rm HO}^{f}\frac{m_{\pi}^3-m_{\pi,\rm phys}^3}{(4\pi f_{\pi})^3}+\tilde{D}_{\rm HO}^{f}a^2\Lambda^2+\tilde{E}_{\rm HO}^{f}a \Lambda +\tilde{G}_{\rm HO}^{f}a^3\Lambda^3\right](w-1). \hspace{6ex} \label{eq:extrapolationHO} 
\end{eqnarray}
No priors were used for the parameters $F^f$, $A^f$, $F_{\rm HO}^{f}$, $A_{\rm HO}^{f}$. The Gaussian priors for the parameters describing the lattice-spacing and pion-mass dependence were chosen as in Ref.~\cite{Meinel:2020owd} except for $E_{\rm HO}^{f}$ and $\tilde{E}_{\rm HO}^{f}$. These coefficients describe the effects of the incomplete $\mathcal{O}(a)$ improvement of the weak currents in Eq.~(\ref{eq:JGamma}), and here we take the prior widths for $E_{\rm HO}^{f}$ and $\tilde{E}_{\rm HO}^{f}$ to be two times larger than in Ref.~\cite{Meinel:2020owd}, based on the observation in Ref.~\cite{Detmold:2015aaa} that these effects may be larger for a heavy-to-heavy current than for a heavy-to-light current. These widths allow for missing $\mathcal{O}(a)$ corrections as large as 10\% at the coarse lattice spacing, motivated by the large $b$-quark momenta used here. In the higher-order fits, we also multiplied the data for each form factor with Gaussian random distributions of central value 1 and appropriate widths to incorporate estimates of systematic uncertainties associated with the residual matching factors $\rho_\Gamma$ (2\% for the vector and axial vector currents, 4.04\% for the tensor currents \cite{Datta:2017aue}) and systematic uncertainties associated with neglecting the down-up quark-mass difference and QED corrections [$\mathcal{O}((m_d-m_u)/\Lambda)\approx 0.8\%$ and $\mathcal{O}(\alpha_{\rm e.m.})\approx 0.7\%$]. Furthermore, to include the scale-setting uncertainty, we also promoted the lattice spacings to fit parameters with Gaussian priors according to the values and uncertainties shown in Table \ref{tab:latticeparams}. All of our lattice calculations were performed with $m_\pi L > 4$, and we therefore expect finite-volume effects to be negligible at least for the heavier pion masses where the $\Lambda_c^*(2595)$ and $\Lambda_c^*(2625)$ are well below strong-decay thresholds. However, we are unable to provide a quantitative estimate of finite-volume effects in the extrapolated form factors.

In the physical limit, the higher-order fits reduce to the same form as in Eq.~(\ref{eq:physicalFF}) but with parameters $F^f_{\rm HO}$ and $A^f_{\rm HO}$. Our results for these parameters are given in the last two columns in Table \ref{tab:FFparams} and also in supplemental files. For any observable $O$, we evaluate the form-factor systematic uncertainty using
\begin{equation}
 \sigma_{O,{\rm syst}} = {\rm max}\left( |O_{\rm HO}-O|,\: \sqrt{|\sigma_{O,{\rm HO}}^2-\sigma_O^2|}  \right), \label{eq:sigmasystII}
\end{equation}
where $O$, $\sigma_O$ denote the central value and uncertainty calculated using $\{F^{f}, A^{f}\}$ and their covariance matrix, and $O_{\rm HO}$, $\sigma_{O,{\rm HO}}^2$ denote the central value and uncertainty calculated using $\{F^{f}_{\rm HO}, A^{f}_{\rm HO}\}$ and their covariance matrix.
%
% To obtain the total uncertainties, the systematic and statistical uncertainties are added in quadrature. The dark-magenta outer bands in Figs.~\ref{fig:FFextrapJ12VA}, \ref{fig:FFextrapJ12T}, \ref{fig:FFextrapJ32VA}, and \ref{fig:FFextrapJ32T} show the total uncertainties of the form factors
%
We find that the (vector and axial-vector) form-factor systematic uncertainties result in an approximately 12 to 13 percent systematic uncertainty in the $\Lambda_b \to \Lambda_c^*(2595)\mu^-\bar{\nu}$ differential decay rate in the kinematic range shown in Sec.~\ref{sec:observables}, and 14 to 18 percent for $\Lambda_b \to \Lambda_c^*(2625)\mu^-\bar{\nu}$. Because the decay rates depend quadratically on the form factors, this corresponds to ``average'' systematic uncertainties of around 6\% in the $\Lambda_b \to \Lambda_c^*(2595)$ vector and axial-vector form factors and around 8\% for $\Lambda_b \to \Lambda_c^*(2625)$.

\begin{table}
 \begin{tabular}{lllllllll}
  \hline\hline
  $f$                & \hspace{2ex} & \hspace{6ex}$F^f$ & & \hspace{4ex}$A^f$  & \hspace{2ex} & \hspace{4ex}$F^f_{\rm HO}$ & & \hspace{4ex}$A^f_{\rm HO}$  \\
  \hline
  $f_0^{(\frac{1}{2}^-)}$                                 &&   $\wm   0.545(64)$   &&   $   -   2.21(66)$   &&   $\wm   0.546(75)$   &&   $   -   2.20(69)$   \\
  $f_+^{(\frac{1}{2}^-)}$                                 &&   $\wm  0.1628(90)$   &&   $\wm    1.16(31)$   &&   $\wm   0.164(14)$   &&   $\wm    1.17(33)$   \\
  $f_{\perp}^{(\frac{1}{2}^-)}$                           &&   $\wm  0.1690(79)$   &&   $\wm    0.57(25)$   &&   $\wm   0.169(13)$   &&   $\wm    0.58(26)$   \\
  $g_0^{(\frac{1}{2}^-)}$                                 &&   $\wm   0.221(11)$   &&   $\wm    0.94(33)$   &&   $\wm   0.221(17)$   &&   $\wm    0.95(35)$   \\
  $g_+^{(\frac{1}{2}^-)}$                                 &&   $\wm   0.582(64)$   &&   $   -   2.24(65)$   &&   $\wm   0.584(76)$   &&   $   -   2.23(68)$   \\
  $g_{\perp}^{(\frac{1}{2}^-)}$                           &&   $\wm    1.22(16)$   &&   $   -    6.1(1.9)$   &&   $\wm    1.22(18)$   &&   $   -    6.1(2.0)$   \\
  $h_+^{(\frac{1}{2}^-)}$                                 &&   $\wm  0.1908(89)$   &&   $\wm    0.47(30)$   &&   $\wm   0.191(14)$   &&   $\wm    0.49(32)$   \\
  $h_{\perp}^{(\frac{1}{2}^-)}$                           &&   $\wm  0.1860(93)$   &&   $\wm    0.98(28)$   &&   $\wm   0.187(15)$   &&   $\wm    0.98(30)$   \\
  $\widetilde{h}_+^{(\frac{1}{2}^-)}$                      &&   $\wm    1.15(16)$   &&   $   -    5.8(1.8)$   &&   $\wm    1.15(18)$   &&   $   -    5.7(1.9)$   \\
  $\widetilde{h}_{\perp}^{(\frac{1}{2}^-)}$                      &&   $\wm   0.543(62)$   &&   $   -   2.12(67)$   &&   $\wm   0.544(75)$   &&   $   -   2.11(71)$   \\
  $f_0^{(\frac{3}{2}^-)}$                                 &&   $\wm    4.29(67)$   &&   $   -   27.3(8.7)$   &&   $\wm    4.31(75)$   &&   $   -   27.0(8.8)$   \\
  $f_+^{(\frac{3}{2}^-)}$                                 &&   $\wm  0.0498(70)$   &&   $\wm    1.28(27)$   &&   $\wm  0.0504(83)$   &&   $\wm    1.29(29)$   \\
  $f_{\perp}^{(\frac{3}{2}^-)}$                           &&   $   -  0.073(14)$   &&   $\wm    2.52(35)$   &&   $   -  0.073(14)$   &&   $\wm    2.54(39)$   \\
  $f_{\perp^{\prime}}^{(\frac{3}{2}^-)}$                      &&   $\wm  0.0687(40)$   &&   $   -  0.280(89)$   &&   $\wm  0.0687(59)$   &&   $   -  0.279(89)$   \\
  $g_0^{(\frac{3}{2}^-)}$                                 &&   $\wm  0.0027(35)$   &&   $\wm    1.23(21)$   &&   $\wm  0.0027(36)$   &&   $\wm    1.23(23)$   \\
  $g_+^{(\frac{3}{2}^-)}$                                 &&   $\wm    3.46(58)$   &&   $   -   24.7(8.1)$   &&   $\wm    3.47(64)$   &&   $   -   24.5(8.1)$   \\
  $g_{\perp}^{(\frac{3}{2}^-)}$                           &&   $\wm    3.47(57)$   &&   $   -   22.6(7.8)$   &&   $\wm    3.49(63)$   &&   $   -   22.4(7.9)$   \\
  $g_{\perp^{\prime}}^{(\frac{3}{2}^-)}$                      &&   $   -  0.062(38)$   &&   $\wm    0.62(57)$   &&   $   -  0.062(37)$   &&   $\wm    0.62(57)$   \\
  $h_+^{(\frac{3}{2}^-)}$                                 &&   $   -  0.124(16)$   &&   $\wm    2.51(32)$   &&   $   -  0.124(18)$   &&   $\wm    2.52(37)$   \\
  $h_{\perp}^{(\frac{3}{2}^-)}$                           &&   $\wm  0.0208(53)$   &&   $\wm    1.22(23)$   &&   $\wm  0.0210(60)$   &&   $\wm    1.22(25)$   \\
  $h_{\perp^{\prime}}^{(\frac{3}{2}^-)}$                      &&   $   - 0.0201(12)$   &&   $\wm   0.040(21)$   &&   $   - 0.0201(19)$   &&   $\wm   0.039(21)$   \\
  $\widetilde{h}_+^{(\frac{3}{2}^-)}$                      &&   $\wm    3.58(59)$   &&   $   -   23.7(8.1)$   &&   $\wm    3.59(66)$   &&   $   -   23.5(8.1)$   \\
  $\widetilde{h}_{\perp}^{(\frac{3}{2}^-)}$                      &&   $\wm    3.72(61)$   &&   $   -   25.1(8.2)$   &&   $\wm    3.74(69)$   &&   $   -   24.8(8.3)$   \\
  $\widetilde{h}_{\perp^{\prime}}^{(\frac{3}{2}^-)}$                      &&   $\wm   0.232(49)$   &&   $   -   0.60(52)$   &&   $\wm   0.235(56)$   &&   $   -   0.60(56)$   \\
  \hline\hline
 \end{tabular}
 \caption{\label{tab:FFparams}The parameters describing the $\Lambda_b \to \Lambda_c^*(2595)$ and $\Lambda_b \to \Lambda_c^*(2625)$ form factors at the physical pion mass and in the continuum limit. The nominal parameters $F^f$ and $A^f$ are used to evaluate the central values and statistical uncertainties, while the ``higher-order'' parameters $F^f_{\rm HO}$ and $A^f_{\rm HO}$ are used in combination with the nominal parameters to evaluate systematic uncertainties as explained in the main text. Files containing the parameter values and the covariance matrices are provided as supplemental material. }
\end{table}

\FloatBarrier
\section{\label{sec:ZRSR}Comparison with zero-recoil sum rules}
\FloatBarrier

At zero recoil ($w=1$), approximate sum-rule bounds on the size of heavy-to-heavy form factors can be derived using the operator product expansion and heavy-quark effective theory \cite{Shifman:1994jh,Bigi:1994ga,Gambino:2010bp,Gambino:2012rd,Mannel:2015osa,Boer:2018vpx}. In Ref.~\cite{Mannel:2015osa}, it was found that the lattice results for the $\Lambda_b \to \Lambda_c$ form factors with the $J^P=\frac12^+$ final state (which constitute the ``elastic'' contribution to the sum rule) almost completely saturate the bounds derived through order $1/m^3$, apparently leaving very little room for ``inelastic'' contributions from other final states such as the $\Lambda_c^*$'s considered here. However, in the case of $B$-meson decays, the size of $1/m^4$ and $1/m^5$ corrections has been found to be approximately 33\% of the size of the $1/m^2$ and $1/m^3$ terms \cite{Gambino:2012rd,Boer:2018vpx}. Allowing for effects of this size also for $\Lambda_b$ decays, the authors of Ref.~\cite{Boer:2018vpx} then obtained estimates of the size of the inelastic contributions, which are expected to be dominated by $\Lambda_b \to \Lambda_c^*(2595)$ and $\Lambda_b \to \Lambda_c^*(2625)$.

When expressed in terms of our form-factor definitions using the relations given in Appendix \ref{sec:FFrelationsBoer}, Eqs.~(46), (48), (50), and (52) of  Ref.~\cite{Boer:2018vpx} become
\begin{eqnarray}
 F_{\rm inel,1/2}&=& \big|f_{+}^{(\frac12^-)}\big|^2_{w=1} + 2\big|f_{\perp}^{(\frac12^-)}\big|^2_{w=1}, \\
 F_{\rm inel,3/2}&=& \frac16\left[ \frac{(m_{\Lambda_b}+m_{\Lambda_{c,3/2}^*})^2}{(m_{\Lambda_b}-m_{\Lambda_{c,3/2}^*})^2} \, \big|f_{+}^{(\frac32^-)}\big|^2 + 2\big|f_{\perp}^{(\frac32^-)}\big|^2+6\big|f_{\perp^\prime}^{(\frac32^-)}\big|^2   \right]_{w=1}, \\
 G_{\rm inel,1/2}&=& \frac13\, \big|g_{0}^{(\frac12^-)}\big|^2_{w=1}, \\
 G_{\rm inel,3/2}&=& \frac{1}{18}\,\frac{(m_{\Lambda_b}+m_{\Lambda_{c,3/2}^*})^2}{(m_{\Lambda_b}-m_{\Lambda_{c,3/2}^*})^2} \,\big|g_{0}^{(\frac32^-)}\big|^2_{w=1}.
\end{eqnarray}
The zero-recoil sum-rule estimate obtained in Ref.~\cite{Boer:2018vpx} is
\begin{eqnarray}
 F_{\rm inel,1/2}+F_{\rm inel,3/2}&\approx& 0.011^{+0.061}_{-0.055},\\
 G_{\rm inel,1/2}+G_{\rm inel,3/2}&\approx& 0.040^{+0.049}_{-0.052}.
\end{eqnarray}
Using our lattice-QCD results for the form factors, we find
\begin{eqnarray}
 F_{\rm inel,1/2}+F_{\rm inel,3/2}&=& 0.093\pm0.009_{\rm stat}\pm0.012_{\rm syst} ,\\
 G_{\rm inel,1/2}+G_{\rm inel,3/2}&=& 0.0162\pm0.0016_{\rm stat}\pm0.0020_{\rm syst}.
\end{eqnarray}
Thus, our result for the axial current falls within the range given in Ref.~\cite{Boer:2018vpx}, while our result for the vector current is slightly above the upper limit.

\FloatBarrier
\section{\texorpdfstring{$\bm{\Lambda_b \to \Lambda_c^*\ell^-\bar{\nu}}$}{Lambdab to Lambdac* l nubar} observables}
\label{sec:observables}
\FloatBarrier

\begin{figure}
 \includegraphics[width=0.47\linewidth]{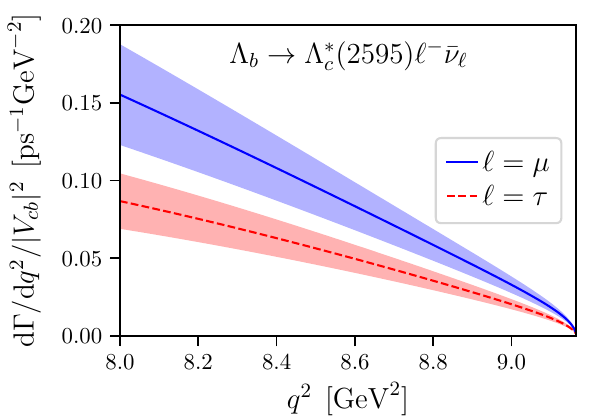} \hfill \includegraphics[width=0.47\linewidth]{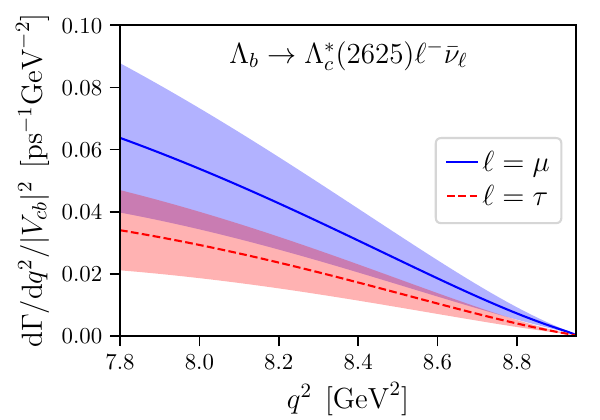}
 
 \includegraphics[width=0.47\linewidth]{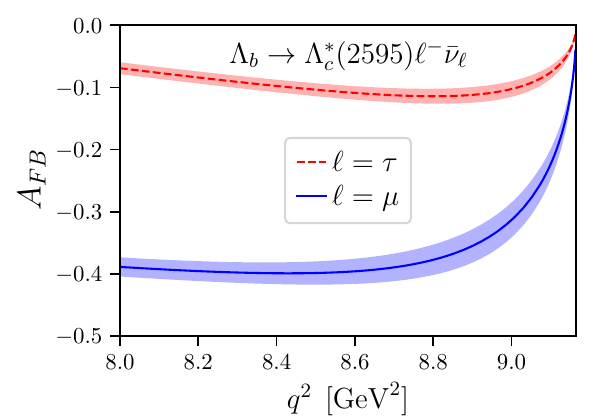} \hfill \includegraphics[width=0.47\linewidth]{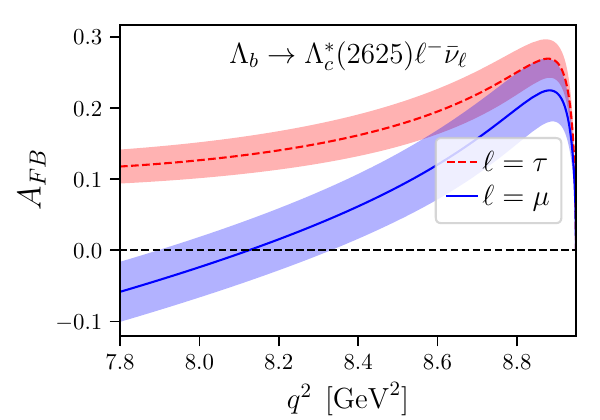}
 
 \includegraphics[width=0.47\linewidth]{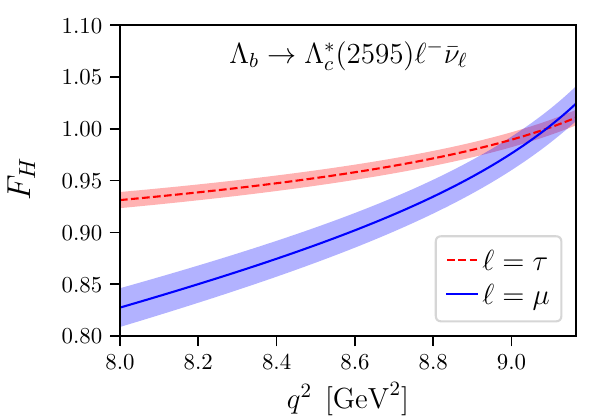} \hfill \includegraphics[width=0.47\linewidth]{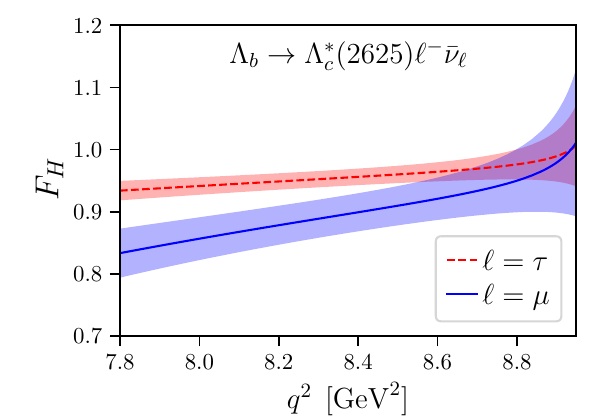}
 
 \caption{\label{fig:observables}$\Lambda_b \to \Lambda_c^*(2595)\ell^-\bar{\nu}$ (left) and $\Lambda_b \to \Lambda_c^*(2625)\ell^-\bar{\nu}$ (right) observables in the high-$q^2$ region calculated in the Standard Model using our form-factor results. From top to bottom: the differential decay rate divided by $|V_{cb}|^2$, the forward-backward asymmetry, and the flat term. In each case, we show results for $\ell=\mu$ and $\ell=\tau$ (the results for $\ell=e$ would look the same as for $\ell=\mu$ in this kinematic region). The bands indicate the total (statistical $+$ systematic) uncertainties.}
\end{figure}

\begin{figure}
 \includegraphics[width=0.6\linewidth]{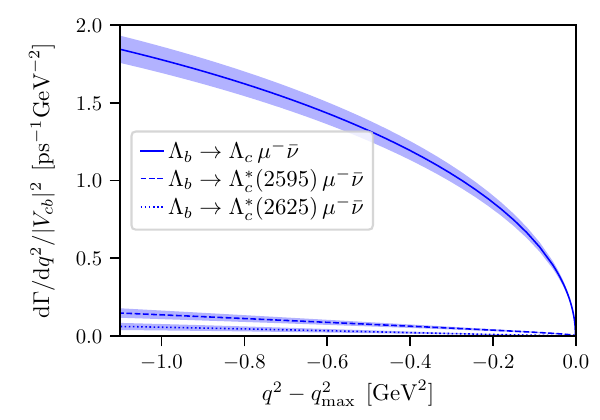}
 
 \caption{\label{fig:observables2}Comparison of the $\Lambda_b \to \Lambda_c\,\mu^-\bar{\nu}$, $\Lambda_b \to \Lambda_c^*(2595)\mu^-\bar{\nu}$, and $\Lambda_b \to \Lambda_c^*(2625)\mu^-\bar{\nu}$ differential decay rates just below $q^2_{\rm max}$, calculated in the Standard Model using the form factors from lattice QCD.}
\end{figure}

The two-fold differential decay rates of $\Lambda_b \to \Lambda_c^*(2595)\ell^-\bar{\nu}$ and $\Lambda_b \to \Lambda_c^*(2625)\ell^-\bar{\nu}$ in the Standard Model
can be written as
\begin{equation}
 \frac{\mathrm{d}^2\Gamma^{(J)}}{\mathrm{d}q^2\:\mathrm{d}\cos\theta_\ell} = A^{(J)} + B^{(J)} \cos\theta_\ell + C^{(J)} \cos^2\theta_\ell,
\end{equation}
where $\theta_\ell$ is the helicity angle of the charged lepton and $A^{(J)}$, $B^{(J)}$, $C^{(J)}$ are functions of $q^2$ only \cite{Boer:2018vpx}. The $J=\frac12,\frac32$ superscript is used to distinguish the $\Lambda_c^*(2595)$ and $\Lambda_c^*(2625)$ final states. The equations for $A^{(J)}$, $B^{(J)}$, and $C^{(J)}$ in terms of the form factors are given in Ref.~\cite{Boer:2018vpx} (where $A^{(J)} = \Gamma_0^{(\ell)}a_\ell^{(J)}$ etc.) and can be converted to our conventions using the relations in Appendix \ref{sec:FFrelationsBoer}. The integral over $\cos\theta_\ell$ yields the $q^2$-differential decay rate
\begin{equation}
 \frac{\mathrm{d}\Gamma^{(J)}}{\mathrm{d}q^2} = 2 A^{(J)} + \frac23 C^{(J)},
\end{equation}
and we also consider two angular observables \cite{Boer:2018vpx}: the forward-backward asymmetry
\begin{equation}
 A_{FB}^{(J)} = \frac{B^{(J)}}{\mathrm{d}\Gamma^{(J)}/\mathrm{d}q^2}
\end{equation}
and the ``flat term''
\begin{equation}
 F_{H}^{(J)} = \frac{2 (A^{(J)} + C^{(J)})}{\mathrm{d}\Gamma^{(J)}/\mathrm{d}q^2}.
\end{equation}
The Standard-Model predictions for $\mathrm{d}\Gamma^{(J)}/\mathrm{d}q^2/|V_{cb}|^2$ and for the angular observables using our form-factor results are shown in Fig.~\ref{fig:observables}. Note that at leading order in heavy-quark effective theory, the differential decay rate for the $J=\frac12$ final state would be a factor 2 smaller than the differential rate for $J=\frac32$, and the lepton-side angular observables considered here would be equal for both final states \cite{Leibovich:1997az,Boer:2018vpx}. In contrast, we find the $J=\frac12$ rate to be approximately 2.5 times larger than the $J=\frac32$ rate, and we find the forward-backward asymmetries to have opposite signs at high $q^2$. Leading-order HQET is of course expected to be inadequate for these decays, in which the light degrees of freedom in the final state have a different angular momentum than in the initial state. The forms of the subleading corrections are known \cite{Leibovich:1997az,Boer:2018vpx}, but we have not been able to obtain an acceptable HQET fit to the full set of form factors even when including these corrections, suggesting that sub-subleading terms may also be significant.

In Fig.~\ref{fig:observables2} we additionally compare the $\Lambda_b \to \Lambda_c^*(2595)\mu^-\bar{\nu}$, and $\Lambda_b \to \Lambda_c^*(2625)\mu^-\bar{\nu}$ differential decay rates with that of $\Lambda_b \to \Lambda_c\,\mu^-\bar{\nu}$, using the form factors from Ref.~\cite{Detmold:2015aaa} for the latter. For example, at $q^2=q^2_{\rm max}-1\:{\rm GeV}^2$, the $\Lambda_b \to \Lambda_c^*(2595)\mu^-\bar{\nu}$ differential decay rate is approximately 13 times smaller than the $\Lambda_b \to \Lambda_c\,\mu^-\bar{\nu}$ differential decay rate. Finally, recall that the CDF Collaboration has measured the total (\emph{i.e.}, integrated over all $q^2$) decay rates, and found the $\Lambda_b \to \Lambda_c^*(2625)\mu^-\bar{\nu}$ total rate to be approximately 1.7 times larger than the $\Lambda_b \to \Lambda_c^*(2595)\mu^-\bar{\nu}$ total rate \cite{Aaltonen:2008eu}. Since our results for the differential decay rates at high $q^2$ show the opposite behavior, the differential rates must cross at some value of $q^2$ lower than considered here.

\FloatBarrier
\section{\label{sec:conclusions}Conclusions}
\FloatBarrier

The decays $\Lambda_b \to \Lambda_c^*(2595)\ell^-\bar{\nu}$ and $\Lambda_b \to \Lambda_c^*(2625)\ell^-\bar{\nu}$ are interesting processes that deserve to be studied in detail, both experimentally and theoretically, to obtain a more complete picture of $b\to c \ell^-\bar{\nu}$ semileptonic decays. This work contributes to this goal by providing the first lattice-QCD determination of the complete set of form factors, albeit only in the vicinity of $q^2_{\rm max}$. The calculation was made possible by the technology developed initially for $\Lambda_b \to \Lambda^*(1520)$ \cite{Meinel:2020owd}: working in the rest frame of the $\Lambda_c^*$ to avoid mixing with unwanted quantum numbers, and using an interpolating field with gauge-covariant spatial derivatives to obtain a good overlap with the $\Lambda_c^*$.

In nature, the $\Lambda_c^*(2595)$ and $\Lambda_c^*(2625)$ are narrow resonances decaying through the strong interaction to $\Lambda_c \pi \pi$, with widths of $2.6(0.6) \:{\rm MeV}$ and $<\!\!0.97 \:{\rm MeV}$, respectively \cite{Zyla:2020zbs}. These values justify the use of the narrow-width approximation. In our lattice calculation with three different pion masses in the range from approximately 300 to 430 MeV, we find that the $\Lambda_c^*$ masses are below all possible strong-decay thresholds, including $\Sigma_c \pi$, except perhaps at the lowest pion mass. Simple chiral-continuum extrapolations of our lattice results for $m_{\Lambda_c^*(2595)}$ and $m_{\Lambda_c^*(2625)}$ yield values in agreement with experiment once systematic uncertainties are taken into account. The hyperfine splittings $m_{\Lambda_c^*(2625)}-m_{\Lambda_c^*(2595)}$ are also found to be consistent with experiment.

We use helicity-based definitions of the $\Lambda_b \to \Lambda_c^*(2595)$ and $\Lambda_b \to \Lambda_c^*(2625)$ form factors. On each ensemble we performed calculations for two different $\Lambda_b$ momenta corresponding to $w\approx 1.01$ and $w\approx 1.03$, where $w=v\cdot v^\prime$. The final results for the form factors, obtained from extrapolations to the continuum limit and physical pion mass, are parametrized as linear functions of $w$. These parametrizations are expected to be accurate only near the kinematic region where we have lattice data. Our results for the form factors at $w=1$ are compatible (albeit only marginally in the case of the vector form factors) with the zero-recoil sum-rules given in Ref.~\cite{Boer:2018vpx}. It will also be interesting to see the impact on unitarity bounds in global analyses of $b\to c \ell^- \bar{\nu}$ form factors \cite{Cohen:2019zev}.

Using our form-factor results, we evaluated the $\Lambda_b \to \Lambda_c^*(2595)\ell^-\bar{\nu}$ and $\Lambda_b \to \Lambda_c^*(2625)\ell^-\bar{\nu}$ differential decay rates, forward-backward asymmetry, and flat term in the Standard Model. We find the $\Lambda_b \to \Lambda_c^*(2595)\ell^-\bar{\nu}$ differential rates to be approximately 2.5 times higher than the $\Lambda_b \to \Lambda_c^*(2625)\ell^-\bar{\nu}$ differential rates (in the kinematic region considered; the CDF measurement of the total rates \cite{Aaltonen:2008eu} suggests that the ordering of the differential rates will switch at lower $q^2$), which is opposite to the behavior predicted by leading-order HQET but consistent with the expectation that subleading contributions in HQET are important for these types of decays. While not discussed in detail in this paper, we also attempted HQET fits at order $1/m$ \cite{Leibovich:1997az,Boer:2018vpx} to our form factor results, but we did not obtain an acceptable description. We expect that $1/m^2$ corrections, which have not yet been studied for these processes, are also large. This will make it more challenging to combine experimental data for the shapes of the muonic decay rates in the entire kinematic range with the lattice results for the form factors near $q^2_{\rm max}$ to obtain Standard-Model predictions for $R(\Lambda_c^*)=\mathcal{B}(\Lambda_b \to \Lambda_c^* \tau^- \bar{\nu})/\mathcal{B}(\Lambda_b \to \Lambda_c^* \mu^- \bar{\nu})$. Lattice calculations at lower $q^2$, while still working in the $\Lambda_c^*$ rest frame, could be performed using finer lattices or using a moving-NRQCD action \cite{Horgan:2009ti} for the $b$ quark. Alternatively, one could use nonzero $\Lambda_c^*$ momenta and explicitly deal with the mixing of quantum numbers by extracting multiple states using larger operator bases; see for example Refs.~\cite{Menadue:2013kfi,Silvi:2021uya}.

\FloatBarrier
\section*{Acknowledgments}
\FloatBarrier

We thank Marzia Bordone and Danny van Dyk for discussions, and the RBC and UKQCD Collaborations for making their gauge field ensembles available. SM is supported by the U.S. Department of Energy, Office of Science, Office of High Energy Physics under Award Number D{E-S}{C0}009913. GR is supported by the U.S. Department of Energy, Office of Science, Office of Nuclear Physics, under Contract No.~D{E-S}C0012704 (BNL). The computations for this work were carried out on facilities at the National Energy Research Scientific Computing Center, a DOE Office of Science User Facility supported by the Office of Science of the U.S. Department of Energy under Contract No. DE-AC02-05CH1123, and on facilities of the Extreme Science and Engineering Discovery Environment (XSEDE) \cite{XSEDE}, which is supported by National Science Foundation grant number ACI-1548562. We acknowledge the use of Chroma \cite{Edwards:2004sx,Chroma}, QPhiX \cite{JOO2015139,QPhiX}, QLUA \cite{QLUA}, MDWF \cite{MDWF}, and related USQCD software \cite{USQCD}.

\appendix

\FloatBarrier
\section{\label{sec:FFrelations}Relations between different form factor definitions}
\FloatBarrier

In this appendix, we provide expressions for the $\Lambda_b\to\Lambda_{c,1/2}^*$ and $\Lambda_b\to\Lambda_{c,3/2}^*$ form factors in other definitions used in the literature (for the vector and axial-vector currents only) in terms of our form factors. Note that the overall sign of the form factors for each decay mode depends on the phase conventions of the states. Thus, in the following relations, only the relative signs among the form factors for a specific final state are well-determined. To make this explicit, we introduce factors of $\sigma^{(J^P)}$ below, which can take on the values $\pm1$.

\subsection{Definition used by Leibovich and Stewart as well as Pervin, Roberts, and Capstick}

\noindent We find that the $\Lambda_b\to\Lambda_{c,1/2}^*$ form factor definitions in Ref.~\cite{Leibovich:1997az} are related to ours as
\begin{eqnarray}
 d_{V_1} &=& \sigma^{(\frac12^-)}_{\rm LS} f_\perp^{(\frac12^-)}, \\
 d_{V_2} &=& - \sigma^{(\frac12^-)}_{\rm LS} m_{\Lambda_b} \left[  \frac{m_{\Lambda_b}+m_{\Lambda_{c,1/2}^*}}{q^2}f_0^{(\frac12^-)} +  \frac{m_{\Lambda_b}-m_{\Lambda_{c,1/2}^*}}{s_-} \left( 1 - \frac{m_{\Lambda_b}^2-m_{\Lambda_{c,1/2}^*}^2}{q^2}  \right)f_+^{(\frac12^-)}  + \frac{2m_{\Lambda_{c,1/2}^*}}{s_-}f_\perp^{(\frac12^-)} \right], \\
 d_{V_3} &=& - \sigma^{(\frac12^-)}_{\rm LS} m_{\Lambda_{c,1/2}^*} \left[  -\frac{m_{\Lambda_b}+m_{\Lambda_{c,1/2}^*}}{q^2}f_0^{(\frac12^-)} +  \frac{m_{\Lambda_b}-m_{\Lambda_{c,1/2}^*}}{s_-} \left( 1 + \frac{m_{\Lambda_b}^2-m_{\Lambda_{c,1/2}^*}^2}{q^2}  \right)f_+^{(\frac12^-)}  - \frac{2m_{\Lambda_b}}{s_-}f_\perp^{(\frac12^-)} \right], \hspace{3ex}
\end{eqnarray}
\begin{eqnarray}
 d_{A_1} &=&  \sigma^{(\frac12^-)}_{\rm LS} g_\perp^{(\frac12^-)}, \\
 d_{A_2} &=& - \sigma^{(\frac12^-)}_{\rm LS}m_{\Lambda_b} \left[  - \frac{m_{\Lambda_b}-m_{\Lambda_{c,1/2}^*}}{q^2}g_0^{(\frac12^-)} -  \frac{m_{\Lambda_b}+m_{\Lambda_{c,1/2}^*}}{s_+} \left( 1 - \frac{m_{\Lambda_b}^2-m_{\Lambda_{c,1/2}^*}^2}{q^2}  \right)g_+^{(\frac12^-)}  + \frac{2m_{\Lambda_{c,1/2}^*}}{s_+}g_\perp^{(\frac12^-)} \right], \\
 d_{A_3} &=& -  \sigma^{(\frac12^-)}_{\rm LS} m_{\Lambda_{c,1/2}^*} \left[  \frac{m_{\Lambda_b}-m_{\Lambda_{c,1/2}^*}}{q^2}g_0^{(\frac12^-)} -  \frac{m_{\Lambda_b}+m_{\Lambda_{c,1/2}^*}}{s_+} \left( 1 + \frac{m_{\Lambda_b}^2-m_{\Lambda_{c,1/2}^*}^2}{q^2}  \right)g_+^{(\frac12^-)}  + \frac{2m_{\Lambda_b}}{s_+}g_\perp^{(\frac12^-)} \right].\hspace{3ex}
\end{eqnarray}
For $\Lambda_b\to\Lambda_{c,3/2}^*$, we find 
\begin{eqnarray}
 l_{V_1}&=&  \sigma^{(\frac32^-)}_{\rm LS} \frac{m_{\Lambda_b}m_{\Lambda_{c,3/2}^*}}{s_-} \bigg[f_\perp^{(\frac32^-)}+f_{\perp^\prime}^{(\frac32^-)}\bigg], \\
 \nonumber l_{V_2}&=& \sigma^{(\frac32^-)}_{\rm LS}\frac{m_{\Lambda_b}^2 m_{\Lambda_{c,3/2}^*}}{q^2 s_+ s_-}  \bigg[ (m_{\Lambda_b}-m_{\Lambda_{c,3/2}^*}) s_- f_0^{(\frac32^-)}  - 2 m_{\Lambda_{c,3/2}^*} q^2 (f_\perp^{(\frac32^-)}- f_{\perp^\prime}^{(\frac32^-)}) \\
 && \hspace{19ex} -\,  (m_{\Lambda_b}+m_{\Lambda_{c,3/2}^*}) \left(m_{\Lambda_b}^2-m_{\Lambda_{c,3/2}^*}^2-q^2\right)f_+^{(\frac32^-)}\bigg], \\
 \nonumber l_{V_3}&=& \sigma^{(\frac32^-)}_{\rm LS} \frac{m_{\Lambda_b}m_{\Lambda_{c,3/2}^*}}{q^2
   s_+ s_-}   \bigg[ -  m_{\Lambda_{c,3/2}^*} (m_{\Lambda_b}-m_{\Lambda_{c,3/2}^*}) s_- f_0^{(\frac32^-)} - 2  m_{\Lambda_b} m_{\Lambda_{c,3/2}^*}
   q^2 f_\perp^{(\frac32^-)} + 2 q^2 (m_{\Lambda_b} m_{\Lambda_{c,3/2}^*}-s_+) f_{\perp^\prime}^{(\frac32^-)} \\
   && \hspace{19ex} +\,  m_{\Lambda_{c,3/2}^*} (m_{\Lambda_b}+m_{\Lambda_{c,3/2}^*}) \left(m_{\Lambda_b}^2-m_{\Lambda_{c,3/2}^*}^2+q^2\right)f_+^{(\frac32^-)}\bigg], \\
 l_{V_4}&=& \sigma^{(\frac32^-)}_{\rm LS} f_{\perp^\prime}^{(\frac32^-)},
\end{eqnarray}
\begin{eqnarray}
 l_{A_1}&=& \sigma^{(\frac32^-)}_{\rm LS} \frac{ m_{\Lambda_b}m_{\Lambda_{c,3/2}^*}}{s_+} \bigg[g_\perp^{(\frac32^-)}+g_{\perp^\prime}^{(\frac32^-)}\bigg], \\
\nonumber l_{A_2}&=& \sigma^{(\frac32^-)}_{\rm LS} \frac{m_{\Lambda_b}^2 m_{\Lambda_{c,3/2}^*}}{q^2 s_+ s_-} \bigg[-(m_{\Lambda_b}+m_{\Lambda_{c,3/2}^*})s_+g_0^{(\frac32^-)} - 2 m_{\Lambda_{c,3/2}^*} q^2 (g_\perp^{(\frac32^-)}-g_{\perp^\prime}^{(\frac32^-)}) \\
&& \hspace{19ex} +\,  (m_{\Lambda_b}-m_{\Lambda_{c,3/2}^*}) \left(m_{\Lambda_b}^2-m_{\Lambda_{c,3/2}^*}^2-q^2\right)g_+^{(\frac32^-)}\bigg], \\
\nonumber l_{A_3}&=& \sigma^{(\frac32^-)}_{\rm LS} \frac{m_{\Lambda_b}m_{\Lambda_{c,3/2}^*}}{q^2 s_+ s_-} \bigg[ m_{\Lambda_{c,3/2}^*}  (m_{\Lambda_b}+m_{\Lambda_{c,3/2}^*})s_+g_0^{(\frac32^-)}  +  2  m_{\Lambda_b} m_{\Lambda_{c,3/2}^*}
   q^2 g_\perp^{(\frac32^-)} - 2 q^2 (m_{\Lambda_b} m_{\Lambda_{c,3/2}^*}+s_-)g_{\perp^\prime}^{(\frac32^-)} \\
   && \hspace{19ex} -\,  m_{\Lambda_{c,3/2}^*} (m_{\Lambda_b}-m_{\Lambda_{c,3/2}^*})\left(m_{\Lambda_b}^2-m_{\Lambda_{c,3/2}^*}^2+q^2\right)g_+^{(\frac32^-)} \bigg], \\
 l_{A_4}&=& \sigma^{(\frac32^-)}_{\rm LS} g_{\perp^\prime}^{(\frac32^-)}.
\end{eqnarray}
Pervin, Roberts, and Capstick \cite{Pervin:2005ve} use the same definitions as Leibovich and Stewart, with the name replacements $d_{V_i} \to F_i$, $d_{A_i} \to G_i$ for the $\frac12^-$ final state and $l_{V_i} \to F_i$, $l_{A_i} \to G_i$ for the $\frac32^-$ final state.

\subsection{Definition used by Gutsche et al.}

\noindent We find that the form factor definitions used in Refs.~\cite{Gutsche:2017wag,Gutsche:2018nks} are related to ours as follows:
\begin{eqnarray}
 F_1^{V(\frac12^-)} &=& \sigma^{(\frac12^-)}_{\rm G} \frac{ (m_{\Lambda_b}-m_{\Lambda_{c,1/2}^*})^2(f_\perp^{(\frac12^-)}-f_+^{(\frac12^-)})}{s_-}-f_\perp^{(\frac12^-)}, \\
 F_2^{V(\frac12^-)} &=& \sigma^{(\frac12^-)}_{\rm G} \frac{m_{\Lambda_b}  (m_{\Lambda_b}-m_{\Lambda_{c,1/2}^*})(f_\perp^{(\frac12^-)}-f_+^{(\frac12^-)})}{s_-}, \\
 F_3^{V(\frac12^-)} &=& \sigma^{(\frac12^-)}_{\rm G} \frac{m_{\Lambda_b} (m_{\Lambda_b}+m_{\Lambda_{c,1/2}^*}) \left( s_-f_0^{(\frac12^-)} +  q^2f_\perp^{(\frac12^-)} -  (m_{\Lambda_b}-m_{\Lambda_{c,1/2}^*})^2 f_+^{(\frac12^-)}\right)}{q^2 s_-}, \\
 F_1^{A(\frac12^-)} &=& \sigma^{(\frac12^-)}_{\rm G}\frac{ (m_{\Lambda_b}+m_{\Lambda_{c,1/2}^*})^2(g_\perp^{(\frac12^-)}-g_+^{(\frac12^-)})}{s_+}-g_\perp^{(\frac12^-)}, \\
 F_2^{A(\frac12^-)} &=& -\sigma^{(\frac12^-)}_{\rm G}\frac{m_{\Lambda_b}  (m_{\Lambda_b}+m_{\Lambda_{c,1/2}^*})(g_\perp^{(\frac12^-)}-g_+^{(\frac12^-)})}{s_+}, \\
 F_3^{A(\frac12^-)} &=& \sigma^{(\frac12^-)}_{\rm G}\frac{m_{\Lambda_b} (m_{\Lambda_b}-m_{\Lambda_{c,1/2}^*}) \left(- s_+g_0^{(\frac12^-)} - q^2g_\perp^{(\frac12^-)} +  (m_{\Lambda_b}+m_{\Lambda_{c,1/2}^*})^2 g_+^{(\frac12^-)}\right)}{q^2 s_+},
\end{eqnarray}
\begin{eqnarray}
 F_1^{V(\frac32^-)} &=& \sigma^{(\frac32^-)}_{\rm G} f_{\perp^\prime}^{(\frac32^-)},  \\
 F_2^{V(\frac32^-)} &=& \sigma^{(\frac32^-)}_{\rm G}\frac{m_{\Lambda_b} m_{\Lambda_{c,3/2}^*} }{s_-}\Big[f_\perp^{(\frac32^-)}+f_{\perp^\prime}^{(\frac32^-)}\Big], \\
\nonumber F_3^{V(\frac32^-)} &=& \sigma^{(\frac32^-)}_{\rm G}\frac{2\, m_{\Lambda_b}^2}{s_- s_+} \Big[- m_{\Lambda_{c,3/2}^*} (m_{\Lambda_b} + m_{\Lambda_{c,3/2}^*})f_\perp^{(\frac32^-)} + m_{\Lambda_{c,3/2}^*} (m_{\Lambda_b} + m_{\Lambda_{c,3/2}^*})(f_{\perp^\prime}^{(\frac32^-)}+f_+^{(\frac32^-)}) - s_+ f_{\perp^\prime}^{(\frac32^-)}\Big],  \\
&& \\
 \nonumber F_4^{V(\frac32^-)} &=& \sigma^{(\frac32^-)}_{\rm G}\frac{m_{\Lambda_b}^2 m_{\Lambda_{c,3/2}^*}}{q^2 s_+ s_-}  \Big[ (m_{\Lambda_b}-m_{\Lambda_{c,3/2}^*}) s_- f_0^{(\frac32^-)}  - 2 m_{\Lambda_{c,3/2}^*} q^2 (f_\perp^{(\frac32^-)}- f_{\perp^\prime}^{(\frac32^-)}), \\
 &&\hspace{18.5ex}-\,  (m_{\Lambda_b}+m_{\Lambda_{c,3/2}^*}) \left(m_{\Lambda_b}^2-m_{\Lambda_{c,3/2}^*}^2-q^2\right)f_+^{(\frac32^-)}\Big],
\end{eqnarray}
\begin{eqnarray}
 F_1^{A(\frac32^-)} &=& \sigma^{(\frac32^-)}_{\rm G}  g_{\perp^\prime}^{(\frac32^-)}  \\
 F_2^{A(\frac32^-)} &=& \sigma^{(\frac32^-)}_{\rm G} \frac{m_{\Lambda_b} m_{\Lambda_{c,3/2}^*} }{s_+}\Big[g_\perp^{(\frac32^-)}+g_{\perp^\prime}^{(\frac32^-)}\Big],  \\
 F_3^{A(\frac32^-)} &=& \sigma^{(\frac32^-)}_{\rm G} \frac{2\, m_{\Lambda_b}^2}{s_- s_+} \Big[m_{\Lambda_{c,3/2}^*} (m_{\Lambda_b}-m_{\Lambda_{c,3/2}^*}) (g_\perp^{(\frac32^-)}-g_{\perp^\prime}^{(\frac32^-)}-g_+^{(\frac32^-)})- s_-g_{\perp^\prime}^{(\frac32^-)}\Big], \\
 \nonumber F_4^{A(\frac32^-)} &=& \sigma^{(\frac32^-)}_{\rm G} \frac{m_{\Lambda_b}^2 m_{\Lambda_{c,3/2}^*}}{q^2 s_+ s_-} \Big[-(m_{\Lambda_b}+m_{\Lambda_{c,3/2}^*})s_+g_0^{(\frac32^-)} - 2 m_{\Lambda_{c,3/2}^*} q^2 (g_\perp^{(\frac32^-)}-g_{\perp^\prime}^{(\frac32^-)}), \\
 &&\hspace{19.5ex} +\, (m_{\Lambda_b}-m_{\Lambda_{c,3/2}^*}) \left(m_{\Lambda_b}^2-m_{\Lambda_{c,3/2}^*}^2-q^2\right)g_+^{(\frac32^-)}\Big].
\end{eqnarray}

\subsection{\label{sec:FFrelationsBoer}Definition used by B\"oer et al.}

\noindent Reference \cite{Boer:2018vpx} also uses a helicity-based definition, which we find to be related to ours as
\begin{eqnarray}
 f_{1/2,t} &=& \sigma^{(\frac12^-)}_{\rm B} \frac12\sqrt{\frac{3 s_-}{m_{\Lambda_b} m_{\Lambda_{c,1/2}^*}}}\frac{m_{\Lambda_b}+m_{\Lambda_{c,1/2}^*}}{m_{\Lambda_b}-m_{\Lambda_{c,1/2}^*}}\,f_0^{(\frac12^-)}, \\
 f_{1/2,0} &=& \sigma^{(\frac12^-)}_{\rm B} \frac12\sqrt{\frac{3 s_+}{m_{\Lambda_b} m_{\Lambda_{c,1/2}^*}}}\frac{m_{\Lambda_b}-m_{\Lambda_{c,1/2}^*}}{m_{\Lambda_b}+m_{\Lambda_{c,1/2}^*}}\,f_+^{(\frac12^-)}, \\
 f_{1/2,\perp} &=& \sigma^{(\frac12^-)}_{\rm B} \frac12\sqrt{\frac{3 s_+}{m_{\Lambda_b} m_{\Lambda_{c,1/2}^*}}}\,f_\perp^{(\frac12^-)}, \\
 g_{1/2,t} &=& \sigma^{(\frac12^-)}_{\rm B} \frac12\sqrt{\frac{3 s_+}{m_{\Lambda_b} m_{\Lambda_{c,1/2}^*}}}\frac{m_{\Lambda_b}-m_{\Lambda_{c,1/2}^*}}{m_{\Lambda_b}+m_{\Lambda_{c,1/2}^*}}\,g_0^{(\frac12^-)}, \\
 g_{1/2,0} &=& \sigma^{(\frac12^-)}_{\rm B} \frac12\sqrt{\frac{3 s_-}{m_{\Lambda_b} m_{\Lambda_{c,1/2}^*}}}\frac{m_{\Lambda_b}+m_{\Lambda_{c,1/2}^*}}{m_{\Lambda_b}-m_{\Lambda_{c,1/2}^*}}\,g_+^{(\frac12^-)}, \\
 g_{1/2,\perp} &=& \sigma^{(\frac12^-)}_{\rm B} \frac12\sqrt{\frac{3 s_-}{m_{\Lambda_b} m_{\Lambda_{c,1/2}^*}}}\,g_\perp^{(\frac12^-)}, 
\end{eqnarray}
\begin{eqnarray}
  F_{(1/2,t)}     &=&  \sigma^{(\frac32^-)}_{\rm B} \frac14\sqrt{\frac{s_-}{m_{\Lambda_b}m_{\Lambda_{c,3/2}^*}}} f_0^{(\frac32^-)} , \\
  F_{(1/2,0)}     &=&  \sigma^{(\frac32^-)}_{\rm B}\frac14\sqrt{\frac{s_+}{m_{\Lambda_b}m_{\Lambda_{c,3/2}^*}}} f_+^{(\frac32^-)}, \\
  F_{(1/2,\perp)} &=&  \sigma^{(\frac32^-)}_{\rm B}\frac14\sqrt{\frac{s_+}{m_{\Lambda_b}m_{\Lambda_{c,3/2}^*}}} f_\perp^{(\frac32^-)}, \\
  F_{(3/2,\perp)} &=& -\sigma^{(\frac32^-)}_{\rm B}\frac14\sqrt{\frac{s_+}{m_{\Lambda_b}m_{\Lambda_{c,3/2}^*}}} f_{\perp^\prime}^{(\frac32^-)}, \\
  G_{(1/2,t)}     &=&  \sigma^{(\frac32^-)}_{\rm B}\frac14\sqrt{\frac{s_+}{m_{\Lambda_b}m_{\Lambda_{c,3/2}^*}}} g_0^{(\frac32^-)} , \\
  G_{(1/2,0)}     &=&  \sigma^{(\frac32^-)}_{\rm B}\frac14\sqrt{\frac{s_-}{m_{\Lambda_b}m_{\Lambda_{c,3/2}^*}}}  g_+^{(\frac32^-)}, \\
  G_{(1/2,\perp)} &=&  \sigma^{(\frac32^-)}_{\rm B}\frac14\sqrt{\frac{s_-}{m_{\Lambda_b}m_{\Lambda_{c,3/2}^*}}} g_\perp^{(\frac32^-)}, \\
  G_{(3/2,\perp)} &=&  \sigma^{(\frac32^-)}_{\rm B}\frac14\sqrt{\frac{s_-}{m_{\Lambda_b}m_{\Lambda_{c,3/2}^*}}} g_{\perp^\prime}^{(\frac32^-)}.
\end{eqnarray}
We also independently derived the Eqs.~(B6) of Ref.~\cite{Boer:2018vpx} (arXiv version 2) which give the relations of the $\Lambda_b\to\Lambda_{c,3/2}^*$ form factors as defined in Ref.~\cite{Boer:2018vpx} to the definition used by Leibovich and Stewart \cite{Leibovich:1997az}. We agree with seven of the eight equations but find the opposite relative sign for $G_{1/2,0}$.

\FloatBarrier

\providecommand{\href}[2]{#2}\begingroup\raggedright\endgroup


\begin{thebibliography}{10}

\bibitem{Gambino:2020jvv}
P.~Gambino {\em et~al.}, ``{Challenges in semileptonic $B$ decays},''
  \href{http://dx.doi.org/10.1140/epjc/s10052-020-08490-x}{Eur. Phys. J. C
  {\bfseries 80} no.~10, (2020) 966},
  \href{http://arxiv.org/abs/2006.07287}{{\ttfamily arXiv:2006.07287
  [hep-ph]}}.

\bibitem{Bifani:2018zmi}
S.~Bifani, S.~Descotes-Genon, A.~Romero~Vidal, and M.-H. Schune, ``{Review of
  Lepton Universality tests in $B$ decays},''
  \href{http://dx.doi.org/10.1088/1361-6471/aaf5de}{J. Phys. G {\bfseries 46}
  no.~2, (2019) 023001}, \href{http://arxiv.org/abs/1809.06229}{{\ttfamily
  arXiv:1809.06229 [hep-ex]}}.

\bibitem{Bernlochner:2021vlv}
F.~U. Bernlochner, M.~F. Sevilla, D.~J. Robinson, and G.~Wormser,
  ``{Semitauonic $b$-hadron decays: A lepton flavor universality laboratory},''
  \href{http://arxiv.org/abs/2101.08326}{{\ttfamily arXiv:2101.08326
  [hep-ex]}}.

\bibitem{Neubert:1993mb}
M.~Neubert, ``{Heavy quark symmetry},''
  \href{http://dx.doi.org/10.1016/0370-1573(94)90091-4}{Phys. Rept. {\bfseries
  245} (1994) 259--396}, \href{http://arxiv.org/abs/hep-ph/9306320}{{\ttfamily
  arXiv:hep-ph/9306320}}.

\bibitem{Detmold:2015aaa}
W.~Detmold, C.~Lehner, and S.~Meinel, ``{$\Lambda_b \to p \ell^-
  \bar{\nu}_\ell$ and $\Lambda_b \to \Lambda_c \ell^- \bar{\nu}_\ell$ form
  factors from lattice QCD with relativistic heavy quarks},''
  \href{http://dx.doi.org/10.1103/PhysRevD.92.034503}{Phys. Rev. {\bfseries
  D92} no.~3, (2015) 034503},
\href{http://arxiv.org/abs/1503.01421}{{\ttfamily arXiv:1503.01421 [hep-lat]}}.
%%CITATION = ARXIV:1503.01421;%%.

\bibitem{Aaij:2015bfa}
{\bfseries LHCb} Collaboration, R.~Aaij {\em et~al.}, ``{Determination of the
  quark coupling strength $|V_{ub}|$ using baryonic decays},''
  \href{http://dx.doi.org/10.1038/nphys3415}{Nature Phys. {\bfseries 11} (2015)
  743--747}, \href{http://arxiv.org/abs/1504.01568}{{\ttfamily arXiv:1504.01568
  [hep-ex]}}.

\bibitem{Dutta:2015ueb}
R.~Dutta, ``{$\Lambda_b \to (\Lambda_c,\,p)\,\tau\,\nu$ decays within standard
  model and beyond},''
  \href{http://dx.doi.org/10.1103/PhysRevD.93.054003}{Phys. Rev. D {\bfseries
  93} no.~5, (2016) 054003}, \href{http://arxiv.org/abs/1512.04034}{{\ttfamily
  arXiv:1512.04034 [hep-ph]}}.

\bibitem{Li:2016pdv}
X.-Q. Li, Y.-D. Yang, and X.~Zhang, ``{$ {\varLambda}_b\to {\varLambda}_c\tau
  {\overline{\nu}}_{\tau } $ decay in scalar and vector leptoquark
  scenarios},'' \href{http://dx.doi.org/10.1007/JHEP02(2017)068}{JHEP
  {\bfseries 02} (2017) 068}, \href{http://arxiv.org/abs/1611.01635}{{\ttfamily
  arXiv:1611.01635 [hep-ph]}}.

\bibitem{Albrecht:2017odf}
J.~Albrecht, F.~Bernlochner, M.~Kenzie, S.~Reichert, D.~Straub, and A.~Tully,
  ``{Future prospects for exploring present day anomalies in flavour physics
  measurements with Belle II and LHCb},''
  \href{http://arxiv.org/abs/1709.10308}{{\ttfamily arXiv:1709.10308
  [hep-ph]}}.

\bibitem{Datta:2017aue}
A.~Datta, S.~Kamali, S.~Meinel, and A.~Rashed, ``{Phenomenology of $
  {\Lambda}_b\to {\Lambda}_c\tau {\overline{\nu}}_{\tau } $ using lattice QCD
  calculations},'' \href{http://dx.doi.org/10.1007/JHEP08(2017)131}{JHEP
  {\bfseries 08} (2017) 131}, \href{http://arxiv.org/abs/1702.02243}{{\ttfamily
  arXiv:1702.02243 [hep-ph]}}.

\bibitem{Alioli:2017ces}
S.~Alioli, V.~Cirigliano, W.~Dekens, J.~de~Vries, and E.~Mereghetti,
  ``{Right-handed charged currents in the era of the Large Hadron Collider},''
  \href{http://dx.doi.org/10.1007/JHEP05(2017)086}{JHEP {\bfseries 05} (2017)
  086}, \href{http://arxiv.org/abs/1703.04751}{{\ttfamily arXiv:1703.04751
  [hep-ph]}}.

\bibitem{Ray:2018hrx}
A.~Ray, S.~Sahoo, and R.~Mohanta, ``{Probing new physics in semileptonic
  $\Lambda_b$ decays},''
  \href{http://dx.doi.org/10.1103/PhysRevD.99.015015}{Phys. Rev. D {\bfseries
  99} no.~1, (2019) 015015}, \href{http://arxiv.org/abs/1812.08314}{{\ttfamily
  arXiv:1812.08314 [hep-ph]}}.

\bibitem{Boer:2019zmp}
{Böer, Philipp and Kokulu, Ahmet and Toelstede, Jan-Niklas and van Dyk,
  Danny}, ``{Angular Analysis of $\Lambda_b\to \Lambda_c (\to \Lambda
  \pi)\ell\bar\nu$},'' \href{http://dx.doi.org/10.1007/JHEP12(2019)082}{JHEP
  {\bfseries 12} (2019) 082}, \href{http://arxiv.org/abs/1907.12554}{{\ttfamily
  arXiv:1907.12554 [hep-ph]}}.

\bibitem{Penalva:2019rgt}
N.~Penalva, E.~Hern\'andez, and J.~Nieves, ``{Further tests of lepton flavour
  universality from the charged lepton energy distribution in $b\to c$
  semileptonic decays: The case of $\Lambda_b\to \Lambda_c \ell
  \bar\nu_\ell$},'' \href{http://dx.doi.org/10.1103/PhysRevD.100.113007}{Phys.
  Rev. D {\bfseries 100} no.~11, (2019) 113007},
  \href{http://arxiv.org/abs/1908.02328}{{\ttfamily arXiv:1908.02328
  [hep-ph]}}.

\bibitem{Ferrillo:2019owd}
M.~Ferrillo, A.~Mathad, P.~Owen, and N.~Serra, ``{Probing effects of new
  physics in $\Lambda^0_{b}\to\Lambda^+_{c}\mu^{-}\bar{\nu}_{\mu}$ decays},''
  \href{http://dx.doi.org/10.1007/JHEP12(2019)148}{JHEP {\bfseries 12} (2019)
  148}, \href{http://arxiv.org/abs/1909.04608}{{\ttfamily arXiv:1909.04608
  [hep-ph]}}.

\bibitem{Mu:2019bin}
X.-L. Mu, Y.~Li, Z.-T. Zou, and B.~Zhu, ``{Investigation of effects of new
  physics in $\Lambda_b\to\Lambda_c \tau\bar\nu_\tau$ decay},''
  \href{http://dx.doi.org/10.1103/PhysRevD.100.113004}{Phys. Rev. D {\bfseries
  100} no.~11, (2019) 113004},
  \href{http://arxiv.org/abs/1909.10769}{{\ttfamily arXiv:1909.10769
  [hep-ph]}}.

\bibitem{Hu:2020axt}
Q.-Y. Hu, X.-Q. Li, Y.-D. Yang, and D.-H. Zheng, ``{The measurable angular
  distribution of $\Lambda_b^0 \to \Lambda_c^+ (\to \Lambda^0 \pi^+)\tau^- (\to
  \pi^- \nu_\tau)\bar{\nu}_\tau$ decay},''
  \href{http://dx.doi.org/10.1007/JHEP02(2021)183}{JHEP {\bfseries 02} (2021)
  183}, \href{http://arxiv.org/abs/2011.05912}{{\ttfamily arXiv:2011.05912
  [hep-ph]}}.

\bibitem{Bowler:1997ej}
{\bfseries UKQCD} Collaboration, K.~C. Bowler, R.~D. Kenway, L.~Lellouch,
  J.~Nieves, O.~Oliveira, D.~G. Richards, C.~T. Sachrajda, N.~Stella, and
  P.~Ueberholz, ``{First lattice study of semileptonic decays of $\Lambda_b$
  and $\Xi_b$ baryons},''
  \href{http://dx.doi.org/10.1103/PhysRevD.57.6948}{Phys. Rev. D {\bfseries 57}
  (1998) 6948--6974}, \href{http://arxiv.org/abs/hep-lat/9709028}{{\ttfamily
  arXiv:hep-lat/9709028}}.

\bibitem{Gottlieb:2003yb}
S.~A. Gottlieb and S.~Tamhankar, ``{A Lattice study of $\Lambda_b$ semileptonic
  decay},'' \href{http://dx.doi.org/10.1016/S0920-5632(03)01612-8}{Nucl. Phys.
  B Proc. Suppl. {\bfseries 119} (2003) 644--646},
  \href{http://arxiv.org/abs/hep-lat/0301022}{{\ttfamily
  arXiv:hep-lat/0301022}}.

\bibitem{Aaij:2017svr}
{\bfseries LHCb} Collaboration, R.~Aaij {\em et~al.}, ``{Measurement of the
  shape of the $\Lambda_b^0\to\Lambda_c^+ \mu^- \overline{\nu}_{\mu}$
  differential decay rate},''
  \href{http://dx.doi.org/10.1103/PhysRevD.96.112005}{Phys. Rev. D {\bfseries
  96} no.~11, (2017) 112005}, \href{http://arxiv.org/abs/1709.01920}{{\ttfamily
  arXiv:1709.01920 [hep-ex]}}.

\bibitem{Bernlochner:2018kxh}
F.~U. Bernlochner, Z.~Ligeti, D.~J. Robinson, and W.~L. Sutcliffe, ``{New
  predictions for $\Lambda_b\to\Lambda_c$ semileptonic decays and tests of
  heavy quark symmetry},''
  \href{http://dx.doi.org/10.1103/PhysRevLett.121.202001}{Phys. Rev. Lett.
  {\bfseries 121} no.~20, (2018) 202001},
  \href{http://arxiv.org/abs/1808.09464}{{\ttfamily arXiv:1808.09464
  [hep-ph]}}.

\bibitem{Bernlochner:2018bfn}
F.~U. Bernlochner, Z.~Ligeti, D.~J. Robinson, and W.~L. Sutcliffe, ``{Precise
  predictions for $\Lambda_b \to \Lambda_c$ semileptonic decays},''
  \href{http://dx.doi.org/10.1103/PhysRevD.99.055008}{Phys. Rev. D {\bfseries
  99} no.~5, (2019) 055008}, \href{http://arxiv.org/abs/1812.07593}{{\ttfamily
  arXiv:1812.07593 [hep-ph]}}.

\bibitem{Aaltonen:2008eu}
{\bfseries CDF} Collaboration, T.~Aaltonen {\em et~al.}, ``{First Measurement
  of the Ratio of Branching Fractions $B(\Lambda^0_b \to \Lambda^+_{c} \mu^{-}
  \bar{\nu}_\mu) / B(\Lambda^0_b \to \Lambda^+_{c} \pi^{-})$},''
  \href{http://dx.doi.org/10.1103/PhysRevD.79.032001}{Phys. Rev. D {\bfseries
  79} (2009) 032001}, \href{http://arxiv.org/abs/0810.3213}{{\ttfamily
  arXiv:0810.3213 [hep-ex]}}.

\bibitem{Zyla:2020zbs}
{\bfseries Particle Data Group} Collaboration, P.~A. Zyla {\em et~al.},
  ``{Review of Particle Physics},''
  \href{http://dx.doi.org/10.1093/ptep/ptaa104}{PTEP {\bfseries 2020} no.~8,
  (2020) 083C01}.

\bibitem{Mannel:2015osa}
T.~Mannel and D.~van Dyk, ``{Zero-recoil sum rules for $\Lambda_b \to
  \Lambda_c$ form factors},''
  \href{http://dx.doi.org/10.1016/j.physletb.2015.10.016}{Phys. Lett. B
  {\bfseries 751} (2015) 48--53},
  \href{http://arxiv.org/abs/1506.08780}{{\ttfamily arXiv:1506.08780
  [hep-ph]}}.

\bibitem{Boer:2018vpx}
P.~Böer, M.~Bordone, E.~Graverini, P.~Owen, M.~Rotondo, and D.~Van~Dyk,
  ``{Testing lepton flavour universality in semileptonic $\Lambda_b \to
  \Lambda_c^*$ decays},'' \href{http://dx.doi.org/10.1007/JHEP06(2018)155}{JHEP
  {\bfseries 06} (2018) 155},
\href{http://arxiv.org/abs/1801.08367}{{\ttfamily arXiv:1801.08367 [hep-ph]}}.
%%CITATION = ARXIV:1801.08367;%%.

\bibitem{Cohen:2019zev}
T.~D. Cohen, H.~Lamm, and R.~F. Lebed, ``{Precision Model-Independent Bounds
  from Global Analysis of $b \to c \ell \nu$ Form Factors},''
  \href{http://dx.doi.org/10.1103/PhysRevD.100.094503}{Phys. Rev. D {\bfseries
  100} no.~9, (2019) 094503}, \href{http://arxiv.org/abs/1909.10691}{{\ttfamily
  arXiv:1909.10691 [hep-ph]}}.

\bibitem{Blechman:2003mq}
A.~E. Blechman, A.~F. Falk, D.~Pirjol, and J.~M. Yelton, ``{Threshold effects
  in excited charmed baryon decays},''
  \href{http://dx.doi.org/10.1103/PhysRevD.67.074033}{Phys. Rev. D {\bfseries
  67} (2003) 074033}, \href{http://arxiv.org/abs/hep-ph/0302040}{{\ttfamily
  arXiv:hep-ph/0302040}}.

\bibitem{Guo:2016wpy}
Z.-H. Guo and J.~A. Oller, ``{Resonance on top of thresholds: the
  $\Lambda_c(2595)^+$ as an extremely fine-tuned state},''
  \href{http://dx.doi.org/10.1103/PhysRevD.93.054014}{Phys. Rev. D {\bfseries
  93} no.~5, (2016) 054014}, \href{http://arxiv.org/abs/1601.00862}{{\ttfamily
  arXiv:1601.00862 [hep-ph]}}.

\bibitem{Arifi:2018yhr}
A.~J. Arifi, H.~Nagahiro, and A.~Hosaka, ``{Three-body decay of
  $\Lambda_c^{*}(2595)$ and $\Lambda_c^{*}(2625)$ with the inclusion of a
  direct two-pion coupling},''
  \href{http://dx.doi.org/10.1103/PhysRevD.98.114007}{Phys. Rev. D {\bfseries
  98} no.~11, (2018) 114007}, \href{http://arxiv.org/abs/1809.10290}{{\ttfamily
  arXiv:1809.10290 [hep-ph]}}.

\bibitem{Nieves:2019kdh}
J.~Nieves, R.~Pavao, and S.~Sakai, ``{$\Lambda _b$ decays into $\Lambda
  _c^*\ell \bar{\nu }_\ell $ and $\Lambda _c^*\pi ^-$ $[\Lambda _c^*=\Lambda
  _c(2595)$ and $\Lambda _c(2625)]$ and heavy quark spin symmetry},''
  \href{http://dx.doi.org/10.1140/epjc/s10052-019-6929-7}{Eur. Phys. J. C
  {\bfseries 79} no.~5, (2019) 417},
  \href{http://arxiv.org/abs/1903.11911}{{\ttfamily arXiv:1903.11911
  [hep-ph]}}.

\bibitem{Nieves:2019nol}
J.~Nieves and R.~Pavao, ``{Nature of the lowest-lying odd parity charmed baryon
  $\Lambda_c(2595)$ and $\Lambda_c(2625)$ resonances},''
  \href{http://dx.doi.org/10.1103/PhysRevD.101.014018}{Phys. Rev. D {\bfseries
  101} no.~1, (2020) 014018}, \href{http://arxiv.org/abs/1907.05747}{{\ttfamily
  arXiv:1907.05747 [hep-ph]}}.

\bibitem{Roberts:1992xm}
W.~Roberts, ``{Semileptonic decays of heavy Lambda's into excited baryons},''
  \href{http://dx.doi.org/10.1016/0550-3213(93)90331-I}{Nucl. Phys. B
  {\bfseries 389} (1993) 549--562}.

\bibitem{Leibovich:1997az}
A.~K. Leibovich and I.~W. Stewart, ``{Semileptonic $\Lambda_b$ decay to excited
  $\Lambda_c$ baryons at order $\Lambda_{\rm QCD} / m_Q$},''
  \href{http://dx.doi.org/10.1103/PhysRevD.57.5620}{Phys. Rev. {\bfseries D57}
  (1998) 5620--5631},
\href{http://arxiv.org/abs/hep-ph/9711257}{{\ttfamily arXiv:hep-ph/9711257
  [hep-ph]}}.
%%CITATION = HEP-PH/9711257;%%.

\bibitem{Pervin:2005ve}
M.~Pervin, W.~Roberts, and S.~Capstick, ``{Semileptonic decays of heavy
  $\Lambda$ baryons in a quark model},''
  \href{http://dx.doi.org/10.1103/PhysRevC.72.035201}{Phys. Rev. {\bfseries
  C72} (2005) 035201},
\href{http://arxiv.org/abs/nucl-th/0503030}{{\ttfamily arXiv:nucl-th/0503030
  [nucl-th]}}.
%%CITATION = NUCL-TH/0503030;%%.

\bibitem{Gutsche:2017wag}
T.~Gutsche, M.~A. Ivanov, J.~G. Körner, V.~E. Lyubovitskij, V.~V. Lyubushkin,
  and P.~Santorelli, ``{Theoretical description of the decays $\Lambda_b \to
  \Lambda^{(\ast)}(\frac12^\pm,\frac32^\pm) + J/\psi$},''
  \href{http://dx.doi.org/10.1103/PhysRevD.96.013003}{Phys. Rev. D {\bfseries
  96} no.~1, (2017) 013003}, \href{http://arxiv.org/abs/1705.07299}{{\ttfamily
  arXiv:1705.07299 [hep-ph]}}.

\bibitem{Gutsche:2018nks}
T.~Gutsche, M.~A. Ivanov, J.~G. Körner, V.~E. Lyubovitskij, P.~Santorelli, and
  C.-T. Tran, ``{Analyzing lepton flavor universality in the decays
  $\Lambda_b\to\Lambda_c^{(\ast)}(\frac12^\pm,\frac32^-) +
  \ell\,\bar\nu_\ell$},''
  \href{http://dx.doi.org/10.1103/PhysRevD.98.053003}{Phys. Rev. D {\bfseries
  98} no.~5, (2018) 053003}, \href{http://arxiv.org/abs/1807.11300}{{\ttfamily
  arXiv:1807.11300 [hep-ph]}}.

\bibitem{Becirevic:2020nmb}
D.~Be\v{c}irevi\'c, A.~Le~Yaouanc, V.~Mor\'enas, and L.~Oliver, ``{Heavy baryon
  wave functions, Bakamjian-Thomas approach to form factors, and observables in
  ${\Lambda_b \to \Lambda_c\left({1 \over 2}^\pm \right) \ell \overline{\nu}}$
  transitions},'' \href{http://dx.doi.org/10.1103/PhysRevD.102.094023}{Phys.
  Rev. D {\bfseries 102} no.~9, (2020) 094023},
  \href{http://arxiv.org/abs/2006.07130}{{\ttfamily arXiv:2006.07130
  [hep-ph]}}.

\bibitem{Meinel:2020owd}
S.~Meinel and G.~Rendon, ``{$\Lambda_b \to \Lambda^*(1520)\ell^+\ell^-$ form
  factors from lattice QCD},''
  \href{http://dx.doi.org/10.1103/PhysRevD.103.074505}{Phys. Rev. D {\bfseries
  103} no.~7, (2021) 074505}, \href{http://arxiv.org/abs/2009.09313}{{\ttfamily
  arXiv:2009.09313 [hep-lat]}}.

\bibitem{Feldmann:2011xf}
T.~Feldmann and M.~W.~Y. Yip, ``{Form factors for $\Lambda_b \to \Lambda$
  transitions in the soft-collinear effective theory},''
  \href{http://dx.doi.org/10.1103/PhysRevD.85.014035}{Phys. Rev. D {\bfseries
  85} (2012) 014035}, \href{http://arxiv.org/abs/1111.1844}{{\ttfamily
  arXiv:1111.1844 [hep-ph]}}. [Erratum: Phys.Rev.D 86, 079901 (2012)].

\bibitem{Aoki:2012xaa}
{\bfseries RBC, UKQCD} Collaboration, Y.~Aoki, N.~H. Christ, J.~M. Flynn,
  T.~Izubuchi, C.~Lehner, M.~Li, H.~Peng, A.~Soni, R.~S. Van~de Water, and
  O.~Witzel, ``{Nonperturbative tuning of an improved relativistic heavy-quark
  action with application to bottom spectroscopy},''
  \href{http://dx.doi.org/10.1103/PhysRevD.86.116003}{Phys. Rev. {\bfseries
  D86} (2012) 116003},
\href{http://arxiv.org/abs/1206.2554}{{\ttfamily arXiv:1206.2554 [hep-lat]}}.
%%CITATION = ARXIV:1206.2554;%%.

\bibitem{Brown:2014ena}
Z.~S. Brown, W.~Detmold, S.~Meinel, and K.~Orginos, ``{Charmed bottom baryon
  spectroscopy from lattice QCD},''
  \href{http://dx.doi.org/10.1103/PhysRevD.90.094507}{Phys. Rev. D {\bfseries
  90} no.~9, (2014) 094507}, \href{http://arxiv.org/abs/1409.0497}{{\ttfamily
  arXiv:1409.0497 [hep-lat]}}.

\bibitem{Aoki:2010dy}
{\bfseries RBC, UKQCD} Collaboration, Y.~Aoki {\em et~al.}, ``{Continuum Limit
  Physics from 2+1 Flavor Domain Wall QCD},''
  \href{http://dx.doi.org/10.1103/PhysRevD.83.074508}{Phys. Rev. {\bfseries
  D83} (2011) 074508},
\href{http://arxiv.org/abs/1011.0892}{{\ttfamily arXiv:1011.0892 [hep-lat]}}.
%%CITATION = ARXIV:1011.0892;%%.

\bibitem{Blum:2014tka}
{\bfseries RBC, UKQCD} Collaboration, T.~Blum {\em et~al.}, ``{Domain wall QCD
  with physical quark masses},''
  \href{http://dx.doi.org/10.1103/PhysRevD.93.074505}{Phys. Rev. {\bfseries
  D93} no.~7, (2016) 074505},
\href{http://arxiv.org/abs/1411.7017}{{\ttfamily arXiv:1411.7017 [hep-lat]}}.
%%CITATION = ARXIV:1411.7017;%%.

\bibitem{Blum:2012uh}
T.~Blum, T.~Izubuchi, and E.~Shintani, ``{New class of variance-reduction
  techniques using lattice symmetries},''
  \href{http://dx.doi.org/10.1103/PhysRevD.88.094503}{Phys. Rev. {\bfseries
  D88} no.~9, (2013) 094503},
\href{http://arxiv.org/abs/1208.4349}{{\ttfamily arXiv:1208.4349 [hep-lat]}}.
%%CITATION = ARXIV:1208.4349;%%.

\bibitem{Shintani:2014vja}
E.~Shintani, R.~Arthur, T.~Blum, T.~Izubuchi, C.~Jung, and C.~Lehner,
  ``{Covariant approximation averaging},''
  \href{http://dx.doi.org/10.1103/PhysRevD.91.114511}{Phys. Rev. {\bfseries
  D91} no.~11, (2015) 114511},
\href{http://arxiv.org/abs/1402.0244}{{\ttfamily arXiv:1402.0244 [hep-lat]}}.
%%CITATION = ARXIV:1402.0244;%%.

\bibitem{Johnson:1982yq}
R.~C. Johnson, ``{Angular momentum on a lattice},''
\href{http://dx.doi.org/10.1016/0370-2693(82)90134-4}{Phys. Lett. {\bfseries
  114B} (1982) 147--151}.
%%CITATION = PHLTA,114B,147;%%.

\bibitem{Oller:2000fj}
J.~A. Oller and U.~G. Meissner, ``{Chiral dynamics in the presence of bound
  states: Kaon nucleon interactions revisited},''
  \href{http://dx.doi.org/10.1016/S0370-2693(01)00078-8}{Phys. Lett. B
  {\bfseries 500} (2001) 263--272},
  \href{http://arxiv.org/abs/hep-ph/0011146}{{\ttfamily arXiv:hep-ph/0011146}}.

\bibitem{Hashimoto:1999yp}
S.~Hashimoto, A.~X. El-Khadra, A.~S. Kronfeld, P.~B. Mackenzie, S.~M. Ryan, and
  J.~N. Simone, ``{Lattice QCD calculation of $\bar{B} \to D \ell \bar{\nu}$
  decay form-factors at zero recoil},''
  \href{http://dx.doi.org/10.1103/PhysRevD.61.014502}{Phys. Rev. {\bfseries
  D61} (1999) 014502},
\href{http://arxiv.org/abs/hep-ph/9906376}{{\ttfamily arXiv:hep-ph/9906376
  [hep-ph]}}.
%%CITATION = HEP-PH/9906376;%%.

\bibitem{ElKhadra:2001rv}
A.~X. El-Khadra, A.~S. Kronfeld, P.~B. Mackenzie, S.~M. Ryan, and J.~N. Simone,
  ``{The Semileptonic decays $B \to \pi \ell \nu$ and $D \to \pi \ell \nu$ from
  lattice QCD},'' \href{http://dx.doi.org/10.1103/PhysRevD.64.014502}{Phys.
  Rev. {\bfseries D64} (2001) 014502},
\href{http://arxiv.org/abs/hep-ph/0101023}{{\ttfamily arXiv:hep-ph/0101023
  [hep-ph]}}.
%%CITATION = HEP-PH/0101023;%%.

\bibitem{Detmold:2016pkz}
W.~Detmold and S.~Meinel, ``{$\Lambda_b \to \Lambda \ell^+ \ell^-$ form
  factors, differential branching fraction, and angular observables from
  lattice QCD with relativistic $b$ quarks},''
  \href{http://dx.doi.org/10.1103/PhysRevD.93.074501}{Phys. Rev. {\bfseries
  D93} no.~7, (2016) 074501},
\href{http://arxiv.org/abs/1602.01399}{{\ttfamily arXiv:1602.01399 [hep-lat]}}.
%%CITATION = ARXIV:1602.01399;%%.

\bibitem{Shifman:1994jh}
M.~A. Shifman, N.~G. Uraltsev, and A.~I. Vainshtein, ``{$|V_{cb}|$ from OPE sum
  rules for heavy flavor transitions},''
  \href{http://dx.doi.org/10.1103/PhysRevD.52.3149}{Phys. Rev. D {\bfseries 51}
  (1995) 2217}, \href{http://arxiv.org/abs/hep-ph/9405207}{{\ttfamily
  arXiv:hep-ph/9405207}}. [Erratum: Phys.Rev.D 52, 3149 (1995)].

\bibitem{Bigi:1994ga}
I.~I.~Y. Bigi, M.~A. Shifman, N.~G. Uraltsev, and A.~I. Vainshtein, ``{Sum
  rules for heavy flavor transitions in the SV limit},''
  \href{http://dx.doi.org/10.1103/PhysRevD.52.196}{Phys. Rev. D {\bfseries 52}
  (1995) 196--235}, \href{http://arxiv.org/abs/hep-ph/9405410}{{\ttfamily
  arXiv:hep-ph/9405410}}.

\bibitem{Gambino:2010bp}
P.~Gambino, T.~Mannel, and N.~Uraltsev, ``{$B \to D^*$ at zero recoil
  revisited},'' \href{http://dx.doi.org/10.1103/PhysRevD.81.113002}{Phys. Rev.
  D {\bfseries 81} (2010) 113002},
  \href{http://arxiv.org/abs/1004.2859}{{\ttfamily arXiv:1004.2859 [hep-ph]}}.

\bibitem{Gambino:2012rd}
P.~Gambino, T.~Mannel, and N.~Uraltsev, ``{$B \to D^*$ Zero-Recoil Formfactor
  and the Heavy Quark Expansion in QCD: A Systematic Study},''
  \href{http://dx.doi.org/10.1007/JHEP10(2012)169}{JHEP {\bfseries 10} (2012)
  169}, \href{http://arxiv.org/abs/1206.2296}{{\ttfamily arXiv:1206.2296
  [hep-ph]}}.

\bibitem{Horgan:2009ti}
R.~Horgan {\em et~al.}, ``{Moving NRQCD for heavy-to-light form factors on the
  lattice},'' \href{http://dx.doi.org/10.1103/PhysRevD.80.074505}{Phys. Rev. D
  {\bfseries 80} (2009) 074505},
  \href{http://arxiv.org/abs/0906.0945}{{\ttfamily arXiv:0906.0945 [hep-lat]}}.

\bibitem{Menadue:2013kfi}
F.~M. Stokes, W.~Kamleh, D.~B. Leinweber, M.~S. Mahbub, B.~J. Menadue, and
  B.~J. Owen, ``{Parity-expanded variational analysis for nonzero momentum},''
  \href{http://dx.doi.org/10.1103/PhysRevD.92.114506}{Phys. Rev. D {\bfseries
  92} no.~11, (2015) 114506}, \href{http://arxiv.org/abs/1302.4152}{{\ttfamily
  arXiv:1302.4152 [hep-lat]}}.

\bibitem{Silvi:2021uya}
G.~Silvi {\em et~al.}, ``{$P$-wave nucleon-pion scattering amplitude in the
  $\Delta$(1232) channel from lattice QCD},''
  \href{http://dx.doi.org/10.1103/PhysRevD.103.094508}{Phys. Rev. D {\bfseries
  103} no.~9, (2021) 094508}, \href{http://arxiv.org/abs/2101.00689}{{\ttfamily
  arXiv:2101.00689 [hep-lat]}}.

\bibitem{XSEDE}
J.~{Towns}, T.~{Cockerill}, M.~{Dahan}, I.~{Foster}, K.~{Gaither},
  A.~{Grimshaw}, V.~{Hazlewood}, S.~{Lathrop}, D.~{Lifka}, G.~D. {Peterson},
  R.~{Roskies}, J.~R. {Scott}, and N.~{Wilkins-Diehr}, ``{XSEDE: Accelerating
  Scientific Discovery},''
  \href{http://dx.doi.org/10.1109/MCSE.2014.80}{Computing in Science
  Engineering {\bfseries 16} no.~5, (2014) 62--74}.

\bibitem{Edwards:2004sx}
{\bfseries SciDAC, LHPC, UKQCD} Collaboration, R.~G. Edwards and B.~Joo, ``{The
  Chroma software system for lattice QCD},''
  \href{http://dx.doi.org/10.1016/j.nuclphysbps.2004.11.254}{Nucl. Phys. B
  Proc. Suppl. {\bfseries 140} (2005) 832},
  \href{http://arxiv.org/abs/hep-lat/0409003}{{\ttfamily
  arXiv:hep-lat/0409003}}.

\bibitem{Chroma}
R.~G. Edwards, B.~Joó, {\em et~al.}, ``{Chroma}.''
\newblock \url{https://github.com/JeffersonLab/chroma}.

\bibitem{JOO2015139}
B.~Joó, M.~Smelyanskiy, D.~D. Kalamkar, and K.~Vaidyanathan,
  \href{http://dx.doi.org/10.1016/B978-0-12-803819-2.00023-9}{``{Chapter 9 -
  Wilson Dslash Kernel From Lattice QCD Optimization},''} in {\em {High
  Performance Parallelism Pearls}}, pp.~139 -- 170.
\newblock Morgan Kaufmann, Boston, 2015.

\bibitem{QPhiX}
B.~Joó {\em et~al.}, ``{QPhiX Dslash and Solver Library}.''
\newblock \url{https://github.com/jeffersonlab/qphix}.

\bibitem{QLUA}
A.~Pochinsky, S.~Syritsyn, {\em et~al.}, ``{QLUA}.''
\newblock \url{https://usqcd.lns.mit.edu/w/index.php/QLUA}.

\bibitem{MDWF}
A.~Pochinsky, S.~Syritsyn, {\em et~al.}, ``{Möbius Domain Wall inverter}.''
\newblock \url{https://github.com/usqcd-software/mdwf}.

\bibitem{USQCD}
{\bfseries USQCD} Collaboration, ``{USQCD Software}.''
\newblock \url{http://usqcd-software.github.io}.

\end{thebibliography}
\end{document}